\documentclass[fleqn,usenatbib]{mnras}
\usepackage[T1]{fontenc}
\usepackage{ae,aecompl}
\usepackage{graphicx}
\usepackage{natbib}
\usepackage{mathtools}
\pdfoutput=1
\usepackage{amsmath}
\usepackage{amssymb}
\usepackage{threeparttable}
\usepackage{float}
\usepackage{epstopdf}
\usepackage{newtxtext,newtxmath}

\let\oldAA\AA
\renewcommand{\AA}{\text{\normalfont\oldAA}}

\newcommand{\lya}{Ly$\alpha$}
\newcommand{\lyas}{Ly$\alpha$\ }
\newcommand{\Lya}{Ly$\alpha$}
\newcommand{\Lyas}{Ly$\alpha$\ }
\newcommand{\fesc}{$f_\mathrm{esc}$}
\newcommand{\fescs}{$f_\mathrm{esc}$\ }
\newcommand{\fcens}{$f_{\rm{cen}}$\ }
\newcommand{\fcen}{$f_{\rm{cen}}$}
\newcommand{\HeIIL}{\hbox{{\rm He}\kern 0.1em{\sc ii}\kern 0.1em{$\lambda1640$}}}
\newcommand{\HeII}{\hbox{{\rm He}\kern 0.1em{\sc ii}}}
\newcommand{\HeIIl}{\hbox{{\rm He}\kern 0.1em{\sc ii}\kern 0.1em{$\lambda1640$}}}
\newcommand{\CIIIL}{\hbox{{\rm C}\kern 0.1em{\sc iii}]\kern 0.1em{$\lambda1907,\lambda1909$}}}
\newcommand{\CIVL}{\hbox{{\rm C}\kern 0.1em{\sc iv}\kern 0.1em{$\lambda1548,\lambda1550$}}}
\newcommand{\CIII}{\hbox{{\rm C}\kern 0.1em{\sc iii}]}}
\newcommand{\MgII}{\hbox{{\rm Mg}\kern 0.1em{\sc ii}}}
\newcommand{\NeIII}{[\hbox{{\rm Ne}\kern 0.1em{\sc iii}}]}

\newcommand{\OIII}{[\hbox{{\rm O}\kern 0.1em{\sc iii}}]}
\newcommand{\OII}{[\hbox{{\rm O}\kern 0.1em{\sc ii}}]}
\newcommand{\CIV}{\hbox{{\rm C}\kern 0.1em{\sc iv}}}
\newcommand{\CII}{[\hbox{{\rm C}\kern 0.1em{\sc ii}}]}

\title[The XLS-$z$2 Survey: Synced Production \& Escape of LyC photons in LAEs]{The Synchrony of Production \& Escape: Half the Bright Ly$\alpha$ Emitters at $z\approx2$ have Lyman Continuum Escape Fractions $\approx50\%$}

\author [R.P.~Naidu \& J.~Matthee et al.]{
\parbox[t]{\textwidth}{
Rohan P. Naidu$^{1}$\thanks{E-mail: rohan.naidu@cfa.harvard.edu, mattheej@phys.ethz.ch}\thanks{These authors contributed equally to this work.}, 
Jorryt Matthee$^{2}$\footnotemark[1]\footnotemark[2]\thanks{Zwicky Fellow}, Pascal A. Oesch$^{3,4}$, Charlie Conroy$^{1}$, David Sobral$^{5}$, Gabriele Pezzulli$^{6}$, Matthew Hayes$^{7}$, Dawn Erb$^{8}$, Ricardo Amor\'in$^{9,10}$, Max Gronke$^{11,12,13}$, Daniel Schaerer$^{3}$, Sandro Tacchella$^{14}$, Josephine Kerutt$^{3}$, Ana Paulino-Afonso$^{15}$, Jo\~ao Calhau$^{16,17}$, Mario Llerena$^{9,10}$ and Huub R\"ottgering$^{18}$}
\vspace*{8pt}\\
$^{1}$Center for Astrophysics $|$ Harvard \& Smithsonian, 60 Garden Street, Cambridge, MA 02138, USA\\
$^{2}$Department of Physics, ETH Z\"urich, Wolfgang-Pauli-Strasse 27, 8093 Z\"urich, Switzerland\\
See Appendix \ref{appendix:affiliations} for the other authors' affiliations.
}

\pubyear{2021}

\begin{document}
\label{firstpage}
\pagerange{\pageref{firstpage}--\pageref{lastpage}}
\maketitle

\begin{abstract}
The ionizing photon escape fraction (LyC $f_{\rm{esc}}$) of star-forming galaxies is the single greatest unknown in the reionization budget. Stochastic sightline effects prohibit the direct separation of LyC leakers from non-leakers at significant redshifts. Here we circumvent this uncertainty by inferring $f_{\rm{esc}}$ using resolved ($R>4000$) Ly$\alpha$ profiles from the X-SHOOTER Ly$\alpha$ survey at $z=2$ (XLS-$z$2). With empirically motivated criteria, we use Ly$\alpha$ profiles to select leakers (\fesc$>20\%$) and non-leakers (\fesc$<5\%$) from a representative sample of $>0.2 L^{*}$ Lyman-$\alpha$ emitters (LAEs). We use median stacked spectra of these subsets over $\lambda_{\rm{rest}}\approx1000-8000\AA$ to investigate the conditions for LyC \fesc. Our stacks show similar mass, metallicity, $M_{\rm{UV}}$, and $\beta_{\rm{UV}}$. We find the following differences between leakers vs. non-leakers: (i) strong nebular CIV and HeII emission vs. non-detections, (ii) \OIII/\OII$\approx8.5$ vs. $\approx3$, (iii) H$\alpha$/H$\beta$ indicating no dust vs. $E(B-V)\approx0.3$, (iv) MgII emission close to the systemic velocity vs. redshifted, optically thick MgII, (v) Ly$\alpha$ \fescs of $\approx50\%$ vs. $\approx10\%$. The extreme EWs in leakers (\OIII+H$\beta\approx1100\ \AA$ rest-frame) constrain the characteristic timescale of LyC escape to $\approx3-10$ Myr bursts when short-lived stars with the hardest ionizing spectra shine. The defining traits of leakers -- extremely ionizing stellar populations, low column densities, a dust-free, high ionization state ISM -- occur simultaneously in the \fesc$>20\%$ stack, suggesting they are causally connected, and motivating why indicators like \OIII/\OII\ may suffice to constrain \fescs at $z>6$ with \textit{JWST}. The leakers comprise half our sample, have a median LyC \fesc$\approx50\%$ (conservative range: $20-55\%$), and an ionising production efficiency $\log({\xi_{\rm{ion}}/\rm{Hz\ erg^{-1}}})\approx25.9$ (conservative range: $25.7-25.9$). These results show LAEs -- the type of galaxies rare at $z\approx2$, but that become the norm at higher redshift -- are highly efficient ionizers, with extreme $\xi_{\rm{ion}}$ and prolific \fescs occurring in sync.
\end{abstract}


\section{Introduction}
\label{sec:intro}
The Epoch of Reionization (EoR) was the last major phase transition of the Universe, when the first stars and galaxies announced their presence by ionizing the vast oceans of neutral Hydrogen (HI) they were born within \citep[e.g,][]{Loeb13}. While the timeline of reionization is increasingly well-constrained ($z\approx6-9$, e.g., \citealt{Fan06,Planck18,Mason19}), the protagonists of reionization remain elusive. Quasars, due to their rapidly fading numbers with increasing redshift, are unlikely to have played a significant role \citep[e.g.,][]{Matsuoka18,Kulkarni19, Shen20}. Star-forming galaxies are the likeliest candidates, but whether the ionizing photon budget arose from a multitude of ultra-faint galaxies (``democratic reionization", e.g., \citealt{Finkelstein19}) or a rarer set of bright galaxies (``reionization by oligarchs", e.g., \citealt{Naidu20}) is a key open question with wide-ranging physical (e.g., reionization topology) and practical (e.g., survey design) implications \citep[][]{Hutter21}.

The ionizing photon budget is typically parametrized as a product of three quantities \citep[e.g.,][]{Madau99,Robertson13,duncan&conselice15} -- the UV star-formation density ($\rho_{\rm{UV}}$), a conversion factor between the UV luminosity and number of ionizing photons ($\xi_{\rm{ion}}$), and the fraction of these photons that make it to the intergalactic medium (IGM) to ionize it (\fesc). $\rho_{\rm{UV}}$ is well-constrained down to $M_{\rm{UV}}\approx-15$ during the EoR \citep[e.g.,][]{Bouwens21} with a clear path to fainter magnitudes with \textit{JWST} \citep[e.g.,][]{Labbe21}. Prospects of constraining $\xi_{\rm{ion}}$ are also bright \citep[e.g.,][]{Chevallard18}. On the other hand, due to the opacity of the intervening intergalactic medium (IGM), \fescs may never be directly observed at $z\gtrapprox4$ \citep[e.g.,][]{Inoue14}. Progress must therefore rely on measuring and understanding \fescs at lower redshifts.

In recent years, direct \fescs studies have largely concentrated on two redshift windows set by available UV instrumentation -- one at $z\approx0.3$ where Lyman continuum (LyC) is accessible to \textit{HST}/COS \citep[e.g.,][]{Izotov16b,Izotov18b, Izotov21,Wang19,Wang21}, and another at $z\approx2-4$ accessible to ground-based facilities and \textit{HST}/WFC3 UVIS \citep[e.g.,][]{Jones18, Smith20, Ji20, Mestric20,Marqueschaves21,Davis21,Prichard21}. The $z\approx0.3$ COS efforts were first undertaken at a time when only a handful of robust LyC leakers had been identified, and it was unclear whether LyC leakage even occurred among the $\gtrapprox 0.5 L^{*}$ galaxies for which \fescs measurements were feasible \citep[e.g.,][]{Izotov16b,Izotov16a}. Their selection functions prioritized rare galaxies with a high theorized probability of non-zero \fescs (e.g., compact, extreme \OIII/\OII, elevated H$\beta$ EW starbursts, i.e., ``Green Peas"). These programs have been remarkably successful in proving \fescs does occur among fairly luminous galaxies and in producing a sample of $\approx20$ galaxies with robust LyC constraints \citep[e.g.,][]{Izotov16b,Izotov18b,Izotov21}. However, the complex selection function and unknown number densities make generalizing these findings to higher redshifts and into the EoR difficult.

LyC studies at higher redshifts ($z\approx2-4$) have simpler selection functions, but are hampered by drastic IGM line of sight variations \citep[e.g.,][]{Inoue14}. Ideally, we would like to perform a controlled comparative experiment by constructing leaker and non-leaker stacks, and then contrasting their features to isolate indicators of \fesc. However, it is generally difficult to decide whether any individual detection/non-detection is due to high/low \fescs or due to a particularly transparent/opaque line of sight. To put numbers to the scale of the problem -- for a randomly sampled IGM sightline, the difference between the 10$^{\rm{th}}$ and 90$^{\rm{th}}$ percentile transmission is $>50\times$ at $z\approx3$ (0.01 vs. 0.60, \citealt{Steidel18}). Selecting apparent high \fescs and low \fescs leakers by applying mean IGM corrections amounts to comparing galaxies lying along transparent sightlines vs. opaque sightlines rather than high \fescs vs. low \fescs sources \citep[e.g.,][]{Bassett21}. These ambiguities due to the IGM transmission are further compounded by viewing angle biases that hydrodynamical simulations show to be important due to the strong anisotropy of LyC \fescs \citep[e.g.,][]{Gnedin08,Wise&Cen09, Wise14,Cen15,Paardekooper15}. For instance, a galaxy may have high \fesc, but via holes pointed away from the observed sightline \citep[e.g.,][]{Fletcher19,Nakajima20,Saxena21}.

Clearly, constructing pure, representative subsamples of leakers and non-leakers from direct LyC observations is challenging at high-$z$. Nonetheless, stacking sufficient ($\gtrsim50$ at $z\approx3$) galaxies from independently sampled sightlines is expected to produce a robust population-averaged \fescs \citep[e.g.,][]{Steidel18}. The current consensus is an average \fesc$\approx5-10\%$ for $\gtrsim0.5L^{*}$ Lyman-break galaxies (LBGs) at $z\approx3$ \citep[e.g.,][]{Marchi17,Naidu18,Pahl21}. The question then is, how do we translate this constraint on $z\approx3$ LBGs to EoR LBGs. These are very different populations, with important properties such as the star-formation surface density ($\Sigma_{\rm{SFR}}$), proposed to be causatively linked to \fescs \citep[e.g.,][]{Heckman11,Sharma16,Naidu20}, rising $\approx0.5-1$ dex higher \citep[e.g.,][]{Oesch10,Shibuya19} as galaxies grow burstier towards the EoR \citep[e.g.,][]{CAFG18, Tacchella20}.

In this work, we propose resolved Lyman-$\alpha$ (Ly$\alpha$) emission-line spectroscopy is the panacea to the challenges around LyC \fesc. The resonant nature of Ly$\alpha$, which makes it highly sensitive to HI in the IGM, is routinely exploited to constrain the timeline of reionization \citep[e.g.,][]{Stark11, Pentericci14, Mason18}. This resonant nature also renders the Ly$\alpha$ line profile sensitive to the HI distribution \textit{within} galaxies. From the emergent sample of LyC leakers it is clear that Ly$\alpha$ profiles are the highest fidelity tracers of \fesc, both at low and high redshifts, and across several dex in physical properties like stellar mass, specific SFR, $\Sigma_{\rm{SFR}}$, and $E(B-V)$ \citep[e.g.,][]{Verhamme17, Izotov18b,Izotov21, Vanzella20}. From a theory point of view, the interpretation is intuitive and well-understood -- line profiles with tightly spaced narrow blue and red peaks, with flux emitted close to the systemic redshift ($z_{\rm{sys}}$), signal a transparent, porous ISM with clear passages for Ly$\alpha$ (and LyC) escape. On the other hand, broad lines, widely separated peaks, and no photons at $z_{\rm{sys}}$ signal an ISM through which Ly$\alpha$ (and LyC) photons struggled to escape \citep[e.g.,][]{Verhamme15,Gronke15b,Dijkstra16,Kimm19,Kakiichi21}.

There are significant advantages to studying LyC with Ly$\alpha$ profiles. The IGM, which severely hampers direct LyC observations, has little effect on Ly$\alpha$ profiles at $z\approx2-3$ \citep[e.g.,][]{Hayes21} so any individual galaxy can be robustly classified as a likely leaker or a non-leaker. Perhaps most importantly, LyC constraints based on Ly$\alpha$ profiles can be extrapolated to higher redshifts with some confidence because LAEs at $z\approx2-6$ are fundamentally similar -- in e.g., their sizes \citep[e.g.,][]{Malhotra12,Paulino-Afonso18}, UV slopes \citep[e.g.,][]{Santos20}, and Ly$\alpha$ line profiles corrected for IGM absorption \citep[e.g.,][]{Hayes21}. Further, Ly$\alpha$ LFs are almost  unevolving across $z\approx2-6$, therefore a luminosity-limited survey at $z\approx2$ would have a similar proportion of bright and faint LAEs as at higher redshifts \citep[e.g.,][]{Sobral18,Herenz19,Ouchi20}.

Realizing the potential of resolved Ly$\alpha$ requires surveys with high spectral resolution at the Ly$\alpha$ wavelength ($R>4000$, e.g., \citealt{Verhamme15}) along with precise $z_{\rm{sys}}$. Further, to ensure the generalizability of the results to higher redshifts, the selection function must be well known and ideally simple. The luminosity-limited ($L_{\rm{Ly\alpha}}>0.2L^{*}$) X-SHOOTER Ly$\alpha$ Survey at $z=2$ (XLS-$z$2, \citealt{Matthee21}), based on the narrow-band CALYMHA Survey \citep{Matthee16,Sobral17}, fulfils exactly these requirements. In this paper we use XLS-$z$2 to extract the first statistical constraints on LyC \fescs via resolved Ly$\alpha$ profiles. In a companion paper (Matthee \& Naidu et al. 2021) we use these constraints to show how LAEs explain the evolution of the cosmic ionizing emissivity from $z\approx2-8$.

A plan for this paper follows -- in \S\ref{sec:data} we describe the XLS-$z$2 sample, in \S\ref{sec:classify} we motivate the \lya-profile based selection criteria for the ``Low Escape" (\fesc$<5\%$) and ``High Escape" (\fesc$>20\%$) stacks that we produce in \S $\ref{sec:spectralstacks}$. In \S\ref{sec:results} we describe the physical conditions for LyC \fescs based on the differences between these stacks, in \S\ref{sec:fesc} we estimate the \fescs of our High Escape stack. We place our results in a broader context, while addressing concerns and caveats in \S\ref{sec:discussion}, and end with a summary in \S\ref{sec:summary}. Throughout this work we reference $L^{*}$, the characteristic luminosity in Schechter function parametrizations of luminosity functions (LFs). In the context of Ly$\alpha$ LFs, the $L^{*}$ is as per the \citet{Sobral18} $z\approx2-6$ consensus LFs ($\log{L_{\rm{Ly\alpha}}/\rm{erg\ s^{-1}}}\approx43$). Magnitudes are in the AB system \citep[e.g.,][]{Oke83}. For summary statistics, unless otherwise mentioned, we report medians with errors on the median from bootstrapping (16$^{\rm{th}}$ and 84$^{\rm{th}}$ percentiles).  We assume a flat $\Lambda$CDM concordance cosmology with $\Omega_M=0.3$, $\Omega_{\Lambda}=0.7$ and $H_0=70$ km s$^{-1}$ Mpc$^{-1}$.

\section{Sample \& Data}
\label{sec:data}
\subsection{$z\approx2$ Sample}
Our sample is drawn from the X-SHOOTER \lyas Survey at $z\approx2$ (XLS-$z2$), which is a deep spectroscopic survey of 35 LAEs \citep{Matthee21}. The sample spans luminosities $0.2-10 \times L^{*}_{\rm Ly\alpha}$ and $0.2-6 \times L^{*}_{\rm UV}$ with a median rest-frame Ly$\alpha$ equivalent width (EW) of 82 {\AA}. The majority of these LAEs have originally been discovered in wide-field narrow-band surveys in well-known extragalactic fields \citep[e.g.][]{Sobral2016}. Active Galactic Nuclei (AGN) were removed based on X-Ray and radio data \citep{Calhau20}, and spectroscopy \citep{Sobral18}. Most of the LAEs in the sample are Ly$\alpha$-flux limited selected. A handful have been pre-selected based on the UV continuum in combination with a high ionisation state (i.e. [OIII]/H$\beta$) or the presence of high ionisation UV lines such as CIII] (see \citealt{Matthee21} for details). However, these properties are ubiquitous among typical $z\approx2$ LAEs so we see no reason to exclude them from our sample. The luminosities of these LAEs are typical for the objects discovered in deep narrow-band surveys \citep[e.g.][]{Gawiser2007,Ouchi2008} and VLT/MUSE observations with $\approx1$ hour depth (e.g. MUSE-wide; \citealt{Herenz19}). 

In this paper we analyse 26 out of the 35 LAEs from XLS-$z2$ (Table \ref{table:sample_selection}). The following objects were excluded from our analysis: XLS-1 because it was identified as an AGN, XLS-9 and XLS-13 as no systemic redshift was measured owing to their faintness, XLS-30 because its data does not cover the H$\alpha$ line, XLS-7, 8, 29, 31 because their Ly$\alpha$ EW is $<25$ {\AA} (the standard definition that Ly$\alpha$ LFs adopt for ``LAE", e.g., \citealt{Sobral18}) and XLS-27 because its Ly$\alpha$ line is significantly offset (by 9 kpc) from the rest-frame optical lines. We split the remaining 26 LAEs in subsets determined by their Ly$\alpha$ line-profile.

The sample of 26 sources we study here is representative of $L_{\rm{Ly\alpha}}>0.2 L^{*}$ LAEs at $z\approx2$. The median \lyas \fescs is $30\pm5\%$, in excellent agreement with measurements of typical LAEs that also find $\approx30\%$ \citep{Hayes10, Song14,Trainor16,Sobral17,Harikane18,Matthee21}. The median \lyas EW is $95\pm16\ \AA$, consistent with published EW distributions at similar redshifts \citep[e.g.,][]{Gronwall07,Hashimoto17,Santos20}. 

\subsection{Data}
The unique feature of XLS$-z2$ is the combination of wide wavelength coverage ($\lambda_{\rm{rest}}\approx1000-8000$ {\AA} at $z\approx2$) with high spectral resolution for the Ly$\alpha$ line ($R\approx4000-5000$) thanks to the X-SHOOTER echelle spectrograph on the VLT \citep{Vernet2011}. The exposure times are $\approx3$ hours on average, which enables simultaneous measurements of systemic redshifts (through the rest-frame optical [OIII] and H$\alpha$ lines) along with sensitive Ly$\alpha$ spectroscopy. Redshift $\approx2$ is the lowest redshift where Ly$\alpha$ can be measured from the ground and the highest redshift where H$\alpha$ falls in the $K$ band, enabling convenient estimates of Balmer decrements and Ly$\alpha$ escape fractions \citep[e.g.,][]{Sobral19}. Another advantage at $z\approx2$ is that the impact of the IGM on Ly$\alpha$ is negligible \citep[e.g.][]{Laursen2011,Hayes21}. The spectral resolution of XLS-$z2$ is a factor $\gtrsim3$ and $\gtrsim5$ higher than the data used by \cite{Kulas2012} and \cite{Trainor15} respectively, who previously studied Ly$\alpha$ profiles at $z\approx2$ and is comparable to the study of a smaller sample ($N=6$) by \cite{Hashimoto2015}.

We use both 1D and 2D spectra in this analysis. Measurements in individual sources and stacks are based on 1D spectra extracted based on the position and size of the UV-continuum. The stacking has been performed in 2D (see \citealt{Matthee21} for details on the data reduction, spectral extraction and stacking procedures).

\subsection{Ly$\alpha$ profile statistics}

The individual Ly$\alpha$ profiles are shown in Appendix \ref{appendix:lyaprofiles}. The typical integrated signal-to-noise ratio of the Ly$\alpha$ line is 20. A multiple peaked Ly$\alpha$ line is detected in 19/26 LAEs (i.e. 73 \%). As the blue peak is in all cases fainter than the red peak (typically containing $\approx$ 17 \% of the total Ly$\alpha$ flux), it is possible, but unlikely given \lyas signal to noise ratio (SNR)>20, that some blue peaks are missed due to their faintness. Out of the multiple peaked systems, 2/19 show three peaks with a clear peak at the systemic velocity (XLS-2 and XLS-21). Two multiple peaked LAEs show additional faint absorption profiles, either in the blue peak or in the red peak (XLS-18, XLS-33).  One of the 7 single-peaked LAEs shows a relatively symmetric Ly$\alpha$ line at the systemic velocity (XLS-20).

\begin{table}
    \centering
    \caption{Ly$\alpha$ properties of our parent sample, split by their inferred LyC \fesc. The Ly$\alpha$ escape fraction ($f_{\rm{esc, Ly\alpha}}$) is computed as $L_{\rm{Ly\alpha}}/8.7\times L_{\rm{H\alpha, int.}}$ (see e.g. \citealt{Hayes15}), where $L_{\rm{H\alpha, int.}}$ is the H$\alpha$ luminosity corrected for dust attenuation using the Balmer decrement and a \citealt{Cardelli89} attenuation law. The peak separation ($v_{\rm{sep}}$) and central escape fraction ($f_{\rm{cen}}$) are discussed in \S\ref{sec:motivate}. $\dagger$ = No Blue Peak detected, $*$ = Triple peak }

    \begin{tabular}{l|r|r|r}
ID & $f_{\rm esc, Ly\alpha}$ & $v_{\rm sep}$/km s$^{-1}$ & $f_{\rm cen}$ \\ \hline
 \textbf{High Escape}  & & & \\ 
 (LyC \fesc$>20\%$)  & & & \\ 
XLS-2 & $0.35^{+0.14}_{-0.23}$ & $424\pm32 *$ & $0.162\pm0.007$ \\
XLS-3 & $0.77^{+0.32}_{-0.45}$ & $184\pm13$ & $0.243\pm0.016$ \\
XLS-11 & $0.55^{+0.38}_{-0.37}$ & $368\pm15$ & $0.287\pm0.005$ \\
XLS-14 & $0.40^{+0.24}_{-0.24}$ & $\dagger$ & $0.135\pm0.037$ \\
XLS-17 & $1.11^{+0.15}_{-0.12}$ & $246\pm15$ & $0.109\pm0.004$ \\
XLS-18 & $0.10^{+0.06}_{-0.04}$ & $\dagger$ & $0.110\pm0.004$ \\
XLS-19 & $0.32^{+0.35}_{-0.19}$ & $445\pm15$ & $0.208\pm0.015$ \\
XLS-20 & $0.27^{+0.37}_{-0.25}$ & $\dagger$ & $0.501\pm0.013$ \\
XLS-21 & $0.16^{+0.04}_{-0.06}$ & $528\pm35 *$ & $0.233\pm0.003$ \\
XLS-23 & $0.41^{+0.19}_{-0.19}$ & $370\pm14$ & $0.314\pm0.002$ \\
XLS-24 & $0.64^{+0.40}_{-0.52}$ & $365\pm17$ & $0.153\pm0.002$ \\
XLS-26 & $0.25^{+0.13}_{-0.13}$ & $389\pm16$ & $0.364\pm0.007$ \\
XLS-28 & $0.35^{+0.09}_{-0.12}$ &  $\dagger$ & $0.357\pm0.013$ \\
 \textbf{Intermediate Escape}  & & & \\ 
 ($5\%<$LyC \fesc$<20\%$)  & & & \\ 
XLS-4 & $0.43^{+0.67}_{-0.39}$ & $\dagger$ & $0.059\pm0.017$ \\
XLS-5 & $0.22^{+0.24}_{-0.17}$ & $336\pm15$ & $-0.006\pm0.008$ \\
XLS-10 & $0.02^{+0.27}_{-0.02}$ & $\dagger$ & $0.062\pm0.014$ \\
XLS-34 & $0.14^{+0.10}_{-0.08}$ & $635\pm15$ & $0.090\pm0.003$ \\
 \textbf{Low Escape}& & & \\
 (LyC \fesc$<5\%$) & & & \\
XLS-6 & $0.71^{+0.36}_{-0.26}$ & $459\pm16$ & $0.006\pm0.008$ \\
XLS-12 & $0.07^{+0.04}_{-0.03}$ & $\dagger$ & $0.047\pm0.007$ \\
XLS-15 & $0.94^{+0.59}_{-0.37}$ & $412\pm18$ & $0.050\pm0.009$ \\
XLS-16 & $0.26^{+0.08}_{-0.09}$ & $611\pm21$ & $-0.020\pm0.018$ \\
XLS-22 & $0.32^{+0.09}_{-0.11}$ & $372\pm15$ & $0.004\pm0.003$ \\
XLS-25 & $0.21^{+0.10}_{-0.06}$ & $560\pm15$ & $0.037\pm0.003$ \\
XLS-32 & $0.04^{+0.05}_{-0.02}$ & $610\pm29$ & $-0.033\pm0.011$ \\
XLS-33 & $0.09^{+0.03}_{-0.03}$ & $417\pm15$ & $0.009\pm0.003$ \\
XLS-35 & $0.13^{+0.03}_{-0.04}$ & $470\pm15$ & $0.039\pm0.002$ \\

    \end{tabular}
    \label{table:sample_selection}
\end{table}

\subsection{Literature Sample of Lyman Continuum leakers}

We design our criteria to select likely leakers and non-leakers based on Ly$\alpha$ profiles by constructing empirical criteria based on literature galaxies which have both direct LyC measurements as well as resolved ($R\gtrsim4000$) Ly$\alpha$ profiles.

The bulk of our literature calibration sample is comprised of $20$ $z\approx0.3$ GPs studied with \textit{HST}/COS compiled in \citet{Izotov21}. LyC is directly measured at $>850\AA$ for these sources along with \lya. The Ly$\alpha$ luminosities of the XLS-$z$2 sample are well-matched to the luminosities of these low-$z$ LyC leakers. Importantly for \lyas comparisons, the physical scale at $z\approx0.3$ probed by the COS apertures is very similar to the physical scale ($\approx7-10$ kpc) at $z\approx2$ probed by the XLS$z$2 slits. This ensures similar central regions of the \lyas emission are being captured. Also note that the spectral resolution for the XLS-$z2$ \lyas profiles are comparable to the resolution of the {\it HST}/COS spectra of the GPs \citep{Orlitova18}, such that there is no {\it differential} effect. These GPs span \fesc$\approx0\%$ to \fesc$\approx70\%$, with four sources showing \fesc$>20\%$.

\begin{figure*}
\centering
\includegraphics[width=\linewidth]{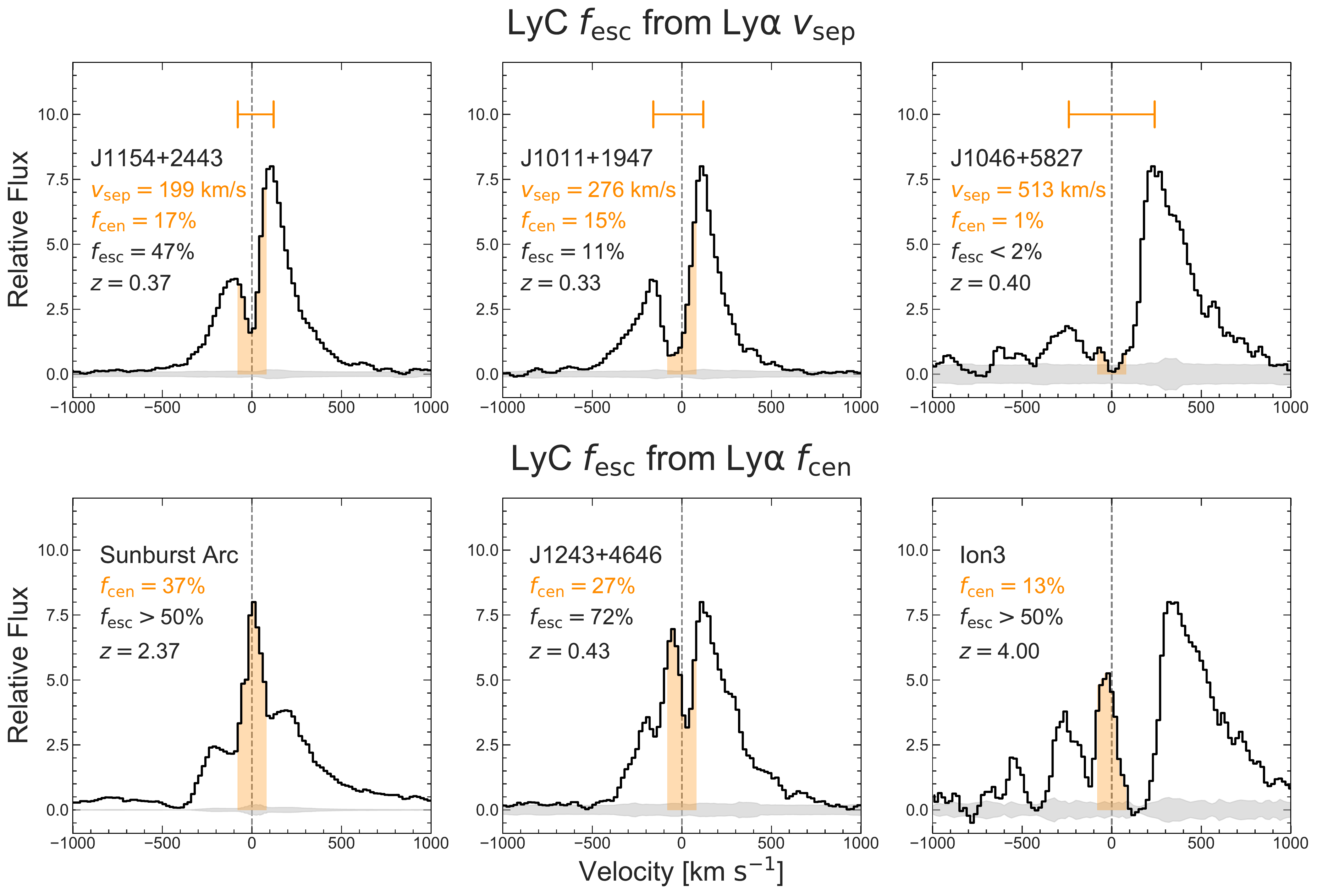}
\caption{\lyas line profiles for a selection of galaxies with direct LyC \fescs measurements. Each panel lists the name of the source, the metric linked to LyC \fescs ($v_{\rm{sep}}$, \fcen), and finally the directly measured LyC \fesc. Measured escape fractions are listed as lower limits for high-$z$ sources due to the unknown IGM transmission along any particular line of sight. \textbf{Top:} The \lyas peak separation ($v_{\rm{sep}}$), depicted as an orange capped line, is an effective predictor of \fescs for the $z\approx0.3$ Green Peas -- the larger the $v_{\rm{sep}}$, the lower the \fesc (Eq. \ref{eqn:izotov}). \textbf{Bottom:} For most LyC leakers with $f_{\rm{esc}}>20\%$, $v_{\rm{sep}}$ is ill-defined and unable to predict \fesc. These sources exhibit direct Ly$\alpha$ escape at the systemic velocity along with narrow lines. To capture this, we introduce a new parameter, the central escape fraction ($f_{\rm{cen}}$), which measures the fraction of the total Ly$\alpha$ flux emitted $\pm100$ km s$^{-1}$ from the systemic velocity (shaded orange region, Eq. \ref{eqn:fcen}).}
\label{fig:litfesc}
\end{figure*}

At higher redshifts, while several LyC leaker candidates have been identified, very few have resolved Ly$\alpha$ measurements. These sources are: Ion2 ($z=3.2$, \citealt{Vanzella16}), Ion3 ($z=4.0$, \citealt{Vanzella18}), Sunburst Arc ($z=2.4$, \citealt{Rivera-Thorsen19}), GS-30668/XLS-26 ($z=2.1$, \citealt{Naidu17}, \citealt{Matthee21}), GS-15601 ($z=3.27$, J. Kerrut, in prep.). While few in number, all these sources show Ly$\alpha$ profiles with prominent emission at the systemic velocity resembling the highest \fescs $z\approx0.3$ Green Peas (see bottom panel of Fig. \ref{fig:litfesc}). This strongly suggests that systemic Ly$\alpha$ emission accompanies high LyC \fesc. A source detected in LyC at $z>2$ despite the stochasticity of IGM transmission is likely to have high LyC \fescs \citep[e.g.,][]{Bassett21}. Indeed, all these sources have an estimated \fesc$>20\%$, thus complementing the GP sample at the high LyC \fescs end.

\section{Classifying Lyman Continuum Leakers and Non-Leakers with \Lyas Profiles}
\label{sec:classify}

\begin{figure*}
\centering
\includegraphics[width=\linewidth]{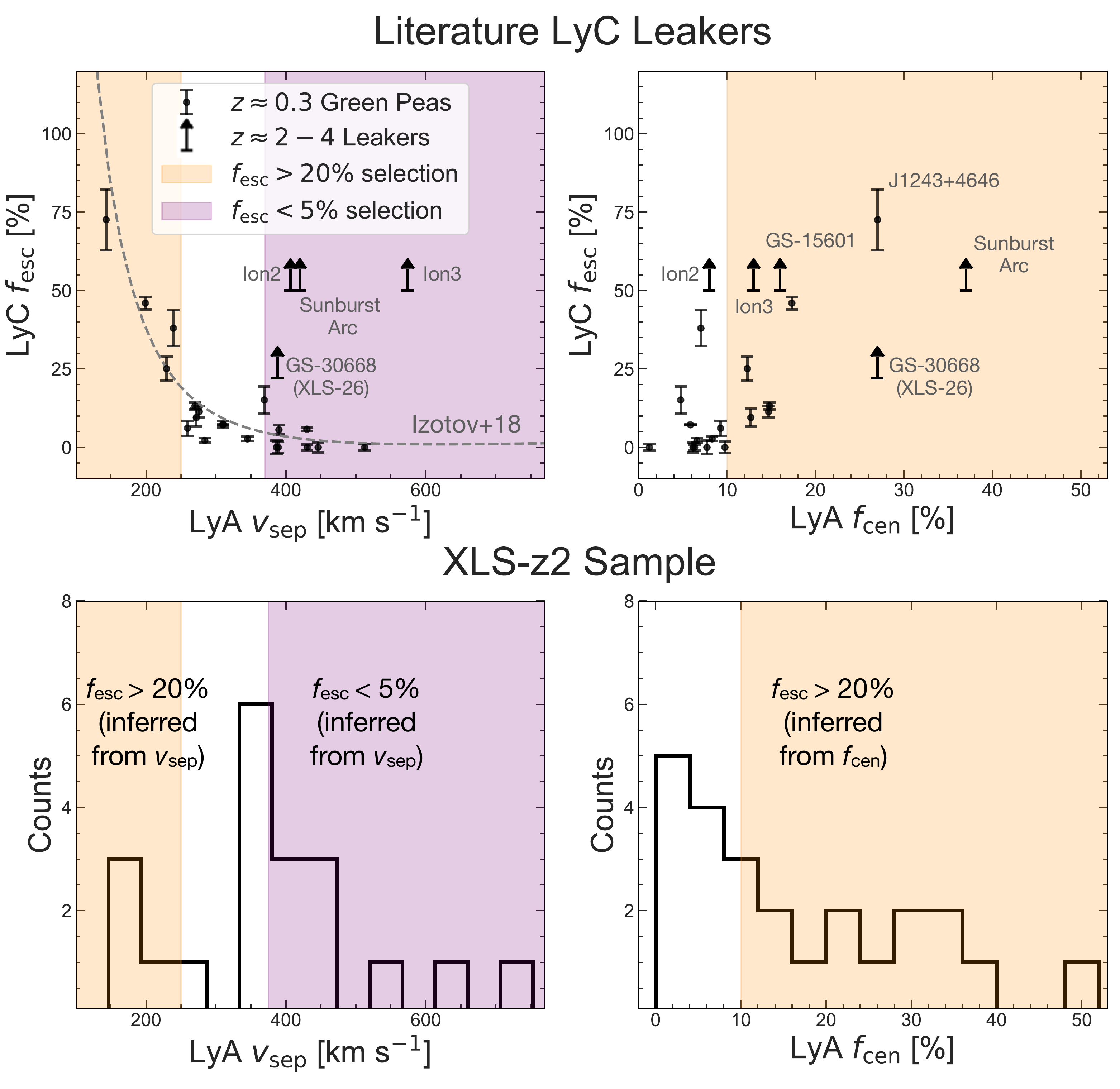}
\caption{Discriminating between leakers and non-leakers using \lyas profiles. \textbf{Top Left:} LyC \fescs as a function of \lyas peak separation ($v_{\rm{sep}}$) for the $z\approx0.3$ \citet{Izotov21} Green Pea compilation and all the $z\approx2-4$ leakers with high-resolution \lyas spectra. The fit from \citet{Izotov18b} is shown with a dashed line. While successful at predicting LyC \fescs for the Green Peas, the \lyas $v_{\rm{sep}}$ fails to identify \textit{all} the $z\approx2-4$ sources as having \fesc$>20\%$ because the systemic Ly$\alpha$ emission in these sources is not captured by $v_{\rm{sep}}$ (Figure \ref{fig:litfesc} bottom). \textbf{Top Right:} To complement $v_{\rm{sep}}$ we introduce the \lyas central escape fraction ($f_{\rm{cen}}$) that measures the fraction of the total Ly$\alpha$ flux emitted at $+/-100$ km s$^{-1}$ around the systemic redshift (\S\ref{sec:motivate}). \fcens selects almost all the $z\approx2-4$ sources missed by $v_{\rm{sep}}$ as having $f_{\rm{esc}}>20\%$. Combined, the \fescs and $v_{\rm{sep}}$ selection of $f_{\rm{esc}}>20\%$ sources (orange) is $\approx90\%$ complete and $\approx80\%$ pure, whereas the $f_{\rm{esc}}<5\%$ selection (purple) is $80\%$ complete and $100\%$ pure. \textbf{Bottom:} Distributions of $v_{\rm{sep}}$ and \fcens for the XLS-$z$2 sample, with selection criteria for the High Escape (purple) and Low Escape (orange) stacks informed by literature sources in the top row.}
\label{fig:LyCclassify}
\end{figure*}

Empirically, the Ly$\alpha$ line profile is the best predictor of LyC \fescs in galaxies in the local Universe \citep{Izotov18b,Izotov21}. We use line profiles to split the XLS-$z$2 sample in subsets -- ``High Escape" (LyC \fesc$>20\%$, ``leakers") and ``Low Escape" (LyC \fesc$<5\%$, ``non-leakers") based on a set of selection criteria that we design in this section. The motivation for \fesc$\approx20\%$ is that this is approximately the average \fescs required for $M_{\rm{UV}}\lesssim-15$ star-forming galaxies to produce reionization in typical calculations \citep[e.g.,][]{Robertson15,Naidu20}, whereas galaxies with \fesc$\approx5\%$ are not relevant to the emissivity since even if all galaxies at e.g., $z\approx7$ had \fesc$\approx5\%$ they would be unable to produce reionization. We do not focus on galaxies with intermediate \fescs between these two limits since we do not have sufficient sources (N=4) to construct stacks with meaningful SNR.

We emphasize that our goal here is to place galaxies in broad \fescs bins that are clean and complete. We do not argue that the \fescs distribution is bimodal, but as \fescs is non-linearly related to (parametrisations of) the shape of the Ly$\alpha$ profile, the expected \fescs of the two stacks are very different. In what follows we describe our newly developed selection criteria and motivate these based on theoretical and empirical grounds.

\subsection{Ly$\alpha$ Peak Separation ($v_{\rm{sep}}$) and Central Escape Fraction ($f_{\rm{cen}}$) as tracers of LyC \fesc: Motivation}
\label{sec:motivate}

Due to resonant scattering, the \Lyas line profile is expected to be a tracer of the kinematics, column density, and distribution of neutral HI within a galaxy \citep[e.g.,][]{Neufeld90,Verhamme06,Gronke15,Dijkstra16,Kakiichi21}. If the ISM is porous with abundant low column density channels, Ly$\alpha$ photons escape with minimal scattering. Radiative transfer simulations show homogeneous, expanding media that cover HII regions with low column densities ($N_{\rm HI}\lesssim10^{18}$ cm$^{-2}$) give rise to narrow, tightly separated red and blue peaks. In clumpy, multi-phase systems with non-unity covering fractions (i.e., so-called riddled ionization-bounded HII regions), Ly$\alpha$ photons escape directly at the systemic velocity across clear lines of sight \citep{Hansen06,Verhamme15,Gronke16,Gronke17}. On the other hand, dense HI distributions force Ly$\alpha$ photons to scatter till they shift out of resonance. This struggle to escape manifests in a broad profile, little flux at the systemic velocity, and widely separated blue and red peaks.

In the sample of $z\approx0.3$ LyC leakers the \Lyas red and blue peak separation ($v_{\rm{sep}}$) has been identified as the most faithful tracer of \fescs \citep[e.g.,][]{Verhamme17,Izotov16b,Izotov16a,Izotov18a,Izotov18b,Izotov21}. This trend is illustrated in the top row of Figure \ref{fig:litfesc} where narrower $v_{\rm{sep}}$ accompanies higher LyC \fesc and quantified as follows \citep{Izotov18b} with $v_{\rm{sep}}$ in km s$^{-1}$: 
\begin{align}
\label{eqn:izotov}
    f_{\rm{esc}} = 3.23\times10^{4} v_{\rm{sep}}^{-2} - 1.05\times10^{2} v_{\rm{sep}}^{-1} + 0.095.
\end{align}

\begin{figure*}
\centering
\includegraphics[width=\linewidth]{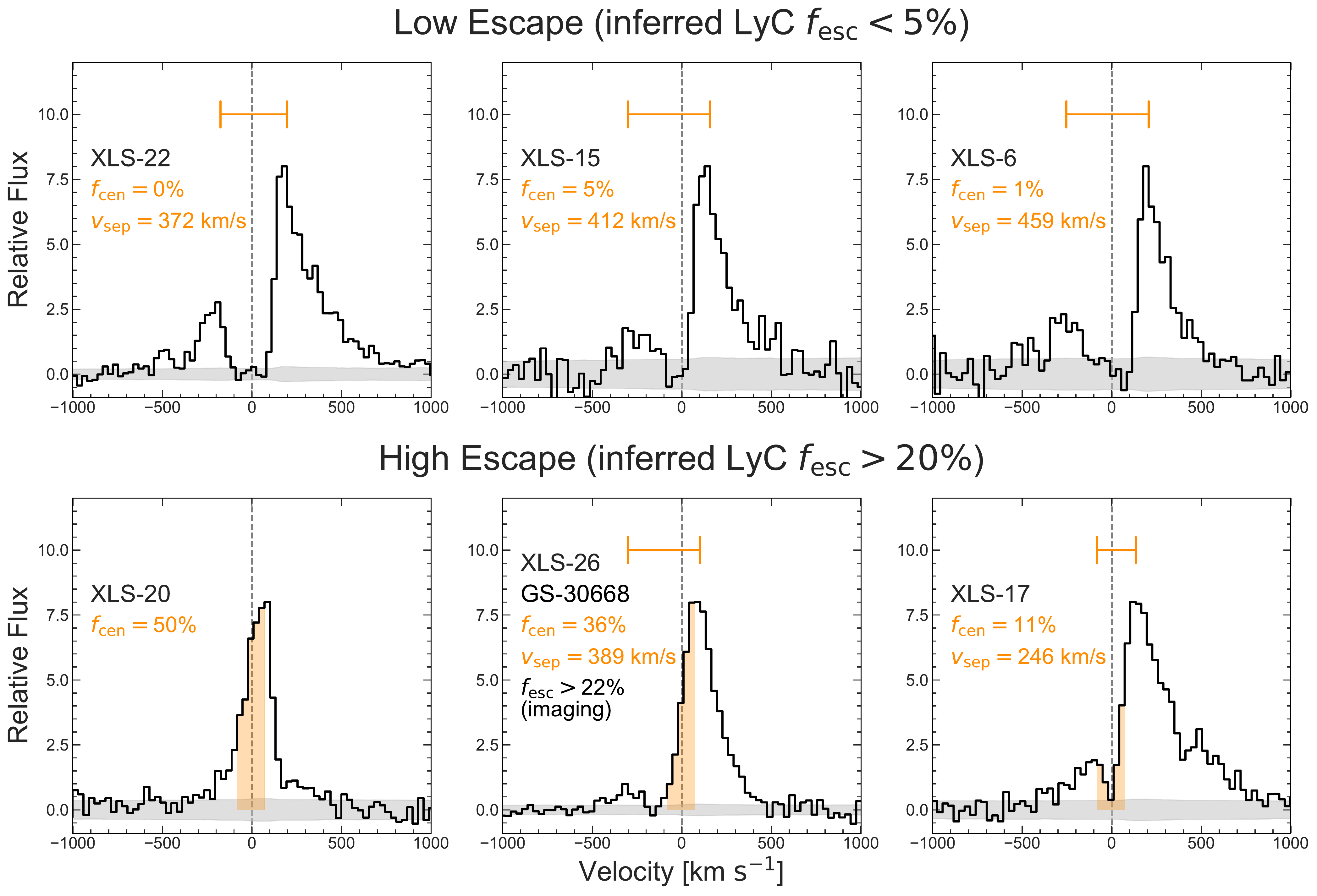}
\caption{Sources from the XLS-$z$2 Low Escape (top) and High Escape (bottom) subsets. The LyC \fescs of these sources is inferred purely based on their \lyas profiles ($v_{\rm{sep}}$ and \fcen). Low Escape sources are characterized by widely separated peaks, broad lines, and little flux emitted around line centre. High Escape sources on the other hand have large \fcens and narrow lines. XLS-26 (bottom-centre) is a known LyC leaker (GS-30668) and acts as a useful cross-check on our selection. Notably, sources like XLS-26/GS-30668 would have been erroneously classed as non-leakers based on $v_{\rm{sep}}$, but highly precise systemic redshifts reveal they have systemic Ly$\alpha$ emission and thus high \fcen.}
\label{fig:XLSfesc}
\end{figure*}

However, $v_{\rm{sep}}$ measurements rely on the detection of a clear red and blue peak -- fainter blue peaks may be missed due to poor SNR. Further, the applicability of $v_{\rm{sep}}$ is ambiguous when multiple peaks or systemic emission are seen in the profile. As a result, $v_{\rm{sep}}$ is an inapplicable  metric for a large fraction of \fesc$>20\%$ leakers. We show this in Figure \ref{fig:LyCclassify}, where we compile $v_{\rm{sep}}$\footnote{In computing $v_{\rm{sep}}$ for a source like the Sunburst Arc (Figure \ref{fig:litfesc}) we set aside the central emission and measure the difference between the closest red and blue peaks. If instead $v_{\rm{sep}}$ was defined as max(flux, $v>0$)$-$max(flux, $v<0$) we would end up with $v_{\rm{sep}}\approx0$ (and an implied unphysical \fesc$\approx100\%$) for all the $z\approx2-4$ sources. This ambiguity is precisely the motivation for introducing \fcen.} for all known LyC leakers with high resolution, ($R\gtrapprox4000$) \lyas measurements and robust systemic redshifts\footnote{The sources J1333+6246, J1442-0209, J1503+3644 are excluded because inspection of their Ly$\alpha$ profiles showed their systemic redshifts to be untrustworthy, see also \citet{Orlitova18}.}. \textit{All} the $z\approx2-4$ LyC leakers observed with high-resolution spectroscopy show complex profiles characterized by flux at line center in addition to red and blue peaks (bottom row, Figure \ref{fig:litfesc}). These sources include GS-30668/XLS-26 ($z=2.2$, \citealt{Naidu17}), the Sunburst Arc\footnote{We note that the Ly$\alpha$ line-profile of the Sunburst Arc corresponds to the profile of the LyC-leaking knot within the galaxy \citep[see also][]{Vanzella21}. It has a Ly$\alpha$ EW of 103 {\AA} (Emil Rivera-Thorsen, private communication).} ($z=2.4$, \citealt{Rivera-Thorsen19}), Ion2 ($z=3.2$, \citealt{Vanzella16}), GS-15601 ($z=3.3$, J. Kerrut, private comm.), and Ion3 ($z=4.0$, \citealt{Vanzella18}). The three Green Peas with \fesc$>20\%$ also show significant line centre emission (e.g., J1243+4646, Figure \ref{fig:litfesc}). Complex profiles that are not just a blue+red peak combination are a routine feature at high \fesc. For these sources $v_{\rm{sep}}$ is ill-defined (e.g., in the Sunburst Arc) and/or drastically underestimates \fescs (e.g., Ion3).  

In these cases $v_{\rm{sep}}$ is a poor tracer of LyC \fescs because the observed Ly$\alpha$ profiles are likely a combination of two distinct modes of \Lyas escape \citep[e.g.,][]{Rivera-Thorsen17}: (i) scattering, resonant escape through relatively higher column density HI that results in red and blue peaks, (ii) direct escape through porous channels that manifests as central \Lya. Since $v_{\rm{sep}}$ is sensitive only to the scattering escape mode, we introduce a new parameter, the ``central escape fraction" ($f_{\rm{cen}}$), that traces the direct escape mode as well. We define \fcens as the fraction of \lyas emission within $\pm100$ km s$^{-1}$ of the systemic velocity, i.e.,
\begin{equation}
\label{eqn:fcen}
    \rm{Central}\ \rm{Escape}\ \rm{Fraction}\ (\textit{f}_{\rm{cen}}) = \frac{\rm{Ly}\alpha\ \rm{flux}\ \rm{at} \pm100\ \rm{km}\ \rm{s}^{-1}}{\rm{Ly}\alpha\ \rm{flux}\  \rm{at} \pm1000\ \rm{km}\ \rm{s}^{-1}},
\end{equation}
where we found that the $\pm1000$ km s$^{-1}$ velocity window captures the total flux for all XLS-$z2$ sources.
Theoretical profiles \citep[e.g.,][]{Behrens14,Verhamme15, Dijkstra16} suggest \fcens should track the relative abundance of low-opacity escape channels which can facilitate prolific LyC \fesc. Note that the denominator in Eq. \ref{eqn:fcen} ensures that if only a small amount of flux is escaping at line centre, the \fcens (and the implied \fesc) is meagre. For instance, if central emission occurs on top of a double-peak profile, their relative weights are accounted for (contrast the $\approx3\times$ higher \fcens of the Sunburst Arc with that of Ion3 in Figure \ref{fig:litfesc}). An advantage of \fcens is that one does not need to identify the exact locations of red or blue peaks, which can be ambiguous for multi-peaked profiles or when the fainter (typically blue) peak is below the detection threshold. 

We caution that the specific choice of $\pm100$ km s$^{-1}$ is empirical and will be resolution dependent. However, the spectral resolution for the XLS-$z2$ observations and the sources used to calibrate the criteria are all very similar. We also caution that for low EW sources continuum subtraction errors can render \fcens uncertain, so this criterion in specifically applicable to LAEs (\lyas EW$>25\AA$). In the following section we provide an empirical verification of the utility of our definition of $f_{\rm{cen}}$.

\subsection{Designing and validating $f_{\rm{cen}}$ \& $v_{\rm{sep}}$ selection criteria}
\label{sec:validatefcen}
Here we use known LyC leakers from the literature to design our joint $f_{\rm{cen}}$ \& $v_{\rm{sep}}$ selection criteria in order to identify High Escape (\fesc$>20\%$) and Low Escape (\fesc$<5\%$) galaxies. We obtained the \lyas spectra for these sources from the \lyas Spectral Database \citep[LASD,][]{Runnholm21} or via private communication. A selection criterion of $f_{\rm{cen}}>10\%$ reliably identifies the bulk (7 out of 9) of literature sources with $f_{\rm{esc}}>20\%$. This adopted \fcens cut not only selects sources with obvious \lyas at line centre like the Sunburst Arc, but also picks up sources with narrow lines and/or tight peak separations (e.g., J1154+2443, Figure \ref{fig:litfesc}). 

When complemented with a $v_{\rm{sep}}<250$ km s$^{-1}$ criterion, corresponding to $f_{\rm{esc}}>20\%$ (Eqn.\ref{eqn:izotov}, top-left, Figure \ref{fig:LyCclassify}), Ion2 is the only source missed (i.e., the only ``false negative"). As for false positives, three Green Peas with marginally lower than expected LyC \fescs $\approx10-15\%$ are picked up -- one of these, J1011+1947, has a \lya \fesc$<20\%$ and so is readily identified as a contaminant. To summarize, the following empirically motivated criterion: 
\begin{equation}
\label{eqn:highfesc}
    \rm{High}\ \rm{Escape}\ (\textit{f}_{\rm{esc}}>20\%): (\textit{f}_{\rm{cen}}>10\%)\ \rm{or}\ (\textit{v}_{\rm{sep}}<250 \ \rm{km}\ \rm{s}^{-1})
\end{equation}
when applied to literature LyC leakers produces a sample of \fesc$>20\%$ sources that is $\approx90\%$ complete and $\approx80\%$ pure.

For selecting galaxies with \fesc$<5\%$, from the top row of Figure \ref{fig:LyCclassify} we observe that once the high \fcens sources are set aside, an entirely pure and $\approx80\%$ complete sample can be selected as follows:
\begin{equation}
\label{eqn:lowfesc}
    \rm{Low}\ \rm{Escape}\ (\textit{f}_{\rm{esc}}<5\%): (\textit{f}_{\rm{cen}}<10\%)\ \rm{and}\ (\textit{v}_{\rm{sep}}>375 \ \rm{km}\ \rm{s}^{-1}).
\end{equation}

It is remarkable that these simple empirical selections based purely on Ly$\alpha$ work so effectively given the intricate, multi-phase physics that drives \fescs \citep[e.g.,][]{Paardekooper15,Ma16,Rosdahl18}. For instance, at first glance, $v_{\rm{sep}}$ and $f_{\rm{cen}}$ appear sensitive only to HI, and not to dust attenuation, which is the other key inhibitor of LyC \fescs (e.g., \citealt{Chisholm18}). However, in \S $\ref{sec:results}$ we argue that low column densities, and low attenuation are likely causally interlinked, and thus $v_{\rm{sep}}$ and $f_{\rm{cen}}$ implicitly select for low dust. In \S $\ref{sec:results}$ we present several such independent spectroscopic points of evidence that inspire confidence in the robustness of the High and Low \fescs selections.

\subsection{Applying $f_{\rm{cen}}$ and $v_{\rm{sep}}$ selections to XLS-$z$2}
\label{sec:selectxls}

We have listed $f_{\rm{cen}}$ and $v_{\rm{sep}}$ measurements for our XLS-$z2$ sample in Table $\ref{table:sample_selection}$ along with Ly$\alpha$ escape fractions measured from the Ly$\alpha$/H$\alpha$ ratio that is dust-corrected via Balmer decrements. Importantly for $f_{\rm{cen}}$, systemic redshifts precise to $<10$ km s$^{-1}$ are measured thanks to the strong [OIII] doublet and its known intrinsic line ratio (see \citealt{Matthee21} for details). H$\alpha$, H$\beta$, and in some cases nebular UV lines such as \HeII \, are further used as a cross-check on the systemic redshift. The peak separation is measured by searching for maxima on either side of the systemic redshift. Faint blue peaks (and hence $v_{\rm{sep}}$) are unreliable or undetected for a few high $f_{\rm{cen}}$ sources. This is not cause for concern -- in \S\ref{sec:validatefcen} we showed all but one of the literature leakers with tight $v_{\rm{sep}}$ were picked up by the $f_{\rm{cen}}$ selection. 

Applying Eqns. \ref{eqn:highfesc} \& \ref{eqn:lowfesc} we construct a sample of 13 High Escape ($f_{\rm{esc}}>20\%$) and 9 Low Escape ($f_{\rm{esc}}<5\%$) sources. 4 sources have intermediate \fescs and are not the subject of this analysis as their stacked spectrum has low SNR due to the small number of stacked sources. All sources in the parent sample are placed in one of these three bins. In Figure \ref{fig:XLSfesc} we show a selection of sources from the Low Escape and High Escape samples. The Low Escape galaxies show little to no flux around line centre and broad, widely separated peaks. In the High Escape sample, XLS-20 is an even more extreme version of the Sunburst Arc, with $\approx50\%$ of its Ly$\alpha$ emitted at line centre, while XLS-17 resembles the Green Peas with tight peak separation.

The four sources classified as High Escape based on $v_{\rm{sep}}$ also have high $f_{\rm{cen}}>10\%$. However, five \fcen-selected sources appear to have relatively wide $v_{\rm{sep}}$ of $\approx400$ km s$^{-1}$ (Table $\ref{table:sample_selection}$). XLS-26 (Figure \ref{fig:XLSfesc}) is the archetype of such sources. We emphasize again that the systemic redshifts for all our \fcens selected sources are highly secure -- e.g., for XLS-26 the $z_{\rm{sys}}$ is confirmed with several lines across multiple X-SHOOTER arms (H$\alpha$, H$\beta$, \OIII$\lambda4960,5008\AA$, \HeIIl, O{\sc iii}]$\lambda1666\AA$). In these five cases we may be  witnessing significant direct Ly$\alpha$ escape, so $v_{\rm{sep}}$ is under-estimating the LyC  \fescs (see \S\ref{sec:validatefcen}). Higher resolution spectra might reveal a clear central peak superimposed on blue and red peaks in these sources. Supporting this interpretation, we note that XLS-26 was identified as a likely LyC leaker with $f_{\rm{esc}}=60^{+40}_{-38}\%$ \citep[``GS-30668" in][]{Naidu17} with direct LyC imaging from the Hubble Deep UV Survey (\citealt{Oesch18b}, which also incorporates earlier UV imaging from \citealt{Rafelski15}). Since its \fescs was based on a probabilistic method (similar to the search that yielded Ion2, \citealt{Vanzella15}), GS-30668 was presented as a likely candidate pending spectroscopic follow-up. Since then, MUSE-HUDF \citep{Bacon17, Nanayakkara19} and X-SHOOTER follow-up (this work) have validated its highly ionizing nature -- it is a \HeII, \CIII, and \CIV\ emitter with \OIII/\OII$>10$ and extreme rest-frame EW([OIII]+H$\beta$)$\approx3400\ \AA$.

Our selection criteria imply $50\pm10$\% (binomial error based on sample size) of the $L>0.2 L^{*}$ Ly$\alpha$ emitters at $z\approx2$ are LyC leakers with $f_{\rm{esc}}>20\%$ (Table \ref{table:sample_selection}). As discussed in \S\ref{sec:data}, the XLS-$z$2 sample is representative of $L>0.2 L^{*}$ LAEs. In the following section we contrast the average properties of \fesc$>20\%$ and \fesc$<5\%$ sources using their stacked rest-frame UV to optical spectra. It is important to note that we are contrasting High Escape \textit{LAEs} and Low Escape \textit{LAEs} -- the differences between High Escape LAEs and non-LAEs (i.e., Ly$\alpha$ EWs$<25\ \AA$) are likely even more pronounced than what we describe here.

\begin{figure*}
\begin{tabular}{cc}
\includegraphics[width=8.5cm]{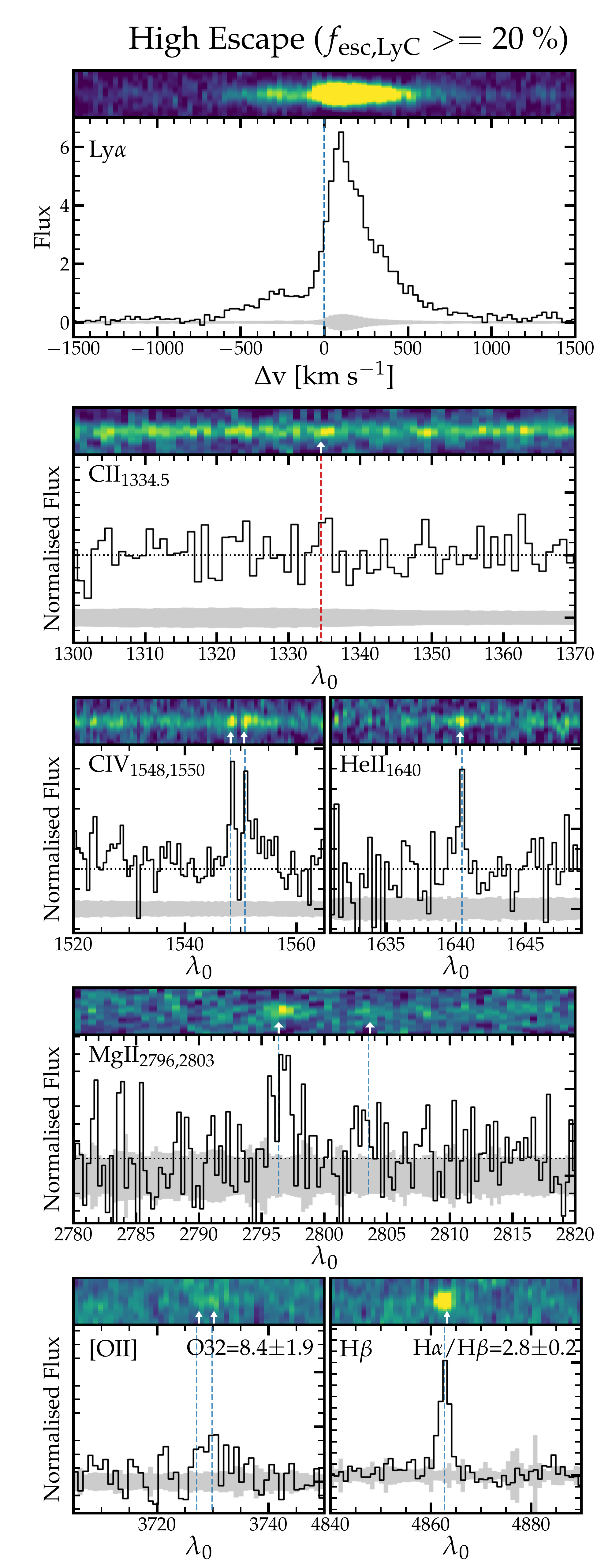} &
\hspace{-0.3cm}\includegraphics[width=8.5cm]{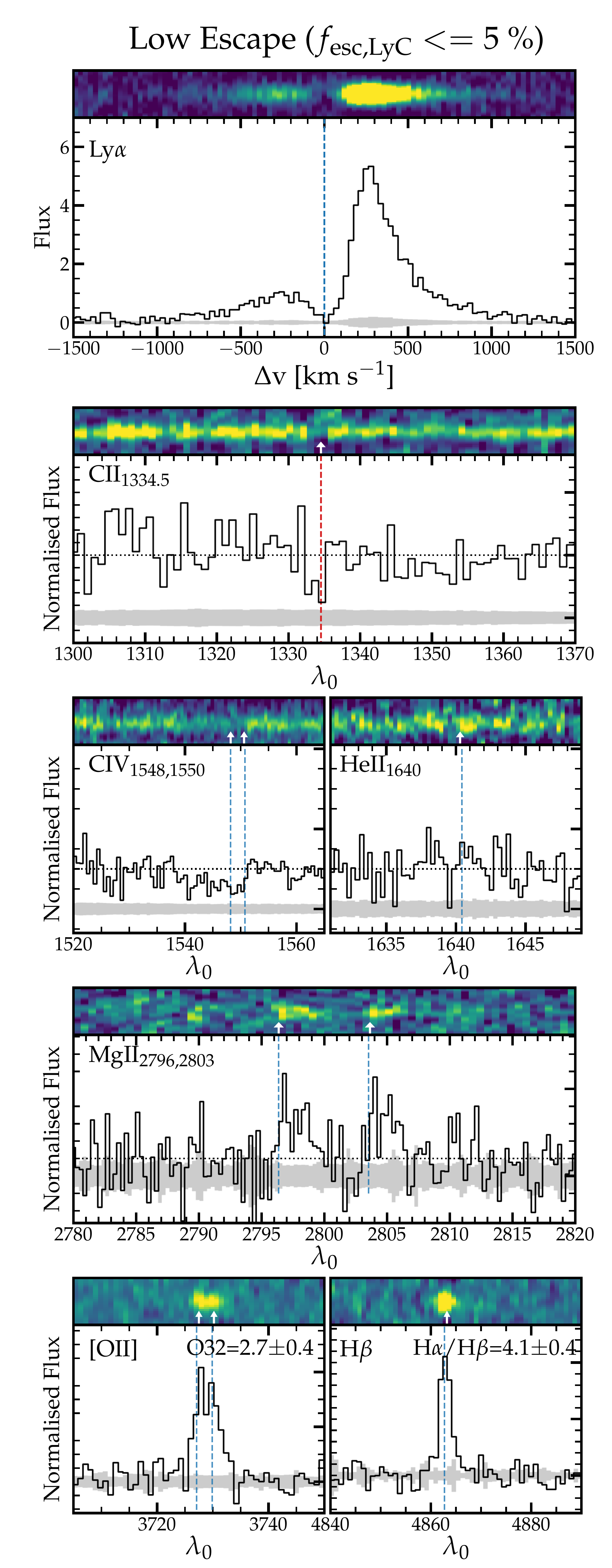}
\end{tabular}
\caption{Median-stacked X-SHOOTER spectra for the High Escape (left) and Low Escape (right) subsets that are selected purely on Ly$\alpha$ line profiles (top row). Each panel is centred on a spectral feature labeled in the top-left corner. Panels are arranged in order of increasing wavelength (top to bottom). All features are nebular emission, except for C{\sc ii} in the second row, which is an interstellar absorption line. Gray shaded regions mark the $1\sigma$ noise level. The locations of emission-lines are marked with vertical blue dashed lines, while red lines mark absorption features. The black horizontal dotted line shows the normalisation level for the second to fourth row. For the fifth (bottom) row, the flux levels are normalised to the [OIII] (left) and H$\beta$ (right) flux, respectively. }
\label{fig:stack}
\end{figure*}

\section{Spectral Stacks}
\label{sec:spectralstacks}

To bring out faint spectral features we construct {\it median}-stacked X-SHOOTER spectra of the High and Low Escape subsamples over the $\lambda_{\rm{rest}}=1000-8000$ {\AA} range. The stacking methodology follows \cite{Matthee21} whose approach we summarise here. Individual 2D spectra are shifted to the rest-frame and centered on the spatial peak of the Ly$\alpha$ line and the flux densities are converted to luminosity densities before the spectra are continuum normalized and median-combined.\footnote{We have also created stacked spectra centred on the UV continuum emission but find only small changes within the uncertainties of the measurements listed in Table $\ref{table:stacks}$. This is because typical spatial offsets between Ly$\alpha$ and the UV continuum are small and our spectral extraction window accounts for variations in the shape of the (stacked) UV continuum light distribution.} An error spectrum is obtained using bootstrap-resampling. The 1D spectrum is then optimally-extracted assuming a gaussian profile. The full width half maximum (FWHM) of the gaussian profile is wavelength-dependent as determined by fits to the spatial extent of the UV continuum and rest-frame optical nebular lines. Key spectral windows of the resulting stacked spectra are presented side by side in Figure \ref{fig:stack} with each row highlighting specific features. 

An array of quantities measured from the stacks are summarized in Table \ref{table:stacks}. The emission-line luminosities, EWs, and line-ratios are derived following \citet{Matthee21}. Due to the complex shape of the Ly$\alpha$ line, its line-luminosity is measured by simply integrating the flux within $\pm1000$ km s$^{-1}$ from the systemic velocity, while the continuum level is estimated over the 1270-1300 {\AA} interval in order to avoid interstellar absorption features. The luminosities of other faint UV lines ($\lambda<3000$ {\AA}) are measured by fitting single gaussian profiles where the FWHM can vary within 50-250 km s$^{-1}$ and the continuum level is estimated using a rolling-median of the flux around the specific emission-line. For doublets, the widths and centroids are fixed to each other. For the rest-frame optical lines where the signal to noise is typically significantly higher, we notice that some lines show complex kinematics, such as a broad component (see \citealt{Matthee21} for examples in individual sources). Therefore we measure their line-luminosities non-parametrically using a curve-of-growth approach by integrating the flux in increasing windows with width $\pm60$ to $\pm400$ km s$^{-1}$ (or until convergence within the uncertainties). Uncertainties on line-luminosities, EWs and luminosity-ratios are obtained by redoing the measurements on data that is perturbed with the propagated noise level 1000 times. For non-detections of UV lines in the Low Escape stack upper limits were estimated by assuming the FWHM of the [OIII]$_{5008}$ line of 150 km s$^{-1}$. These widths are validated in the UV lines that are detected. Due to their low detection S/N, absorption line EWs are measured non-parametrically by integrating the flux in a window between $-500<\Delta v<+100$ km s$^{-1}$ from the systemic redshift. This window is determined based on the typical velocity profile of absorption lines in deeper stacks of LAEs at $z\approx2$ \citep{Trainor15,Matthee21}.

Stellar masses and rest-frame UV luminosities are obtained from spectral energy distribution (SED) fitting using the \texttt{MAGPHYS} code \citep{dacunha08} applied to aperture-corrected photometry from \citet{Santos20} spanning $\approx0.3-5\ \mu$m in the well-studied COSMOS field. Nebular attenuation, $E(B-V)$, is estimated from the Balmer decrement based on H$\alpha$/H$\beta$ following \citet{Reddy20}. For further details we refer readers to \citet{Matthee21}.

\section{Results: The Conditions for Lyman Continuum Escape} \label{sec:results}
In this section we explore the physical differences between the High Escape and Low Escape stacks. First, we point out similarities: within errors, the stellar mass ($M_{\rm{\star}}$), UV luminosity ($M_{\rm{1500}}$), and UV slope ($\beta_{\rm{UV}}$) show no significant difference (Table \ref{table:stacks}). This implies that for LAEs the Ly$\alpha$ line-profile and the inferred \fescs do not strongly depend on these properties. Now, based on the differences between the stacks we aim to understand the differing physical conditions between leakers and non-leakers. In what follows we split the results in groups of features pertaining to the production of ionizing photons, the ISM they are radiated into, and finally the ease with which they are able to escape their parent galaxy.

\begin{table}
    \centering
    \caption{Summary of measured properties for the High Escape and Low Escape stacks. We report medians and bootstrapped errors on medians ($16^{\rm{th}}$ and $84^{\rm{th}}$ percentile). Upper limits are $99^{\rm{th}}$ percentile values from bootstrapping. EWs are in the rest-frame. All emission-line ratios are dust-corrected.}

    \begin{tabular}{l|c|c}
 \textbf{Basic Properties}  & \textbf{High Escape} & \textbf{Low Escape} \\ 
   & \textbf{ ($f_{\rm{esc}}>20\%$)} & \textbf{ ($f_{\rm{esc}}<5\%$)} \\ \hline
log$_{10}$(M$_{\rm \star}$/M$_{\odot}$) & $9.2\pm0.2$  & $9.4\pm0.2$ \\
M$_{1500}$ & $-19.7\pm0.3$ & $-20.2\pm0.3$ \\
$\beta$ & $-2.10\pm0.21$ & $-1.95\pm0.16$ \\
& & \\
\textbf{Production} & & \\ \hline
  EW$_{\rm HeII1640}$/{\AA} & $1.9^{+0.8}_{-0.5}$ & $<0.9$ \\
  EW$_{\rm OIII]1661+1666}$/{\AA} & $2.2^{+0.8}_{-0.7}$ & $1.6\pm0.5$ \\
  EW$_{\rm CIII]1907+1909}$/{\AA} & $6.8^{+3.3}_{-2.1}$ & $6.4^{+1.9}_{-1.8}$ \\
EW$_{\rm [OIII]4960+5008}$/{\AA} & $820\pm260$ & $670\pm160$ \\
EW$_{\rm H\alpha}$/{\AA} & $720\pm200$ & $430\pm110$ \\
 log$_{10}(\xi_{\rm ion}$/Hz erg$^{-1}$)  & $25.57^{+0.03}_{-0.03}$ $_{(f_{\rm esc}=0.0)}$ & $25.55^{+0.06}_{-0.07}$ \\ 
    & $25.87^{+0.03}_{-0.03}$ $_{(f_{\rm esc}=0.5)}$ &  \\

  & & \\
 \textbf{Escape} & & \\ \hline
 $f_{\rm esc, Ly\alpha} = \frac{L_{\rm Ly\alpha}}{8.7 L_{\rm H\alpha, int}}$ & $47^{+3}_{-8}$ \% & $9^{+2}_{-2}$ \% \\
  $R_{\rm MgII}= \frac{\rm MgII_{2796}}{\rm MgII_{2803}}$ & - & $0.9^{+0.4}_{-0.3}$ \\ 
    $v_{\rm Ly\alpha, red}$/km s$^{-1}$ & $+106\pm3$ & $+254\pm4$ \\
  $v_{\rm CIV}$/km s$^{-1}$ & $+60\pm20$ & - \\
  $v_{\rm MgII}$/km s$^{-1}$ & $+30\pm10$ & $+130\pm10$ \\
 & & \\

  \textbf{Production \& Escape} & & \\ \hline

  EW$_{\rm Ly\alpha}$/{\AA} & $111\pm6$ & $61\pm3$ \\
 EW$_{\rm CIV1548+1550}$/{\AA} & $2.0\pm0.4$ & $<1.4$ \\
 
 EW$_{\rm MgII 2796}$/{\AA} & $6.7^{+2.5}_{-2.0}$ & $5.7\pm1.8$ \\
 EW$_{\rm MgII 2803}$/{\AA} & $<3.4$ & $6.1^{+1.7}_{-1.5}$ \\
& & \\

 \textbf{ISM conditions} & & \\ \hline
 E$(B-V)$ & $0.00^{+0.07}_{-0.00}$ & $0.34^{+0.10}_{-0.09}$ \\
 O32 = ${\rm \frac{[OIII]_{5008}}{[OII]_{3727,3729}}}$ & $8.4^{+2.2}_{-1.6}$ & $2.7^{+0.4}_{-0.3}$ \\
O3Hb = ${\rm \frac{[OIII]_{5008}}{H\beta}}$ & $4.3^{+0.6}_{-0.5}$ & $6.0^{+0.5}_{-0.5}$ \\
Ne3O2 = ${\rm \frac{[NeIII]_{3870}}{[OII]_{3727,3729}}}$ & $0.7^{+0.3}_{-0.2}$ & $0.3^{+0.1}_{-0.1}$ \\
R23 = ${\rm \frac{[OIII]_{4960, 5008} + [OII]_{3727,3729}}{H\beta}}$  & $6.3^{+0.9}_{-0.7}$ & $10.3^{+1.2}_{-1.0}$ \\
N2Ha = ${\rm \frac{[NII]_{6584}}{H\alpha}}$& $<0.08$ & $<0.03$ \\ 
12+log(O/H) & $8.2\pm0.3$ & $8.1\pm0.1$\\
    \end{tabular}
    \label{table:stacks}
\end{table}

\subsection{Production: High Escape accompanies extreme $\xi_{\rm{ion}}$ and hard ionizing spectra revealed by \HeII\ and \CIV\ emission}
\label{sec:production}

Here we focus on the ionizing photons produced by the stellar populations powering our stacks \textit{before} they make it into the ISM. Based on rest-frame optical and UV line ratios (e.g., \OIII/H$\beta$, \CIV/\CIII, \CIV/\HeIIL), we find that both stacks have emission-lines that are photoionised by young stars and not by AGN activity \citep[e.g.,][]{Kauffmann03,Juneau14,Feltre16}.

For stellar populations, a canonical quantity in the context of ionizing photon production is $\xi_{\rm{ion}}$, the Hydrogen ionizing photon production efficiency, which is cast in terms of the rate of Hydrogen ionizing photons ($\mathrm{N(H^{0})}$) produced per unit (intrinsic) UV luminosity ($L_{1500}$). We derive this quantity in terms of the dust-corrected H$\alpha$ and UV luminosities using the Balmer decrement \citep[e.g.,][]{Bouwens16b,Matthee17,Shivaei18}:
\begin{equation}
    \xi_{\mathrm{ion}} = \frac{N(H^0)}{L_{1500}} =
    \frac{L(\rm{H}\alpha)}{(1-f_{\rm{esc}})}\frac{1}{L_{1500}}
    [\mathrm{7.4\times10^{11} s^{-1}/erg\  s^{-1}Hz^{-1}}].
\end{equation}

We measure a $\log({\xi_{\rm{ion}}/\rm{Hz\ erg^{-1}}})=25.55^{+0.06}_{-0.06}$ in the low \fescs stack (assuming LyC \fesc$=0\%$), and $\log({\xi_{\rm{ion}}/\rm{Hz\ erg^{-1}}})=25.87^{+0.03}_{-0.03}$ in the high \fescs stack (assuming LyC \fesc$=50\%$, see \S\ref{sec:fesc}; for \fesc$=0\%$ we find $25.57^{+0.03}_{-0.03}$). For the low escape stack (E$(B-V)\approx0.3$) we caution the dust correction is uncertain on a 0.3 dex level due to the unknown stellar-to-nebular attenuation and differences across dust curves \citep[e.g.,][]{Shivaei18}, while this is not a concern for the leakers which have E$(B-V)\approx0$.

The High Escape stack also appears to have a harder ionizing spectrum. Prominent narrow \CIVL\ and \HeIIl\ emission is detected at SNR of 4.7 and 3.5, respectively, while there is no sign of these lines among the non-leakers (third row, Figure \ref{fig:stack}; Figure \ref{fig:FescProps}). The HeII EW is at least a factor $2\times$ higher among the leakers. Strong \HeII\ emission is clear evidence for the production of photons with $>54.4$ eV \citep[e.g.,][]{Shirazi12,Berg2018,Nanayakkara19,Saxena20}. That these features are seen in the \textit{median} stack implies such hard ionizing photons occur routinely among LyC leakers. To put the shape of the SED in perspective, the effective ionizing spectral slope, $\alpha$\footnote{$\alpha$ is defined such that integrating $f_{\rm{\nu}}\propto \nu^{\alpha}$ for $<912\AA$ matches the total number of $<912\AA$ ionizing photons computed by integrating the model SED \citep[e.g.,][]{Becker13}.}, of the BPASS burst SEDs that produce the observed H$\alpha$ EWs accounting for an \fesc$\approx50\%$ \citep[e.g.,][]{Stanway20} is \textit{shallower} than the slopes typically adopted for AGN ($\alpha\approx-1.3$ vs. $\alpha\approx-1.7$, e.g., \citealt{Becker13}). However, note that $\alpha$ only effectively captures the total number of Hydrogen ionizing photons, and typical quasar SEDs \citep[e.g.,][]{Lusso15} still produce a higher number of Helium ionizing photons at fixed $M_{\rm{UV}}$.

\subsection{Production \& Escape: hints from \CIV}
\label{sec:civ}

The \textit{simultaneous} detection of nebular \CIV\ emission alongside \HeII\ (\S\ref{sec:production}) in leakers is evidence that  $<260 \AA$ photons are not only being produced but might also be escaping the ISM \citep[][]{Berg19}. The resonant \CIV\ line, analogous to \lya, is sensitive to the column density of high-ionization gas. This imprint of the column density on \CIV\ may be seen among the MUSE \HeII\ emitters \citep{Nanayakkara19}, only a small fraction of which show \CIV\ in emission while the majority show interstellar absorption and/or stellar wind features \citep[e.g.][]{Plat2019}. This is despite \CIV\ requiring a lower ionization energy than \HeII\ (47.9 eV versus 54.4 eV, \citealt{Draine11}), and despite the presence of sufficient Carbon in the ISM (\CIII\ is detected). In these sources \CIV\ may be suffering significant absorption and scattering -- $<260\AA$ photons are being produced but likely fail to escape the ISM. However, tellingly, one of the highest EW \CIV\ emitters in the MUSE \HeII\ emitter sample is the $z\approx2.2$ LyC leaker XLS-26/GS-30668/MUSE-1273 that we discussed earlier in the context of central Ly$\alpha$ escape in \S\ref{sec:classify} (Figure \ref{fig:XLSfesc}). A similar scenario as in XLS-26 occurs in the High Escape stack, where nebular \CIV\ emission appears alongside \HeII. The \CIV\ doublet ratio indicates some absorption in the ISM -- the blue line is weaker than the expectation for pure emission based on the relative oscillator strengths. The line is observed relatively close to the systemic velocity ($+60\pm20$ km s$^{-1}$) implying little scattering.

\subsection{ISM: large differences in attenuation at comparable metallicity}
\label{sec:dust}

Once the ionizing photons leave their sites of production their fate is decided by the contents, density, and geometry of the ISM they encounter. The gas-phase metallicities of our stacks are similar within errors : 12 + $\log{\rm{O/H}}$ of $8.2\pm0.3$ versus $8.1\pm0.1$. We measure this with a composite of strong-line indicators -- R$_{\rm{23}}$, \OIII/\OII, \OIII/H$\beta$, \NeIII/\OII\ -- calibrated on high-redshift analogues in the local Universe \citep{Bian2018}. We caution that for the High Escape group, R$_{\rm{23}}$ and \OIII/H$\beta$ yield much (0.5 dex) higher metallicity (12 + $\log{\rm{O/H}}\approx8.4$) compared to the the other two indicators. This is potentially because these two indicators are bi-valued and lose sensitivity around sub-solar ($\lesssim20\%$) metallicity ranges \citep[e.g.,][]{Perez-Montero21}. For the Low Escape group the indicators are in better agreement.

Dust is expected to be a key inhibitor of LyC \fescs \citep[e.g.,][]{Inoue01,Chisholm18}. The Low Escape stack has a Balmer decrement of $4.1\pm0.4$ indicating widespread dust attenuation among the non-leakers. On the other hand, in the High Escape stack the Balmer decrement is indistinguishable from the expected value for case B recombination ($2.8\pm0.2$) in gas with electron temperatures 10-15 kK indicating essentially dust-free pathways for LyC escape, at least outside the HII regions in which the ionizing photons were produced, as the Balmer decrement is insensitive to the attenuation law at $\lambda<912$ {\AA} \citep{Israel80,Reines2008}. Understanding the dust law at $\lambda<912$ {\AA} is important, now that we know LyC leakers produce copious photons far below the Lyman edge (see \S\ref{sec:feschard}). We emphasize that there is no explicit information on the attenuation in our stacking criteria (\lyas $v_{\rm{sep}}$ and $f_{\rm{cen}}$), so the Balmer decrements are a clear, independent validation that High Escape is associated with low (negligible) attenuation. 

\subsection{ISM: super star cluster-like extreme ionization state in leakers revealed by elevated \OIII/\OII$>8$}
\label{sec:o32}

The ionization parameter ($U$) -- the ratio of the number density of ionizing photons to the number density of Hydrogen gas -- is typically used to characterize the state of photoionized gas in galaxies \citep[e.g.,][]{Sanders15}. In our stacks, Ne3O2 ($\rm{[NeIII]_{3870}/[OII]_{3727,3729}}$) and \OIII/\OII\ are tracers of the ionization parameter \citep[e.g.,][]{Levesque14,Strom17,Maiolino19}. 

\begin{figure}
\vspace{-0.4cm}
\centering
\includegraphics[width=\linewidth]{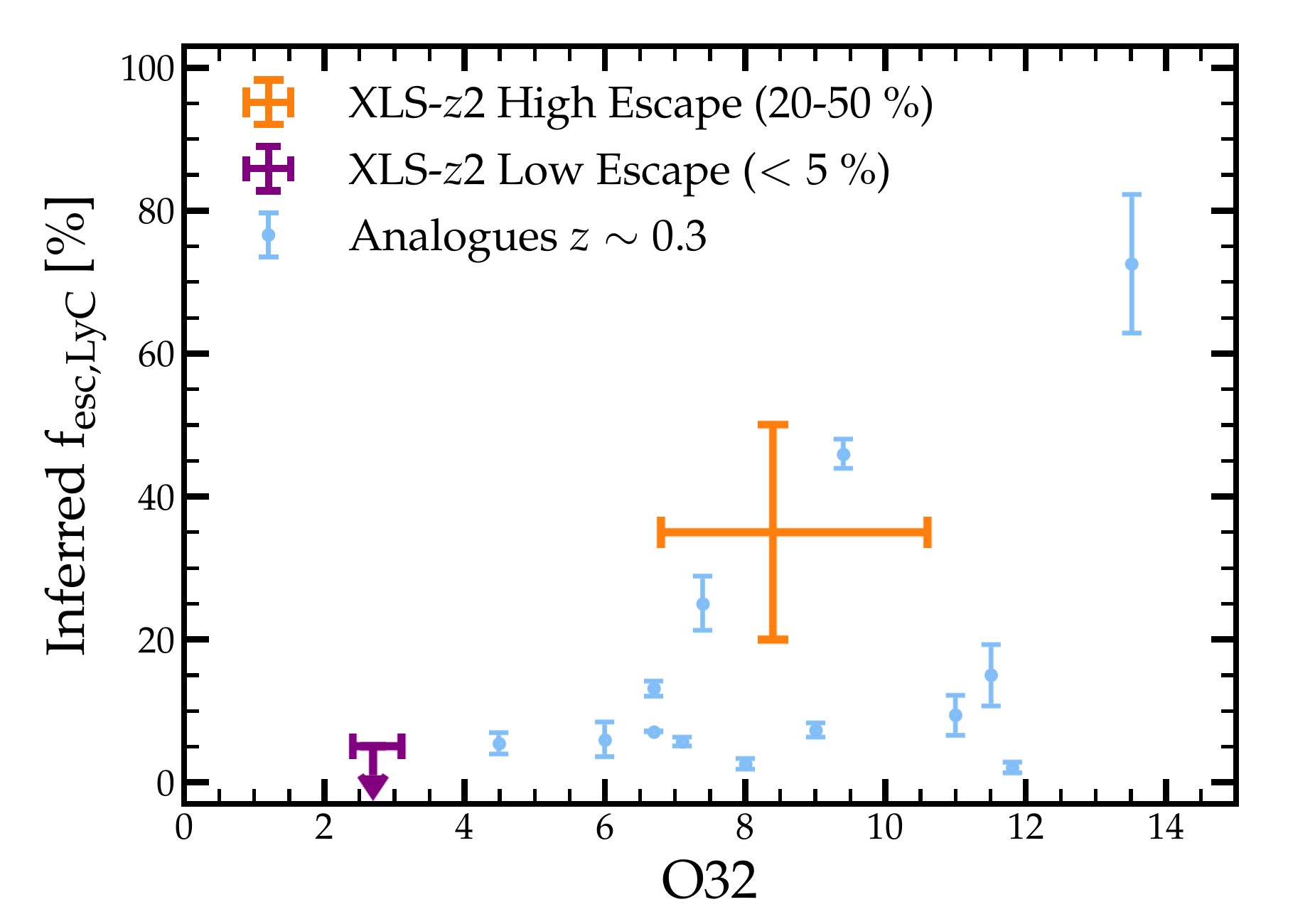}\\
\includegraphics[width=\linewidth]{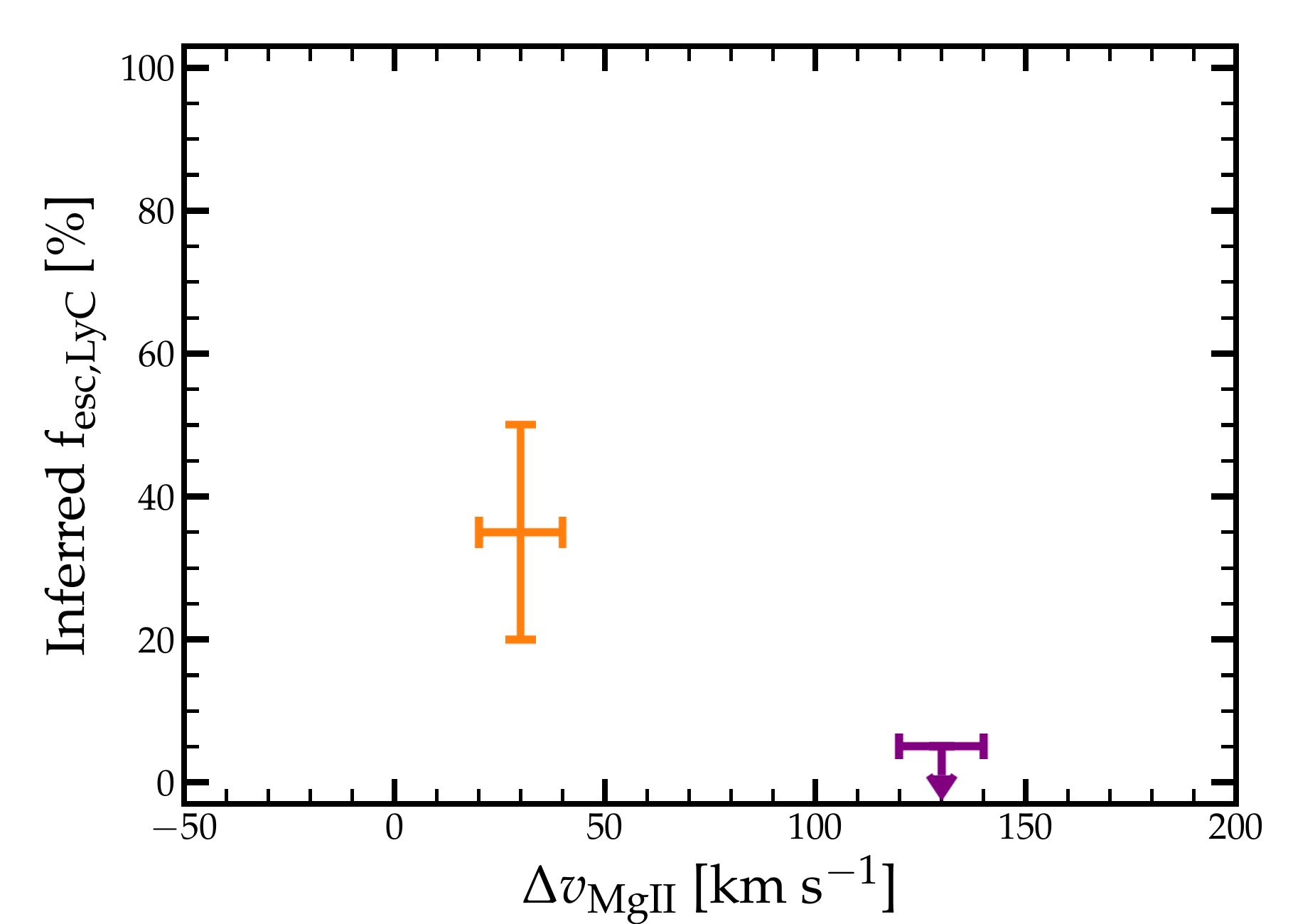}\\
\includegraphics[width=\linewidth]{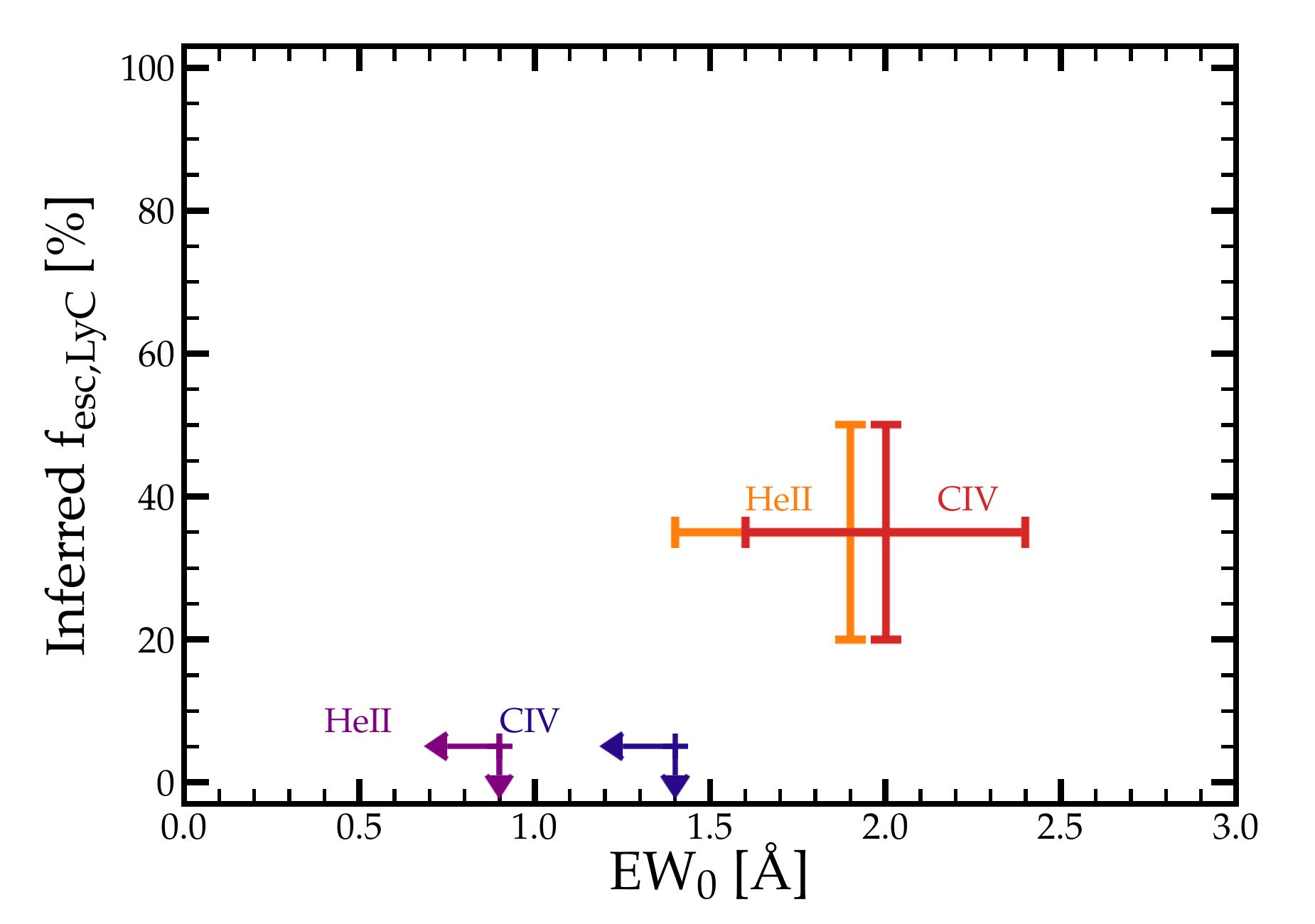}\\
\caption{Spectroscopic tracers of LyC \fescs identified in this work that will be easily accessible with \textit{JWST} -- the \OIII/\OII\ ratio (top), the MgII velocity offset (middle), and the rest-frame EWs of high ionization lines (\HeII, \CIV; bottom). The High Escape stack is represented in shades of orange, while the Low Escape stack is shown in shades of purple. We depict the full conservative range (LyC \fesc$\approx20-50\%$) for the High Escape stack (see Table \ref{table:fesc_summary}). For comparison, individual \OIII/\OII\ measurements for the $z\approx0.3$ \citet{Izotov21} Green Peas are shown in the top panel -- note that most of the GPs were selected for follow-up because they have high \OIII/\OII. \MgII\ offsets and UV emission line EWs are not yet available for these sources.}
\label{fig:FescProps}
\end{figure}

One of the most striking differences between the two stacks is the \OIII/\OII\ ratio (bottom row, Figure \ref{fig:stack}; Figure \ref{fig:FescProps}) -- $8.4^{+2.2}_{-1.6}$ in the High Escape stack versus $2.7^{+0.4}_{-0.3}$. This translates to a $\log{U}=-2.3\ (-2.6)$ for the High (Low) Escape stacks adopting the \citet{Strom18} calibration appropriate for $z\approx2$. Likewise, the Ne3O2 ratio also implies $\log{U}=-2.3\ (-2.5)$ for the High (low) Escape subsets. The $\log{U}=-2.3$ of our High Escape stack is among the highest observed for a population, close to the theoretical/observational threshold in the $z\approx0$ Universe \citep{Dopita06,Kewley19, Perez-Montero21}, comparable to confirmed LyC leakers at $z\approx0.3$ \citep{Guseva2020}, and $\approx0.5$ dex higher than continuum-selected samples at $z\approx2-3$ \citep[e.g.,][]{Strom18, Topping20}.

As a population, the elevated ionization state of the High Escape stack is comparable to the largest star clusters \citep[][]{Kewley19}, the so-called ``super star clusters",  which routinely show $\log{U}\approx-2.3$ \citep[e.g.,][]{Indebetouw09,Leitherer18,Micheva19}. These compact (order 10 pc), $\approx10^{6}M_{\star}$ complexes of young (order $10$ Myrs), massive stars may be the key sites for LyC production and escape \citep[e.g.,][]{Vanzella19,Vanzella21,Ostlin21}.

\subsection{Escape: Optically thin gas traced by \lyas \fescs}
\label{sec:lya}

The stacked \lyas profiles (top panel, Figure \ref{fig:stack}), by design, show the expected signatures -- the High Escape sources have an \fcen$\approx 0.27$ and a red-peak that is $\approx100$ km s$^{-1}$ from line centre (see Table $\ref{table:stacks}$), indicating low column densities, ionized channels, and modest scattering (\S\ref{sec:motivate}). The Low Escape stack on the other hand has little flux emitted at line center (\fcen$\approx 0.03$), and its red peak is observed at $\approx250$ km s$^{-1}$ from the systemic velocity, indicating considerably more scattering than the High Escape stack. The peak separation ($v_{\rm{sep}}$) is not well-defined in either case because the location of the blue peak is ambiguous.

Now we dwell on other aspects of the \lyas line that did not go into our selection of the subsets. The \lyas \fesc, which is computed by comparing the observed \lyas luminosity with the dust-corrected H$\alpha$ luminosity, provides a useful upper bound on the LyC \fesc. This is clear empirically \citep[e.g.,][]{Gazagnes20,Izotov21}, through radiative transfer simulations \citep[e.g.,][]{Dijkstra16, Kimm19}, and makes intuitive sense: \lyas photons can scatter and escape through somewhat higher column density gas, while LyC cannot. The \lyas \fescs for our High Escape stack is $47^{+3}_{-8}\%$ whereas for the Low Escape stack we find $9^{+2}_{-2}\%$ (Table $\ref{table:stacks}$). This is a strong cross-check on the robustness of our stacks -- the Low Escape stack is ruled out from having a LyC \fesc$>10\%$ while the High Escape stack may have an \fescs as high as $\approx50\%$. 

We find that the leakers have a \lyas EW $\approx2\times$ higher than the non-leakers. However, note that the $61\pm3\AA$ EW in the Low Escape stack is also substantial and demonstrates that \lyas EW by itself is an impure predictor of \fesc. This is discussed further in \S\ref{sec:EWs}.

\subsection{Escape: Optically thin gas traced by \MgII}
\label{sec:mgii}

The \MgII\ doublet has a similar ionization potential to that of H$^{0}$ (15 eV), resonantly scatters like \lya, and will be within the grasp of \textit{JWST} at $z>6$ when \lyas is damped by the neutral IGM \citep[e.g.,][]{Henry18,Feltre18,Chisholm20}. Our stacks show that the \MgII\ doublet can be used as an indirect tracer of HI column density (fourth row, Figure \ref{fig:stack}; Figure \ref{fig:FescProps}). The Low Escape stack shows redshifted ($+130\pm10$ km s$^{-1}$) \MgII\ emission while in the High Escape stack \MgII\ emission arises much closer to the line-centre ($+30\pm10$ km s$^{-1}$). Further, the line ratio of the \MgII\ doublet ($R_{\rm MgII 2796/2803}$) is in agreement with recent results from \cite{Chisholm20} who argued the column density of neutral Hydrogen is proportional to $R_{\rm MgII 2796/2803}$ in the optically thin regime. Indeed, in the Low Escape stack $R_{\rm MgII 2796/2803}\approx1$ whereas in the High Escape stack the redder line in the doublet is undetected, implying a higher line ratio. 

\subsection{Escape: Low covering fraction in leakers revealed by CII absorption}
\label{sec:cii}

The covering fraction of neutral gas is a measure of how riddled with ionized channels (``holes") the ISM is. Covering fractions, as inferred from ISM absorption lines (both metal lines as well as Hydrogen lines), are expected to trace LyC \fesc, with a higher covering fraction corresponding to lower \fescs \citep[e.g.,][]{Reddy16b,Gazagnes18, Gazagnes20,Mauerhofer21}. Our sensitivity for detecting HI absorption lines blue-wards of Ly$\alpha$ is low. In stacks of LAEs the strongest low-ionisation interstellar absorption lines are typically SiII and CII \citep{Trainor15}. We clearly detect CII absorption in the Low Escape stack (with an EW$=-1.6\pm0.2$ {\AA}), and no such absorption feature in the High Escape stack (with a 2$\sigma$ limiting EW $>-0.6$ {\AA}; Figure \ref{fig:stack}). No significant SiII absorption is detected in any of the stacks (2$\sigma$ limiting EWs $>-0.8$ {\AA} and $>-0.6$ {\AA} for the High and Low escape stack, respectively). The difference between SiII and CII is likely of instrumental origin as our sensitivity is a factor $\approx1.4$ better around CII then around SiII. The differences between the CII absorption strengths in the stacks is another line of evidence that a porous ISM conducive to high LyC \fescs occurs in the High Escape stack, whereas the Low Escape sources do not have such pathways. 

\section{Results: The Escape Fraction of the High Escape stack}
\label{sec:fesc}

The High Escape stack was constructed purely based on Ly$\alpha$ profiles to have LyC $f_{\rm{esc}}>20\%$, and contains $50\pm10\%$ (binomial error from sample size) of the sample studied in this work (Table \ref{table:sample_selection}). Through multiple spectroscopic indicators ($\S\ref{sec:results}$) we have verified this stack is indeed probing high \fesc. We now estimate what the \fescs of this sample is likely to be.

\subsection{Constraints from the Ly$\alpha$ escape fraction}
\label{sec:fesclya}

A strict upper bound on the LyC \fescs is due to the Ly$\alpha$ \fesc=$47^{+3}_{-8}\%$, the $95^{\rm{th}}$ percentile of which is $\approx55\%$. Both empirical \citep[e.g.,][]{Gazagnes20, Izotov21} and theoretical \citep[e.g.,][]{Dijkstra16,Kimm19,Kimm2021} work show that the \lyas \fesc $\geq$LyC \fesc. Ly$\alpha$ and LyC likely emanate from the same production sites powered by young stars, and the resonant scattering of Ly$\alpha$ gives it an added advantage when it comes to escaping the ISM. We deem this broad range ($20-55\%$) our ``conservative" estimate since it encompasses the entire realm of possibility for our stack.

We make a finer estimate of the LyC \fescs by observing that in the \citet{Izotov21} Green Pea compilation, the higher the LyC \fesc, the closer it is to the \lyas \fesc. This trend is supported by the \citet{Kimm19} simulations in which turbulent clouds with LyC \fesc$>20\%$ have $f_{\rm{esc,LyC}}/f_{\rm{esc, Ly\alpha}}\approx1$. Indeed, for the seven \citet{Izotov21} GPs that satisfy our High Escape selection criteria (Eq. \ref{eqn:highfesc}) we calculate a bootstrapped ratio of $f_{\rm{esc,LyC}}/f_{\rm{esc, Ly\alpha}} = 0.82^{+0.16}_{-0.15}$. This ratio produces an LyC \fescs of $38^{+9}_{-8}\%$ for the High Escape stack. 

The Ly$\alpha$ \fescs we use for this estimate is calculated via the same assumptions as the \citet{Izotov21} compilation (e.g., the intrinsic Ly$\alpha$/H$\alpha$ ratio is matched). Further, the spatial scale probed by the \textit{HST}/COS apertures in the Green Pea studies ($\approx1.3''$ radius, $7-10$ kpc at $z\approx0.3$) is comparable to our 1'' slits at $z\approx2$. This ensures consistent Ly$\alpha$ \fescs comparison, given the spatially extended nature of Ly$\alpha$ emission \citep[e.g.,][]{Hayes13,Wisotzki15}. We also clarify that this argument does not imply that high \lyas \fescs selects for high LyC \fescs -- the point is that at high LyC \fesc, $f_{\rm{esc,LyC}}/f_{\rm{esc, Ly\alpha}} \approx 1$ which is supported by these $f_{\rm{esc}}>20\%$ systems being in ``density bounded nebulae" that are transparent to LyC along all lines of sight \citep[e.g.,][]{Ramambason20}, thus diminishing the resonance advantage of Ly$\alpha$ over LyC. 

\subsection{Constraints from the Ly$\alpha$ red peak and HI covering fraction}
\label{sec:fescredpeak}

The HI covering fraction ($f_{\rm{cov}}$) -- the fraction of high column density (N(HI)$>10^{16} \rm{cm^{-2}}$) channels --  has been used as a successful predictor of the LyC escape fraction \citep[e.g.,][]{Reddy16b,Reddy21,Chisholm18} as follows:

\begin{align}
\label{eqn:fcov_ion}
    f_{\rm{esc}} &= \left(1-f_{\rm{cov}}\right)\times10^{-0.4 A(\lambda=912\AA)}.
\end{align}

In our case the attenuation is negligible, so we set $A(\rm{\lambda=912\AA})=0$. To estimate $f_{\rm{cov}}$ we exploit the $3\sigma$ correlation with the Ly$\alpha$ red peak velocity ($V_{\rm{red}} - V_{\rm{trough}}$) reported in \citet{Gazagnes20}. Note that the red peak velocity is measured with respect to the Ly$\alpha$ ``trough", i.e., the minima between the red and blue peak in typical double-peaked profiles. The \citet{Gazagnes20} sample, mostly drawn from the \citet{Izotov21} compilation, has very similar Ly$\alpha$ resolution to the sample studied here, so differential effects are limited. The significant, albeit noisy, relationship (S. Gazagnes, private comm.) is as follows:

\begin{align}
    f_{\rm{cov}} = \left(0.29\pm0.10\right)\times\left(V_{\rm{red}} - V_{\rm{trough}}\right)/(100\ \rm{km\ s^{-1}}) + 0.14\pm0.22.
\end{align}

We cannot apply this metric directly to the stack since there is no clear trough in the profile -- instead, we apply it object by object to each individual source and compute the median $f_{\rm{cov}}$. In the three cases where there is no trough in the profile we either set aside the source (XLS-24) or assume $V_{\rm{red}} - V_{\rm{trough}}=0$ (XLS-20, XLS-28). The adopted values for all galaxies are shown along with their profiles in Appendix \ref{appendix:lyaprofiles}. The median $V_{\rm{red}} - V_{\rm{trough}}$ for our sample is $147^{+28}_{-28}$ km s$^{-1}$ which translates to $1-f_{\rm{cov}} = 43^{+26}_{-26}\%$.

We have verified that for the seven GPs that satisfy the High Escape criteria computing $1-f_{\rm{cov}}$ in this manner results in a number slightly higher than their mean LyC \fescs ($48\%$ vs $35\%$) -- this is due to significant dust attenuation in these sources, i.e., $A(\rm{\lambda=912\AA})$ is not zero. It is important to note that Eqn. \ref{eqn:fcov_ion} assumes an ``ionization bounded nebula" -- i.e., the ionization front is surrounded by an impermeable layer of high column density gas ($N_{\rm{HI}}\gg10^{18}$ cm$^{-2}$) that is perforated by a smattering of low column density channels whose proportion is $\approx1-f_{\rm{cov}}$. However, as we discuss in the following section, \fesc$>20\%$ leakers likely deviate from this physical picture.

\subsection{The difference between Lyman edge \fescs (850-912 \AA) and total \fescs (0-912 \AA)}
\label{sec:feschard}

The optical line ratios and covering fractions of prolific LyC leakers (\fesc$>20\%$) imply they are best described as ``density bounded nebulae" \citep{Ramambason20}. That is, the ionization front is surrounded by low column density gas ($10^{16}-10^{18}$ cm$^{-2}$) punctuated by entirely transparent ($<10^{16}$ cm$^{-2}$) channels (see Fig. 16 of \citealt{Gazagnes20} for an excellent schematic). Modifying Eqn. \ref{eqn:fcov_ion} for this situation and assuming no dust attenuation we have:

\begin{align}
\label{eqn:fcov_ion2}
    f_{\rm{esc, LyC}} &= \underbracket[0.8pt]{\left(1-f_{\rm{cov}}\right)\times1}_{<10^{16}\ \rm{cm}^{-2}} + \underbracket[0.8pt]{f_{\rm{cov}}\times f_{\rm{cov,esc}}}_{10^{16}-10^{18}\ \rm{cm}^{-2}}.
\end{align}

This equation expresses the view that there is a fraction (1-$f_{\rm{cov}}$) of entirely transparent channels with $f_{\rm{esc, LyC}}=100\%$ through which the ionizing continuum emerges as is. However, there is also a fraction of channels ($f_{\rm{cov}}$) that is not entirely transparent ($10^{16}-10^{18}$ cm$^{-2}$) but is permeable to ionizing photons which have an effective escape fraction of $f_{\rm{cov,esc}}$. It is in the context of $f_{\rm{cov,esc}}$ that the $<850\AA$ photons powering our High Escape stack become important.

All the LyC \fescs measurements we have discussed in this paper, devised our selections around, and used to empirically estimate the LyC \fescs in the previous sections were made at the Lyman edge ($\approx850-912\rm{\AA}$). However, photons below the Lyman edge ($<850\rm{\AA}$) are produced in copious amounts in the High Escape sources, as testified by the presence of strong optical line EWs (e.g., rest-frame [OIII]$_{4960,5008}$+H$\beta\approx$1100 \AA), as well as  \HeIIL\ and \CIVL\ emission. The $<850\AA$ photons have much lower photoionization cross-sections compared to those at $>850\AA$. Thus, there may be significant differences between the total escape fraction measured across the entire ionizing continuum ("total \fesc", $0-912\rm{\AA}$) compared to the escape fraction measured only at the Lyman edge (``edge \fesc", $850-912\rm{\AA}$) \citep[e.g.,][]{Gnedin08,Inoue10,Haardt12, McCandliss17,Kimm19,Berg19}. Since $<850\AA$ photons are an ubiquitous feature of LyC \fesc, occurring in our \textit{median} stack, literature LyC \fescs estimates and the empirical scaling relations we used in this section may be systematically underestimating the total LyC \fesc. And it is the total \fescs that ultimately matters for reionization calculations.

For the High Escape stack, with an $f_{\rm{cov}}\approx60\%$ (\S\ref{sec:fescredpeak}), our estimates from the previous sections are roughly underestimated by $\gtrapprox10\%$ for covering column densities of N(HI)$<10^{18}$ cm$^{-2}$ expected in the density-bound scenario, i.e., a total \fesc$\gtrapprox50\%$ (Figure \ref{fig:hardphotons}). Independently, the \citet{Gnedin08} hydrodynamical simulations provide an explicit scaling of \fesc(0-912$\AA)\approx1.25$\fesc(912$\AA$), which also results in an \fesc(0-912$\AA)\approx50\%$ for our stack.

From the arguments in this section it might seem our Low Escape stack (edge LyC \fesc$<5\%$, \lyas \fesc$<10\%$) must also have a higher total \fesc. However, it displays significant CII absorption implying high column densities (i.e., it is likely ionization bounded as further suggested by its lower O32 ratio). The difference between edge and total \fescs applies only to the $f_{\rm{cov}}\times f_{\rm{cov,esc}}$ term in Eq. \ref{eqn:fcov_ion2} when the covering gas is also transparent ($10^{16}-10^{18}$ cm$^{-2}$). Further, the Low Escape stack lacks the ionizing sources producing $<850\AA$ photons (e.g., no \HeII\ and \CIV\ emission). Due to the correlated nature of low column densities, low dust, and ionizing stellar populations, the difference between edge and total \fescs must be thought of as a ``high edge \fescs implies higher total \fesc'' effect. We also note that a contribution from free-bound emission of H which peaks shortward of the LyC limit may mean that reported edge escape fractions are conversely somewhat overestimated (see \citealt{Inoue10}). Here we only seek to argue that \textit{some} difference likely exists between the edge and total \fesc, and have provided an approximate estimate -- more sophisticated modeling that accounts for e.g., realistic ionization and density structure is warranted. 

\subsection{Consistency with existing LBG and LAE escape fraction constraints}

A back-of-the-envelope consistency check for our estimated LyC \fescs comes from recent stacked \fescs measurements of $M_{\rm{UV}}\lesssim-19$ LBGs at $z\approx2.5-4$ that all find an average \fesc$\approx5-10\%$ \citep{Marchi17, Steidel18,Pahl21}. At these redshifts and for comparable $M_{\rm{UV}}$ the fraction of $L>0.2 L^{*}$ LAEs (i.e., our survey faint limit) in LBG samples is $\approx30\%$ \citep[e.g.,][]{Santos21}. Importantly, the entire $p$($M_{\rm{UV}}| L_{\rm{Ly\alpha}}$) distribution for $L>0.2 L^{*}$ LAEs is contained at $M_{\rm{UV}}<-19$ as seen via the XLS-$z$2 sample \citep{Matthee21}. If half of these LAEs (i.e., $15\%$ of LBGs) have an average \fesc$\approx40\%$ at the Lyman edge, and all other galaxies have \fesc$\approx0\%$ then the stacked \fescs for an $M_{\rm{UV}}\lesssim-19$ LBG sample should be $\approx5-10\%$, in excellent agreement with literature estimates. Another cross check comes from the fraction of individually detected leakers in the LBG samples -- for the Keck Lyman Continuum Survey (KLCS) this fraction is $\approx10\%$ \citep{Pahl21}, which is consistent with the $\approx15\%$ ($<15\%$ with IGM damping) implied by our results. For a more sophisticated exploration of how our LAE constraints translate to the overall LBG population, we refer readers to \S5 of our companion paper (Matthee \& Naidu et al. 2021).

As for LyC studies of LAEs, \citet{Oesch21} and \citet{Bian2020} stacked direct LyC imaging at $z\approx3$ from the Hubble Deep UV Survey \citep[HDUV,][]{Oesch18b} with samples dominated by $M_{\rm{UV}}\lesssim-18$ LAEs from the MUSE-Wide \citep{Urrutia19} and MUSE-HUDF \citep{Bacon17} surveys. These authors report $2\sigma$ upper limits of $\lesssim20\%$ on the $\approx900\AA$ \fescs (see also \citealt{Japelj17} who report consistent estimates using a smaller sample of MUSE LAEs with shallower imaging). On the other hand, the LAE subsample of KLCS ($M_{\rm{UV}}<-19$) has reported \fesc$\approx20\%$, albeit with considerable IGM transmission uncertainties due to their sample size of 26 \citep{Pahl21}. As we argue that only half the LAEs are in the leaking-phase, the imaging constraints are marginally inconsistent and the spectroscopic constraints in excellent agreement with our results that expect these studies to find $f_{\rm esc} \approx20$\% (half the LAEs have edge \fescs of $\approx40\%$).

\begin{table}
    \centering
    \caption{Summary of $f_{\rm{esc}}$ constraints for the High Escape stack from a variety of arguments. Assuming our sample is representative, half the $>0.2L^{*}$ LAEs at $z\approx2$ have this median \fesc.}
    \begin{tabular}{l|r|}
    \hline
Conservative range by selection and $f_{\rm{esc,LyC}}<f_{\rm{esc,Ly\alpha}}$ (\S\ref{sec:fesclya}) & $20-55\%$  \\
Empirical $f_{\rm{esc,LyC}}/f_{\rm{esc,Ly\alpha}}$ ratio for $f_{\rm{esc,LyC}}>20\%$ (\S\ref{sec:fesclya}) & $38\pm9\%$\\
Ly$\alpha$ red peak and covering fraction correlation (\S\ref{sec:fescredpeak}) & $43\pm26\%$\\
Accounting for difference between edge and total \fescs (\S\ref{sec:feschard}) & $\approx50\%$ \\
    \end{tabular}
    \label{table:fesc_summary}
\end{table}

\section{Discussion}
\label{sec:discussion}
\subsection{Implications for constraining LyC \fescs with \textit{JWST}: can strong lines do it all?}
\label{sec:jwst}

A prime directive of LyC studies at $z<4$ is to identify indirect estimators of LyC that are easily accessible during the EoR. The pressing question is, which spectroscopic features must be targeted in future observations at the highest redshifts? Our High and Low Escape stacks show promising and significant median differences in lines like MgII and CII expected to be tightly linked to the HI column density \citep[e.g.,][]{Gazagnes18,Gazagnes20,Mauerhofer21,Henry18,Chisholm20}. This bodes well for programs pursuing these faint features with \textit{JWST}.

Furthermore, highly ionizing stellar populations (seen via \HeII, \CIV\ and extreme optical line EWs), a dust-free, high-ionization state ISM (seen via H$\alpha$/H$\beta$, \OIII/\OII) occur, on average, \textit{simultaneously} in the High Escape stack. The LAEs with a low inferred \fescs on the other hand are on average dusty, have a less ionised ISM and weaker ionizing populations. Therefore, the correlated nature of these conditions may mean that selecting for any one (or any combination) of these properties dramatically increases the chances of selecting for a low HI column density as well. The possible underlying physics driving the correlated conditions is that highly ionizing stellar populations in super star clusters may be carving porous channels in the HI as well as destroying dust while producing a high ionization state ISM.

At first glance, this may imply that a handful of relatively easily observed strong emission lines may be used to implicitly chart \fescs without the need for detecting faint, explicitly HI-linked features such as MgII or [SII] which will likely be measured only for a small fraction of galaxies at $z>6$. However, there are two important subtleties to this picture.

The first subtlety is that our stacking analysis gives us the \textit{average} picture of leakers and non-leakers. Across individual galaxies there is substantial scatter in any given property among our identified correlated conditions (e.g., high ionization state ISM). For example, the top panel in Fig. $\ref{fig:FescProps}$ suggests substantial scatter between \fescs and \OIII/\OII\ in low-redshift analogues (discussed further in \S $\ref{sec:O32}$). Some individual LAEs that are part of the Low Escape stack are known to show nebular HeII and CIV emission (e.g. XLS-22; see \citealt{Amorin17}). Therefore, our stacks should ideally be compared not to any individual galaxies, but to well-defined stacks of galaxies to constrain the \textit{average} \fesc, which is ultimately the key quantity relevant to reionization.

The second subtlety is that our parent sample consists of LAEs with strong Ly$\alpha$ emission. The escape of Ly$\alpha$ emission is likely dependent on the viewing direction \citep[e.g.][]{Behrens2019,Smith19,Smith21}. It is plausible that a \lya-selected sample picks out only galaxies viewed at favorable angles, which might minimize the scatter between \fescs and galaxy properties reported in previous studies \citep[e.g.,][]{Fletcher19,Nakajima20,Tang21}. Alternatively, it is possible that the Ly$\alpha$ pre-selection selects for ``hidden'' parameters that otherwise add scatter to such correlations as well, e.g., the presence of outflows or the relatively low mass of LAEs \citep[e.g.][]{Matthee21}. Thus, the findings of this work apply only to LAEs, and not to the overall galaxy population.

However, the fraction of UV-selected galaxies that are LAEs increases strongly with redshift \citep[e.g.][]{Stark10,Kusakabe20}. At redshifts $z\sim6$, the LAE fraction of bright $M_{\rm{UV}}\gtrsim-19$ galaxies is about 40\%, while it is about 10\% at $z\approx3$ (see \citealt{Ouchi20} for a review). Thus, our findings about LAEs at $z\approx2$ might apply to a significant fraction of the galaxy population at higher redshifts. Naively extrapolating our results, at $z\approx8$ when the LAE fraction among $M_{\rm{UV}}\lesssim-18$ galaxies may approach $\approx100$\%, we would expect half the galaxies to be dust-free leakers with average LyC \fesc$\approx50\%$ and \OIII/\OII$\approx8.5$, while the other half would have \fesc$<5\%$, \OIII/\OII$\approx3$ as well as high E$(B-V)$ evident from Balmer lines. We refer to our companion paper (Matthee \& Naidu et al. 2021) for an analysis how the population averaged \fescs of UV-selected galaxies may evolve with redshift based on what we know about the evolution of Ly$\alpha$ emission from galaxies over $z\approx2-6$.

\subsection{A case for optimism about \OIII/\OII\ as a LyC \fescs predictor at high-redshift} \label{sec:O32}
The \OIII/\OII\ line-ratio has been considered a promising indicator of LyC leakage as it might trace density-bounded nebulae \citep[e.g.,][]{JaskotOey13,Nakajima14} and has been extensively studied due to the relative brightness of both emission-lines. However, tests of this indicator have produced mixed results \citep[e.g.,][]{Naidu18,Chisholm18,Jaskot19,Bassett19,Nakajima20}. On the other hand, we identify a stark contrast in \OIII/\OII\ across our stacks of LAEs (\S $\ref{sec:results}$). 

We consider three effects that may explain the difference between our results and previous work. First, the LAE selection may help reduce the scatter arising from viewing angle effects \citep[e.g. akin to those expected for LyC escape,][]{Gnedin08, Paardekooper15, Cen15} by honing in on galaxies at similar, favorable viewing angles to begin with. The LAE selection also selects galaxies with relatively low masses, compact sizes, high specific star formation rate, and elevated star-formation surface density - conditions that may correlate with \fesc, and thereby further reduce scatter \citep[e.g.][]{Heckman11,Marchi18,Cen20,Naidu20,Matthee21}. Second, the absence of a strong correlation between \fescs and \OIII/\OII\ in high-redshift analogues at $z\approx0.3$ that are all LAEs \citep[e.g.,][]{Izotov21} may be due to physical sources of scatter that are absent at higher-redshift. These could, for example, be diverse star formation histories on longer timescales that drive differences in chemical abundances in the ionising stellar populations. Third, the absence of strong differences in \OIII/\OII\ between LAEs classed as LyC leakers or non-leakers using direct imaging experiments \citep[e.g.,][]{Fletcher19} may be explained by stochasticity in the IGM transmission \citep[e.g.][]{Steidel18}. Differences in properties such as \OIII/\OII\ are obscured if average IGM transmission values are applied to compute \fescs for individual sources -- such samples are then effectively split by IGM transmission and not by whether sources are genuine leakers or non-leakers \citep[e.g.,][]{Bassett21}. 

Our Ly$\alpha$ line-profile based strategy likely minimizes viewing angle effects, bypasses IGM transmission stochasticity, and thus helps clearly identify the physical conditions in galaxies associated with LyC escape at high-redshift. The increased prevalence of LAEs with increasing redshift, and the clear variations in \OIII/\OII\ among LAEs with high and low \fescs are likely cause for optimism for \OIII/\OII\ as a stand-alone indicator of the average LyC \fescs for galaxies at the highest redshifts.

\begin{figure}
\centering
\includegraphics[width=1.02\linewidth]{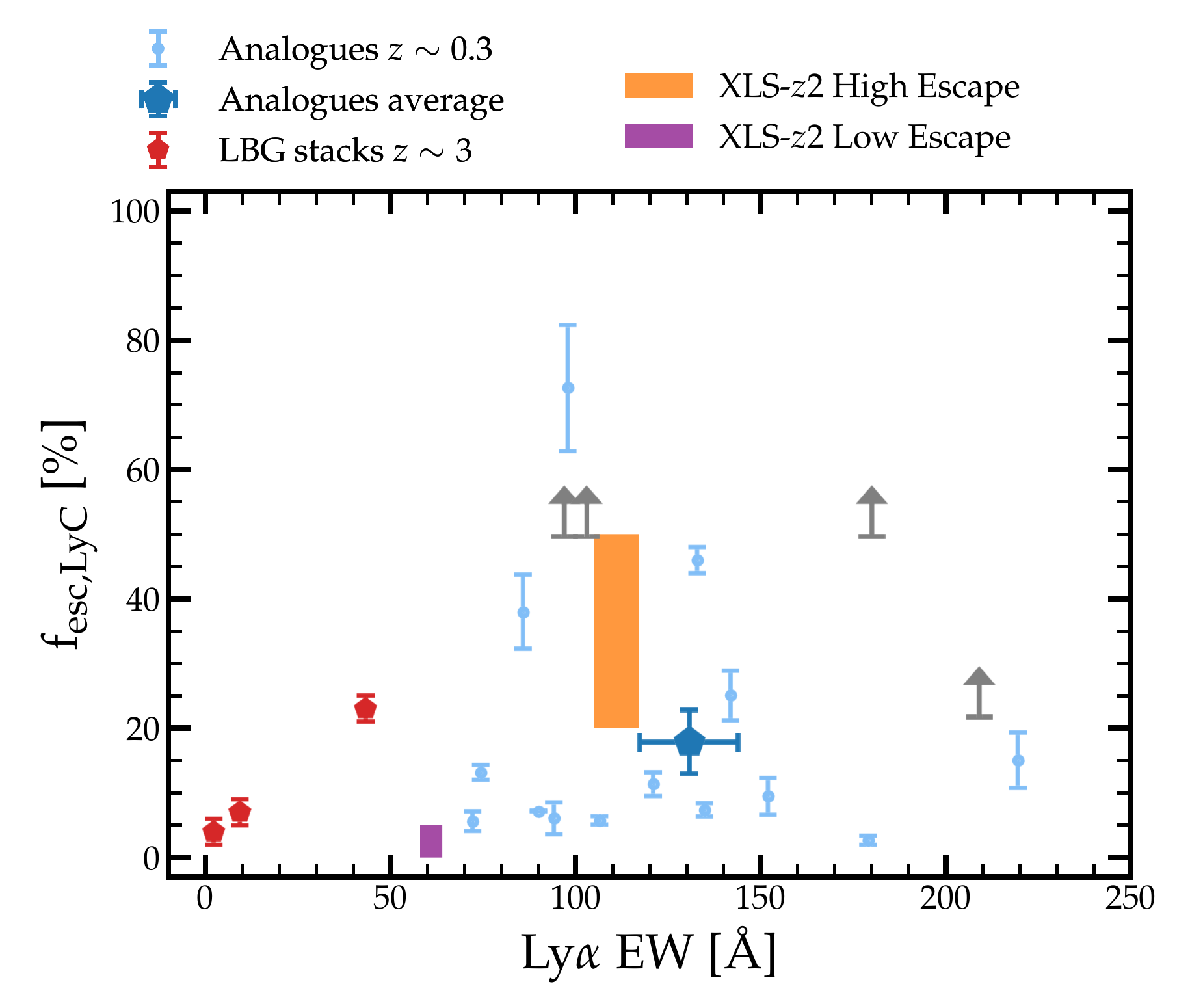} \vspace{-0.3cm}
\caption{Compilation of observed LyC \fesc and observed rest-frame Ly$\alpha$ EWs from stacks of Lyman-break galaxies at $z\sim3$ (\citealt{Steidel18,Pahl21}), low-redshift Green Pea galaxies that are analogues to high-redshift galaxies (\citealt{Izotov16a,Izotov16b,Izotov18a,Izotov18b}) and a handful of individual high-redshift leakers (GS30668, Ion2, Sunburst Arc, see \S\ref{sec:classify}). The \fescs values in high-redshift sources are displayed as lower limits due to their uncertain IGM correction. The orange and purple shaded regions show the Ly$\alpha$ EWs and full conservative range of the estimated escape fractions of the High ($20-50\%$) and Low Escape ($<5\%$) XLS-$z2$ stacks.}
\label{fig:LyaEWFesc}
\end{figure}

\subsection{The interplay between Ly$\alpha$ EW and LyC $f_{\rm{esc}}$: high \fescs does not imply low EWs}
\label{sec:EWs}

It may seem intuitive that high LyC \fescs sources must have weak emission lines since large fractions of ionizing photons are lost to the IGM without exciting emission in the ISM. Thus, a potential concern underlying this work is that the Ly$\alpha$ line luminosity may decrease with increasing \fesc, and the most prolific LyC leakers are missed in our Ly$\alpha$-selected sample. On the other hand, it is expected \citep[e.g.,][]{Dijkstra16} as well as observed \citep[e.g.,][]{Izotov20} that the Ly$\alpha$ escape fraction (and thus the emerging Ly$\alpha$ luminosity) is correlated with the LyC escape fraction as both are sensitive to the HI column density. 

The Ly$\alpha$ EW that emerges from a galaxy is plausibly proportional to the intrinsic \lyas EW associated with a stellar population and the Ly$\alpha$ escape fraction. As the Ly$\alpha$ and LyC escape fractions are correlated \citep[e.g.][]{Izotov21,Kimm2021}, there is a regime of small LyC \fescs where variations in Ly$\alpha$ EWs correlate with $f_{\rm esc,LyC}$ producing a linear relation \citep[e.g.][]{Steidel18}. It is expected that the correlation between Ly$\alpha$ EW and $f_{\rm esc,LyC}$ breaks or flattens eventually \citep[e.g.][]{Nakajima14}, as very high escape fractions impact the Ly$\alpha$ source term too much. The question is where this break happens, which is complicated as galaxies show varying intrinsic Ly$\alpha$ EWs (i.e. due to differences in stellar ages and metallicities).

We investigate these effects in Fig. $\ref{fig:LyaEWFesc}$, where we compiled measured Ly$\alpha$ EWs and LyC \fesc, and also illustrate the conservative range of \fescs (Table \ref{table:fesc_summary}) for the High and Low Escape LAEs. The LyC leaker GPs at $z\approx0.3$ are all accompanied by high observed Ly$\alpha$ EWs $>70$ {\AA} \citep[e.g.][]{Izotov16a,Izotov16b,Izotov18a,Izotov18b}, without the Ly$\alpha$ EW being a criterion in their sample selection. For the leaking GPs, the average LyC \fescs is $\approx$ 20 \% and the average Ly$\alpha$ EW $\approx130$ {\AA}. This suggests that these GPs are experiencing particularly young bursts \citep{Chisholm2019} leading to high intrinsic EWs. None of the confirmed leakers with \fescs $>10$ \% have a Ly$\alpha$ EW that is below 20-25 {\AA} (i.e. the typical selection thresholds for LAE surveys). On the other hand, there are several galaxies with a Ly$\alpha$ EW $>100$ {\AA}, but a moderate $<10$\% escape fraction. 

There are two key takeaways from this compilation. The first is that the conditions that produce high \fescs (e.g., the presence of young, extremely ionizing stellar populations) often also produce very high intrinsic EWs such that even the weakened EWs due to the (1-\fesc) in the source term are high (e.g., $\approx110\AA$ for Ly$\alpha$ in our High Escape stack compared to $\approx60\AA$ for the Low Escape stack). The second takeaway is that there is considerable scatter in the relation between Ly$\alpha$ EW and \fesc, both in individual sources and in our two High and Low Escape stacks, which both show relatively high Ly$\alpha$ EW. High LyC leakage is found in galaxies with high emerging Ly$\alpha$ EWs, suggesting that Ly$\alpha$ EW may serve as a complete pre-selector of LyC leaking galaxies. However, Ly$\alpha$ EW alone will also select a high number of false-positives. 

\subsection{The coincidence of high \fescs with high $\xi_{\rm{ion}}$ and timing the LyC escape phase to $\approx2-10$ Myrs} \label{sec:coincidence}

\begin{figure}
\centering
\includegraphics[width=\linewidth]{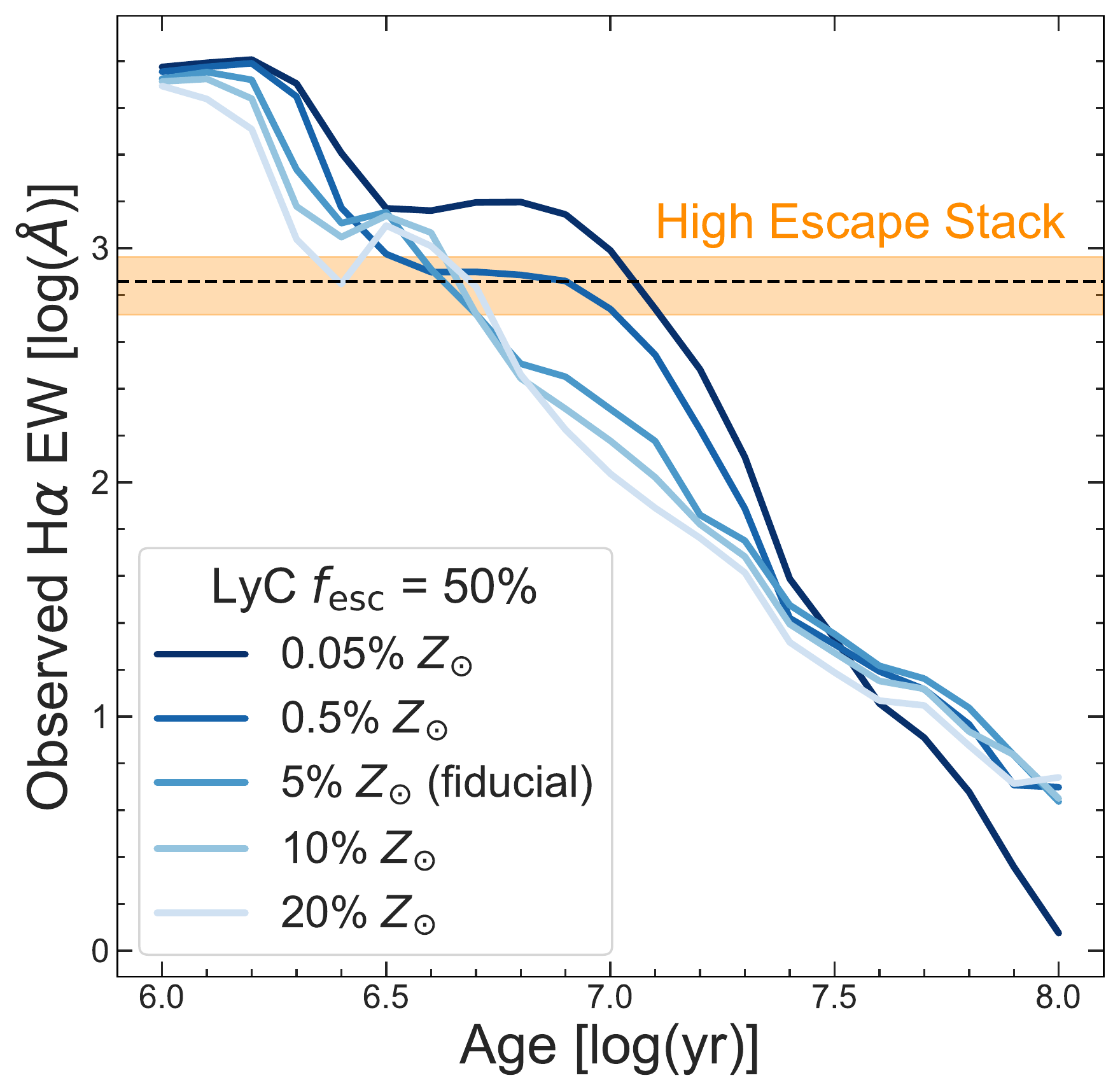}
\caption{Comparison of observed H$\alpha$ EWs with predictions from \texttt{BPASS} burst SEDs as a function of age and metallicity. The EW from the burst SEDs are adjusted for an \fesc$\approx50\%$ that we estimated for the High Escape stack (\S\ref{sec:fesc}). We see that for stellar metallicities consistent with the High Escape stack ($\approx5\%Z_{\rm{\odot}}$) only $\approx2-10$ Myr bursts match the observed H$\alpha$ EW.}
\label{fig:timing}
\end{figure}

A main concern of \fescs simulations (e.g., \citealt{Ma15}) is that short-lived massive stars with highly ionizing spectra are also the stars that need to explode and clear the ISM. The paucity of these massive stars once the ISM is cleared may lead to poor ionizing output. That is, high \fescs periods and high $\xi_{\rm{ion}}$ periods may be out of phase. Our results show that galaxies with high LyC \fescs are also the ones with hard ionizing spectra and elevated $\xi_{\rm{ion}}$.

What could cause this? Hydrodynamical simulations have emphasized the importance of bursty feedback from young, massive stars in driving LyC \fescs \citep[e.g.,][]{Rosdahl18,Kimm19,Ma20} where the LyC \fescs is expected to be highly stochastic, varying rapidly on $\approx10$ Myr timescales \citep[e.g.,][]{Trebitsch17,Barrow20}. We can test this by estimating the burstiness of the recent star formation history in the stacks using EWs of recombination lines like H$\alpha$ that are, to first order, sensitive to the relative number of very hot stars and therefore to the age of stellar populations. These EWs have the added advantage of only mild dependencies on the initial mass function, metallicity and fine-grained properties like binarity and rotation velocity \citep[e.g.][]{Leitherer1999,GrafenerVink2015}. 

Following \citet{JaskotOey13}, we estimate the age of the star-bursts in the High and Low Escape stacks based on H$\alpha$ EWs from \texttt{BPASS} \citep{Stanway18} and \texttt{Starburst99} models \citep{Leitherer1999}. We assume a stellar metallicity $Z=5\%Z_{\rm{\odot}}$ motivated by the inferred gas-phase metallicity \citep[see also][]{Matthee21}. The High Escape stack has an H$\alpha$ EW of $\approx1400$ {\AA} (corrected for 1/(1-\fesc))\footnote{In the lower limiting case \fesc=20 \%, the intrinsic H$\alpha$ EW of the High Escape stack is $\approx900$ {\AA} which implies ages $\lesssim7-30$ Myr.}, which requires a very young age of $2-10$ Myr for a single burst and $\lesssim10$ Myr for continuous star formation (see top panel of Figure \ref{fig:timing}). The H$\alpha$ EW of the Low Escape stack is compatible with a much larger age spread: $\approx10$ Myr for a single burst, but $<200$ Myr for continuous star formation. Confirming the emerging picture from simulations, a key distinguishing feature of High Escape galaxies is that they have undergone a very recent ($\lesssim10$ Myr) burst.

A bursty SFH does not necessarily yield coherence in the phases of high ionising photon production {\it and} escape. Feedback first needs to clear the birth clouds before LyC \fescs can occur. The effects of binary star evolution have been proposed as a remedy to this issue, since they yield significant LyC production even after the most massive stars have exploded \citep[e.g.,][]{Ma16,Rosdahl18,Doughty21}. The EWs of nebular \HeII\ and \CIV\ emission can help test this scenario since they are sensitive to hotter stars than HI-ionising stars, and may probe Myr timescales \citep[e.g.,][]{Gotberg19,Stanway20,Senchyna21}. Knowledge of the origin of these strong nebular lines could in the future further help timing the LyC leaking phases. In addition, more sensitive spectroscopy that can measure P Cygni stellar wind features from e.g. NV and OIV could be useful in constraining the ages of the most massive stars \citep[e.g.][]{Izotov18b,Chisholm2019}.

\begin{figure*}
\centering
\includegraphics[width=\linewidth]{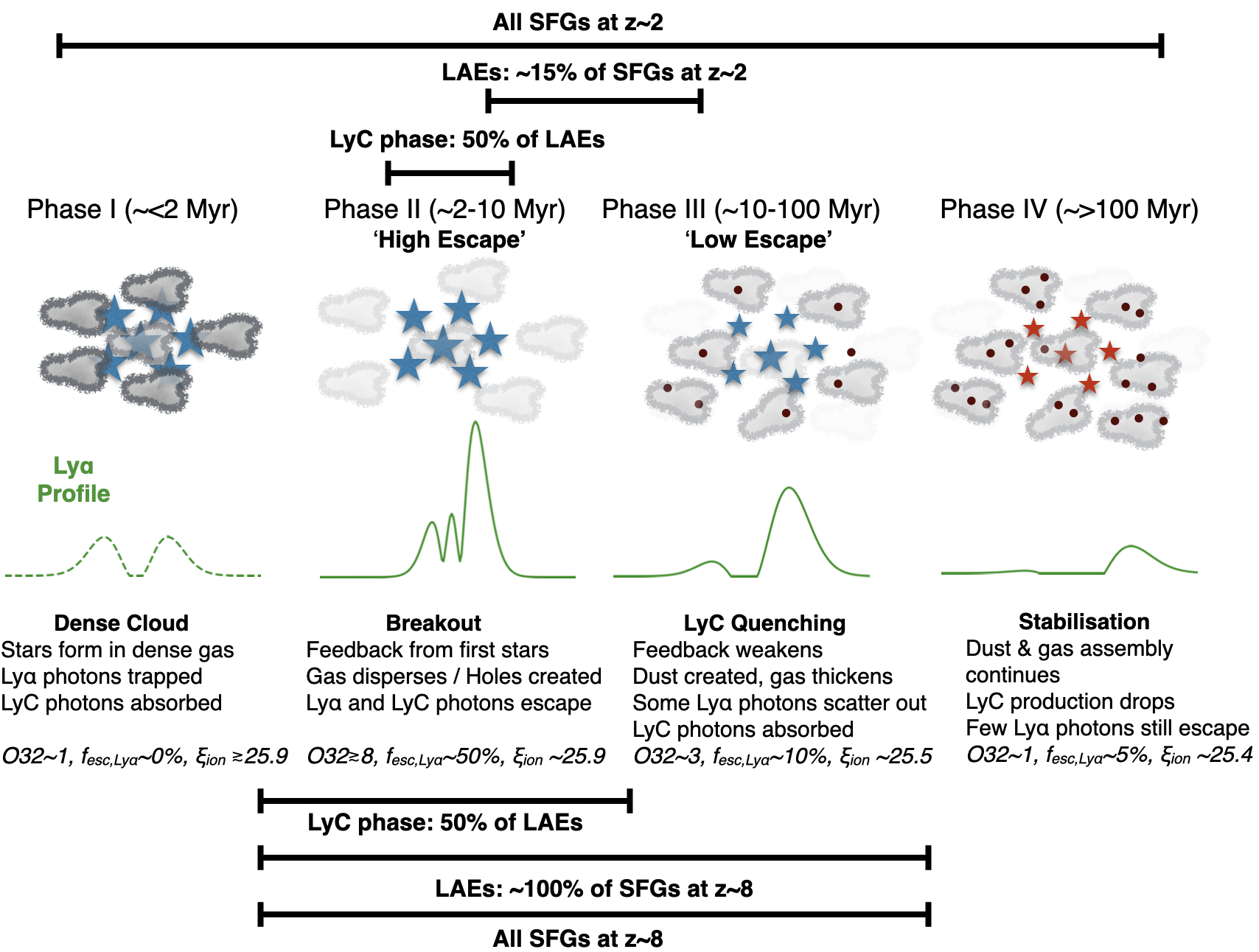}
\caption{Schematic of the LyC duty cycle. Young stellar populations are depicted as blue stars (sizes scaled by stellar mass), old ones in red, HI is shown as gray clouds (colors scaled with density), and dust is shown as dark brown points. We divide the duty cycle into four phases and show the expected emergent Ly$\alpha$ profile for each phase in green. The proportion of galaxies observed and expected in each phase is indicated at the top and bottom of the plot for $z\approx2$ and $z\approx8$ respectively. We note that the fraction of galaxies observed to be in a specific phase is, in addition to the relevant timescales, also likely modulated by the covering fraction of favourable sight-lines between the clusters and the observer. In Phase I, massive stars are born embedded in dense birth clouds -- despite their hard ionizing spectra there is little LyC \fesc. Phase II, corresponding to the LAEs observed in the High Escape stack, is when the dense birth clouds have been destroyed by feedback, and when high LyC \fescs occurs through a transparent ISM -- the duration is likely short, because the most ionizing stars are short-lived. In Phase III, corresponding to our Low Escape stack, young stars are still present but the most massive stars have already exploded as core collapse supernovae -- the emergent ionizing spectrum is not as hard, and dust/HI in the ISM begin to clog pathways to \fescs once again. The majority of star-forming galaxies at $z\approx2$ are in Phase IV (non-LAE LBGs), and are characterized by older stellar populations and a dusty, high column HI ISM. At any given time, $<10\%$ of the star-forming galaxy (SFG) population at $z\approx2$ (half of the LAEs) is in the LyC leaking phase, but this fraction is expected to rise to $\approx40-50\%$ during the EoR as the LAE fraction among the total SFG population rises. 
}
\label{fig:sketch}
\end{figure*}

\subsection{A unified scheme for LyC Escape and extrapolating $z\approx2$ results to $z>6$}
\label{sec:schematic}

In Figure \ref{fig:sketch} we present a simple unified picture of LyC \fescs across all star-forming galaxies by synthesizing our findings, the relation between LAEs and the general galaxy population, and results from recent hydrodynamical simulations \citep[e.g.,][]{Trebitsch17,Rosdahl18,Kimm19,Ma20,Barrow20,Kimm2021}. Our evolutionary sequence for LyC \fescs follows in the footsteps of \citet{Tenorio-Tagle99, Mas-Hesse03,Mao07} who presented such sequences for the emergence of Ly$\alpha$ from star-forming galaxies.

We distinguish galaxies as being in one of four phases. In Phase I, super star cluster like objects form in vigorous starbursts. It takes $\approx2-3$ Myrs for the massive stars in them to destroy their birth clouds and clear channels through the ISM via feedback (e.g., winds, supernovae). Phase I might explain the persisting mystery as to why \fescs constraints (more precisely, N(HI) constraints) from long Gamma Ray Bursts \citep[GRBs; e.g.,][]{Vielfaure20} across a wide range of redshifts find low average \fescs (e.g., $0.5\%$, \citealt{Tanvir19}), at odds with LBG stacks \citep[e.g.][]{Pahl21}. A possible explanation is that the short-lived ($<5$ Myr) progenitors of long GRBs ($>40$ M$_{\rm{\odot}}$ stars, e.g., \citealt{Levan16}) preferentially explode while the birth clouds and ISM are yet to be cleared.

In Phase II, which describes the phase of the galaxies in our High Escape stack, the birth clouds are plausibly disrupted and channels in the ISM have been carved for LyC \fesc (\S $\ref{sec:coincidence}$). Binary products, in addition to young stars that continue forming within the ionized shells cleared by feedback, radiate photons into the IGM with a very high ionising photon production efficiency. In Phase III, corresponding to our Low Escape stack, the ISM is more opaque to Ly$\alpha$ and LyC as feedback weakens since the most massive stars have already exploded  -- relatively young populations that can produce Ly$\alpha$ are nonetheless still present. Phase IV describes the `steady-state' LBGs in which the ISM is opaque to dust and HI and the ionising photon production efficiency is about $\approx3$ times lower than during Phase II (e.g. \citealt{Bouwens16b} and \S $\ref{sec:results}$).

Remarkably, we found that half the LAEs are in Phase II despite its short characteristic timescale ($<10$ Myrs) compared to Phase III (10-100 Myrs). This implies that the likelihood of observing a galaxy to be in a specific phase is not solely determined by the duration of these phases. There might be two important physical effects at play. First, the fraction of favorable viewing angles such that a Phase II galaxy is observed as a \lyas emitter is likely higher compared to a Phase III galaxy. This may be the case when the large scale ISM around young Phase II clusters has a larger fraction of low column density and dust-poor sight-lines compared to Phase III bursts (as supported by \S\ref{sec:results}). Second, the duty cycle of Phase II might be rapid, with consecutive bursts occurring in quick succession. \citet{Ma20} detail such a scenario: an accelerating supernova superbubble sweeps up material in its wake, and consecutive generations of young stars form \textit{inside} the cleared out bubble -- a kpc-scale bubble is expected to take $\approx20-40$ Myrs to expand during which it would support Phase II conditions \citep{kjb19}.

At $z\approx2$, the majority of LBGs are not detected as LAEs \citep[e.g.,][]{Cassata2015,Kusakabe2018} and Phase IV is therefore the most common. Galaxies in any phase can go back to being in Phase I and II when they undergo starbursts. However, the more massive a galaxy is, the more unlikely it is that young starbursts can dominate significant fraction of the light. This could be either because they are out-shined by older star-forming regions, or because the fraction of favourable sight-lines through the galaxy decreases with increasing mass. This may be reflected by observations that find that the Ly$\alpha$ escape fraction generally decreases with increasing mass for galaxies selected irrespective of their Ly$\alpha$ strength \citep[e.g.][]{Matthee16,Oyarzun17}. Our line-profile statistics (\S $\ref{sec:selectxls}$) suggest that Phase II is concentrated in a minority ($<10\%$, i.e., half the LAEs) of the overall galaxy population ($\gtrsim0.5L^{*}$ LBGs) at $z\approx2$. 

This framework helps explain the rarity of LyC leakers at lower redshifts and their probable increasing incidence at higher redshifts \citep[e.g.][]{FaucherGiguere20} in terms of the growing LAE fraction (specifically Phase II fraction) among LBGs. At higher redshifts, the LAE population forms an increasing fraction of the LBG population \citep[e.g.][]{Stark10}, and therefore a higher fraction of the total galaxy population occupies Phases II-III sketched in Fig. $\ref{fig:sketch}$. There are already strong hints that Phase II conditions (e.g., high \OIII/\OII, extreme [OIII] EWs) grow increasingly common towards the EoR in lockstep with the rising LAE fraction \citep[e.g.,][]{Labbe13,debarros19,Endsley21}. At $z\approx2$ when the LAE fraction among $M_{\rm{UV}}<-18$ LBGs is $\approx0.1$, the average LyC \fescs is $\approx0.05$, whereas at $z\approx8$ when the LAE fraction may be $\approx1$, we expect an average \fescs of 0.25 (half the LAEs have \fesc$\approx50\%$, the other half have \fesc$\approx0$, non-LAEs have \fesc$\approx0$). Indeed, the Ly$\alpha$ escape fraction measured in LAEs at $z\approx2$ is comparable to the LBG population-averaged Ly$\alpha$ escape fraction at $z>6$ \citep{Hayes2011,Sobral2016,Matthee21}. This implies that -- if the Ly$\alpha$ to LyC connection is not evolving -- the average LyC escape fraction of LAEs at $z\approx2$ is comparable to the average LyC escape fraction of the star-forming galaxy population in the Epoch of Reionisation. 

It is therefore plausible that we can extrapolate our \fescs results on LAEs at $z\approx2$ to LAEs at $z>3$ as LAEs show redshift-invariance of various properties relevant to \fescs -- sizes, $\Sigma_{\rm{SFR}}$, line profiles, $\beta_{\rm{UV}}$ slopes, and luminosity functions \citep[e.g.,][]{Malhotra12,Paulino-Afonso18, Herenz19, Santos20, Hayes21}, but direct tests -- such as the evolution of the distribution of Ly$\alpha$ line-profiles with redshift -- would be able to verify this. In a companion paper (Matthee \& Naidu et al. 2021), we quantify the implications of this framework by showing the ionizing emissivity from bright LAEs is sufficient to explain the cosmic ionizing background from $z\approx2-8$.

\subsection{Caveats \& Limitations}

Here we discuss caveats and limitations around our results and ways to address them. 

While theoretically well-motivated, and validated by multiple independent spectroscopic indicators (\S\ref{sec:results}), our \lyas profile-based LyC selection criteria were designed based on a small sample of $\approx25$ sources that have both high-resolution \lyas profiles as well as direct LyC \fescs measurements. These criteria must be further validated with larger samples -- e.g., there are now $\approx20$ LyC leaker candidates at $z\approx2-4$ awaiting high-resolution \lyas measurements \citep[e.g.,][]{Bian17,Steidel18, Fletcher19}. 

A $\approx4\times$ larger sample would help confirm our results and reduce the error on the fraction of LyC leaking LAEs by half (currently we report $50\pm10\%$). However, obtaining high-resolution spectra spanning the entire rest-frame UV to optical wavelength range that we have analyzed here is challenging. Very few datasets currently exist at \textit{any} redshift with such coverage. We hope the validation of the \lya-based approach in this study spurs greater investments in large surveys designed to measure the bare minimum high-resolution Ly$\alpha$ coupled with a precise systemic redshift for galaxies drawn representatively from \lyas LFs (see Matthee \& Naidu et al. 2021 for a \lya-LF based framework for the emissivity).

The high ionization lines (\CIVL, \HeIIL) detected only in the \fesc$>20\%$ stack may prove very informative -- they show extremely ionizing ($>54.4$ eV) photons are produced during periods of elevated LyC \fesc. However, latest stellar population models are unable to match the observed EWs of these lines (bottom panel of Figure \ref{fig:timing}), and so quantitative details relying on these models such as the exact time when \HeIIL\ production peaks after a burst are uncertain. We now have yet another motivating reason -- understanding LyC \fescs -- to unravel the origins of nebular \HeII\, \citep[e.g.,][]{Stanway20,Senchyna21,Simmonds21,Olivier21}.

Finally we comment on the generalizability of our results to lower and higher redshifts. To first order, the framework presented in Figure \ref{fig:sketch} applies to any redshift -- what changes is the fraction of galaxies that are in each phase. However, it must also be acknowledged that despite all their similarities, $z\approx2$ LAEs may have different star-formation histories, stellar abundances, and interstellar media compared to $z\approx0$ LAEs or $z\approx6$ LAEs. In our framework these differences would manifest in the details and duration of each phase. For instance, Phase II (the LyC phase), may be even more leaky and extended at $z\approx6$ given possibly lower metallicities and thus harder ionizing spectra (bottom panel of Fig. \ref{fig:timing}). To clarify this issue we need a systematic study of the detailed ISM and stellar populations of LAEs across redshift in the style of \citet{Izotov21analog} who focused on compact star-forming galaxies.

\section{Summary \& Outlook}
\label{sec:summary}

We seek to isolate the physical conditions for LyC \fescs by comparing samples of LyC leakers (inferred \fesc$>20\%$, High Escape) against a control sample of non-leakers (inferred \fesc$<5\%$, Low Escape). Such a controlled study has been difficult to perform at high-$z$ due to sightline effects and at low-$z$ due to complex selection functions. Here we circumvent these hurdles by using resolved Ly$\alpha$ profiles from the luminosity-limited XLS-$z$2 survey to select leakers and non-leakers. Our empirically motivated selection criteria using the Ly$\alpha$ peak separation ($v_{\rm{sep}}$) and central fraction ($f_{\rm{cen}}$) are based on literature sources with \fescs measurements at $z\approx0-4$, and have solid theoretical grounding in decades of radiative transfer simulations [\S\ref{sec:classify}, Figs. \ref{fig:litfesc}, \ref{fig:LyCclassify}, \ref{fig:XLSfesc}, Table \ref{table:sample_selection}]. By contrasting stacked spectra of the High Escape and Low Escape samples we find the following [Figure \ref{fig:stack}, Table \ref{table:stacks}] :

\begin{itemize}\itemsep1em
  \item The robustness of our stacks -- that they do separate leakers (\fesc$>20\%$) from non-leakers (\fesc$<5\%$) -- is confirmed by half a dozen independent spectroscopic indicators sensitive to HI and dust, the two chief regulators of \fesc. In the High Escape stack \MgII\ is observed close to the systemic velocity, the resonant \CIV\ is in emission, the Ly$\alpha$ \fescs which is $\geq$ LyC \fescs is $\approx50\%$, CII absorption that is a hallmark of dense HI columns is undetected, and the H$\alpha$/H$\beta\approx2.8$ reveals a dust-free ISM. On the other hand, the low escape stack shows redshifted \MgII\ (+130 km s$^{-1}$), no \CIV\ emission, $f_{\rm{esc, LyC}}\leq f_{\rm{esc, Ly\alpha}}\approx10\%$, strong CII absorption, and H$\alpha$/H$\beta\approx4$ implying $E(B-V)\approx0.3$. [\S\ref{sec:dust}-\S\ref{sec:cii}, Figure \ref{fig:stack}, Table \ref{table:stacks}]

  \item The leakers show strong nebular \HeII\ and \CIV\ emission in the \textit{median} stack, signaling the ubiquity of hard ionizing spectra along with a $\log({\xi_{\rm{ion}}/\rm{Hz\ erg^{-1}}})\approx25.7-25.9$ (for \fesc$=20-50\%$). These high-ionization features and extreme EWs (e.g., \OIII+H$\beta$ rest-frame EW$\approx1100\AA$) can be produced only by young ($<10$ Myr), low metallicity ($<10\% Z_{\rm{\odot}}$) stellar populations in theoretical burst SEDs. Low Escape sources have similar $M_{\rm{UV}}$ and a high $\log({\xi_{\rm{ion}}/\rm{Hz\ erg^{-1}}})\approx25.6$ but lack the extreme EWs and \HeII\ ($>54.4$ eV). That is, the star-formation in leakers and non-leakers is similar on timescales of $\approx100$ Myrs but not on $\approx10$ Myrs. [\S\ref{sec:production}, Figure \ref{fig:stack}, Table \ref{table:stacks}] 
  
  \item The massive star-formation in the High Escape sources is occurring in an extreme ionization state ISM (\OIII/\OII$\approx8.5$, $\log{(U)}\approx-2.3$), comparable to local super star cluster complexes. Non-leakers have a less ionized ISM with \OIII/\OII$\approx3$. [\S\ref{sec:o32}, Figure \ref{fig:stack}, Table \ref{table:stacks}]
  
  \item The LyC \fescs at $850-912\AA$ (``Lyman edge \fesc'') of the High Escape stack is $20-55\%$ -- the lower bound is by the Ly$\alpha$ profile-based selection and the upper bound is by requiring $f_{\rm{esc,Ly\alpha}}\geq f_{\rm{esc,LyC}}$. We make a finer estimate of the edge LyC \fescs ($\approx40\%$) by exploiting empirical correlations with the Ly$\alpha$ \fescs and the Ly$\alpha$ red-peak velocity. Since the $<850\AA$ photons that are ubiquitous among the leakers escape easily compared to edge photons, we estimate the total \fescs ($0-912\AA$) that matters for reionization calculations to be $\approx50\%$. [\S\ref{sec:feschard}, Table \ref{table:fesc_summary}, Figure \ref{fig:hardphotons}]

  \item With \textit{JWST}, the LyC \fescs for $z>6$ galaxies may be constrained with a handful of strong emission lines. The defining characteristics of leakers -- low column densities, hard ionizing spectra, a dust-free, high-ionization state ISM -- occur \textit{simultaneously} in the \fesc$>20\%$ stack of LAEs. That is, these properties are highly correlated, and selecting for one of them increases the chances of selecting the others on \textit{average} -- significant scatter across individual galaxies still exists. So even though \OIII/\OII\ and Balmer decrements are not explicitly sensitive to HI, they may be sufficient to implicitly estimate the average \fesc. This result is derived for LAEs, but is applicable to the EoR since the majority of $z>6$ galaxies are expected to be LAEs. [\S\ref{sec:jwst}, Figure \ref{fig:FescProps}]
 
  \item Observed emission lines need not be weak when LyC \fescs is high. Based on our stacks as well as a literature compilation of leakers we argue high \fescs occurs during a period of prolific ionizing photon production and so the intrinsic emission line EWs are so high that the observed EWs (e.g., $\approx110\AA$ for Ly$\alpha$) are still $\approx2\times$ higher than non-leakers. [\S\ref{sec:EWs}, Figure \ref{fig:LyaEWFesc}]

  \item We chart the highly non-linear relationship between observed Ly$\alpha$ EW and LyC \fesc. The Ly$\alpha$ EW serves as a complete but highly impure selector of high \fescs galaxies. For instance, our \fesc$<5\%$ stack has a $\approx60\AA$ EW, which according to the linear Ly$\alpha$ EW - LyC \fescs relation in \citet{Steidel18, Pahl21} implies an LyC \fesc$\approx30\%$ that is strongly ruled out by the indicators discussed above (e.g., $f_{\rm{esc, Ly\alpha}}\approx10\%\geq f_{\rm{esc, LyC}}$). [\S\ref{sec:EWs}, Figure \ref{fig:LyaEWFesc}]
  
  \item We synthesize our findings in the following physical picture that confirms several aspects of recent hydrodynamical simulations. LyC leakers are galaxies that have undergone recent ($<10$ Myr) episodes of vigorous star-formation. The super star clusters born out of these episodes have produced spatially concentrated feedback. This feedback has carved channels through the ISM, thus clearing paths for LyC. Crucially, for reionization, even after the ISM is cleared a reservoir of highly ionizing sources is still available to stream photons into the IGM. The galaxies with properties that favour a high escape of ionizing photons are also the galaxies that emit copious amounts of those photons at the right time -- production and escape occur in sync. In sharp contrast, non-leakers (despite being strong Ly$\alpha$ emitters in our sample) have a relatively dusty, opaque ISM that has not been cleared out by feedback, most likely linked to their dearth of the youngest, most massive stars. [\S \ref{sec:coincidence} - \ref{sec:schematic}, Figures $\ref{fig:timing}$, $ \ref{fig:sketch}$]
  
\end{itemize}

An important contribution of this work is the statistics of LyC \fescs among LAEs. Half the LAEs have \fesc$\approx50\%$, the other half have \fesc$<5\%$, and non-LAE LBGs have \fesc$\approx0\%$ (LyC \fesc$\leq$ Ly$\alpha$ \fesc $\approx0$). Since fundamental LAE properties are redshift-invariant, we can, with some confidence, extrapolate the constraints derived here to LAEs at higher redshifts. While comprising a minuscule fraction of the overall galaxy population at $z\approx2$, the LAE fraction strongly evolves such that almost every $L_{\rm{UV}}>0.1 L^{*}$ galaxy at $z\approx8$ is perhaps an LAE. This work motivates and forms the basis of a Ly$\alpha$-based formalism for the cosmic ionizing emissivity that uses Ly$\alpha$ luminosity functions instead of UV luminosity functions. The $L_{\rm{UV}}$ that varies on $\approx100$ Myr timescales and is insensitive to the HI column densities is replaced with $L_{\rm{Ly\alpha}}$ that is intimately tied to the bursty, stochastic LyC \fescs that fluctuates on Myr timescales. Developing this Ly$\alpha$-anchored formalism for reionization is the focus of our companion paper. 

\section*{Acknowledgements}

We thank Stephan McCandliss and John O'Meara for their generous counsel on matters bluer than $912\AA$ and sharing code to generate LyC transmission curves in Figure \ref{fig:hardphotons}. It is a pleasure to thank the organizers of SAZERAC2.0, a conference where we received helpful affirmation from the community that we were onto something meaningful, particularly from Aaron Smith, John Chisholm, Steve Finkelstein, Joki Rosdahl, Peter Senchyna, Simon Gazagnes, and Andy Bunker. This project owes its existence to the Leiden/ESA Astrophysics Program for Summer Students (LEAPS) 2017, one of the few truly open-armed research opportunities for undergrads from developing countries, through which RPN and JM were first connected. We thank the Conroy group at Harvard, particularly Ana Bonaca, for their enthusiasm for all things $z=0$ and $z>0$. RPN thanks the Peters family for their hospitality while several aspects of this paper were being worked out, particularly \S\ref{sec:fesc} that was aided by the Hefeweizen of Idletyme Brewing Co.

RPN gratefully acknowledges an Ashford Fellowship granted by Harvard University. PO and JK acknowledge support from the Swiss National Science Foundation through the SNSF Professorship grant 190079.
The Cosmic Dawn Center (DAWN) is funded by the Danish National Research Foundation under grant No.\ 140. CC acknowledges funding from the Packard foundation.
GP acknowledges support from the Netherlands Research School for Astronomy (NOVA). MH is fellow of the Knut and Alice Wallenberg Foundation. RA acknowledges support from Fondecyt Regular Grant 1202007. MG was supported by NASA through the NASA Hubble Fellowship grant HST-HF2-51409.
ST is supported by the 2021 Research Fund 1.210134.01 of UNIST (Ulsan National Institute of Science \& Technology). JC acknowledges support from the Spanish Ministry of Science and Innovation, project PID2019-107408GB-C43 (ESTALLIDOS) and from Gobierno de Canarias through EU FEDER funding, project PID2020010050. MLl acknowledges support from the ANID/Scholarship Program/Doctorado Nacional/2019-21191036.

We made extensive use of the \lyas Spectral Database (LASD\footnote{\url{http://lasd.lyman-alpha.com}}, \citealt{Runnholm21}) without which compiling literature Ly$\alpha$ profiles would have been an infinitely more miserable task. Other software used in this work include: \texttt{matplotlib} \citep{matplotlib}, \texttt{jupyter} \citep{jupyter}, \texttt{IPython} \citep{ipython}, \texttt{numpy} \citep{numpy}, \texttt{scipy} \citep{scipy}, \texttt{TOPCAT} \citep{topcat}, and \texttt{Astropy} \citep{astropy}.

\section*{Data availability}
The VLT/X-SHOOTER data underlying this article were accessed from the ESO archive. The raw ESO data can be accessed through http://archive.eso.org/cms.html. The {\it HST}/COS spectra were accessed through the Lyman-$\alpha$ Spectral Database accessible through http://lasd.lyman-alpha.com/. The derived data generated in this research will be shared on reasonable request to the corresponding authors.

\bibliographystyle{mnras}
\bibliography{MasterBiblio.bib}

\appendix 
\renewcommand\thetable{A.\arabic{table}}    
\renewcommand\thefigure{A.\arabic{figure}}

\section{Ly$\alpha$ profiles of individual galaxies} \label{appendix:lyaprofiles}
\begin{figure*}
    \begin{tabular}{ccc}
    \includegraphics[width=5.3cm]{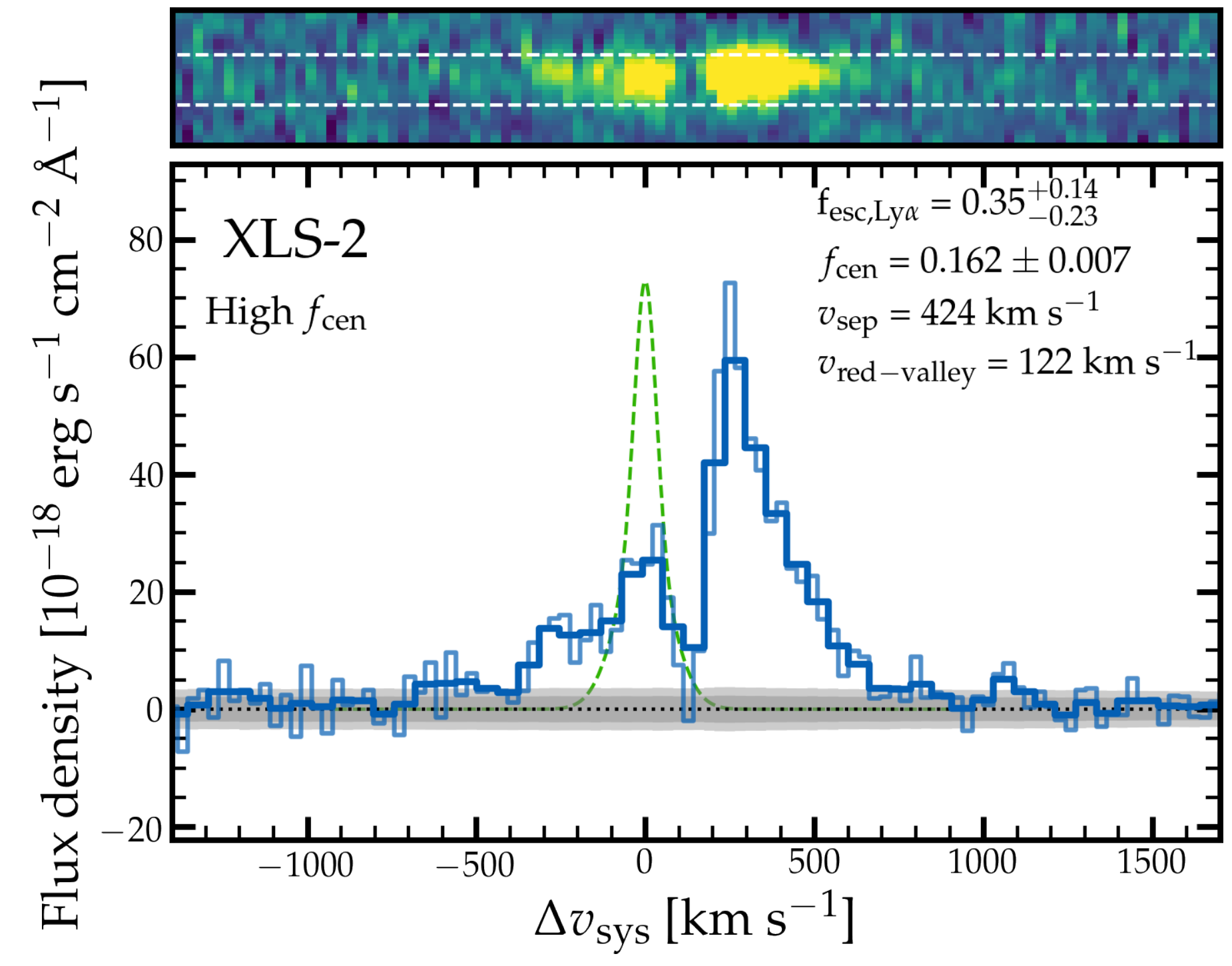} &
    \includegraphics[width=5.3cm]{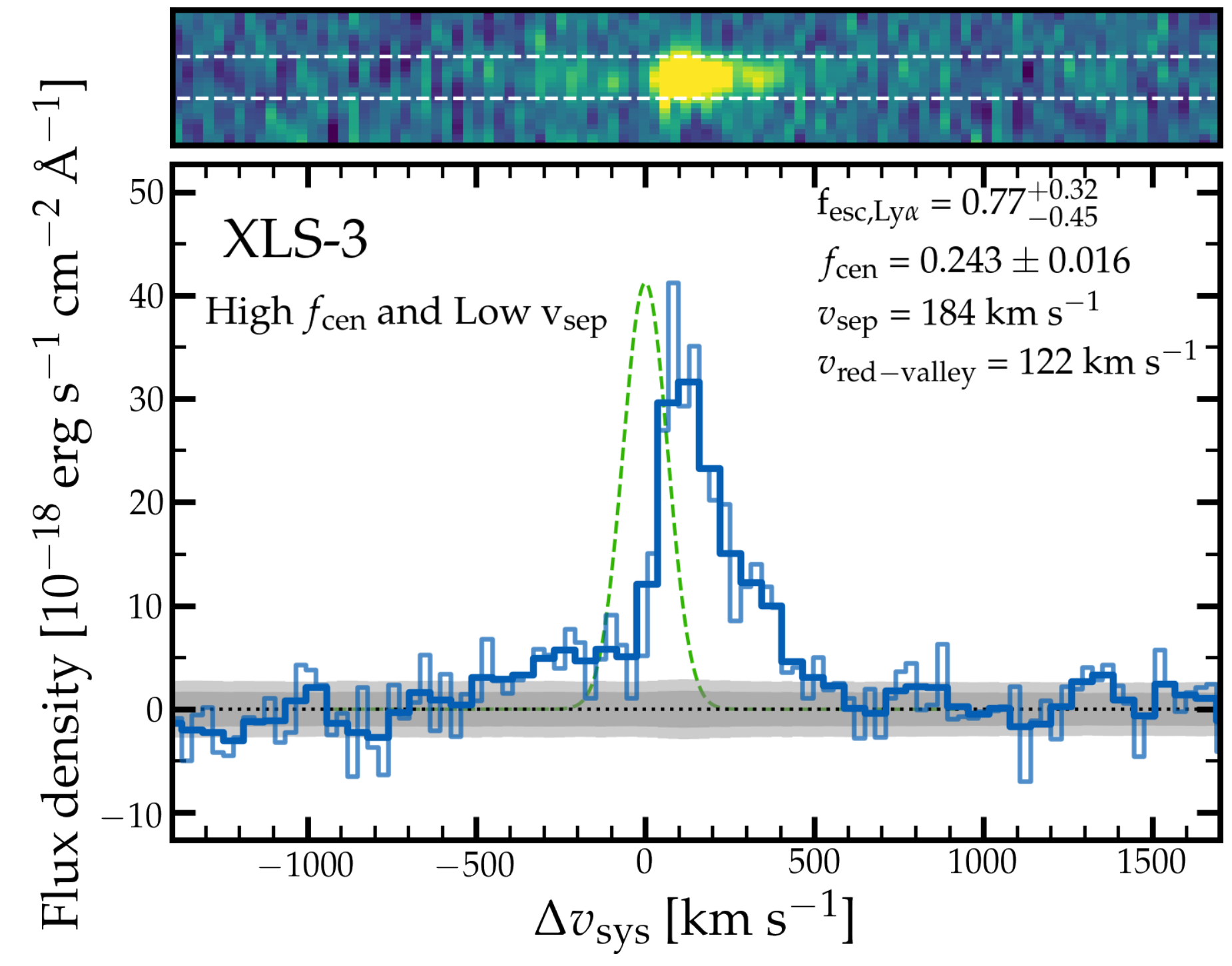} &
    \includegraphics[width=5.3cm]{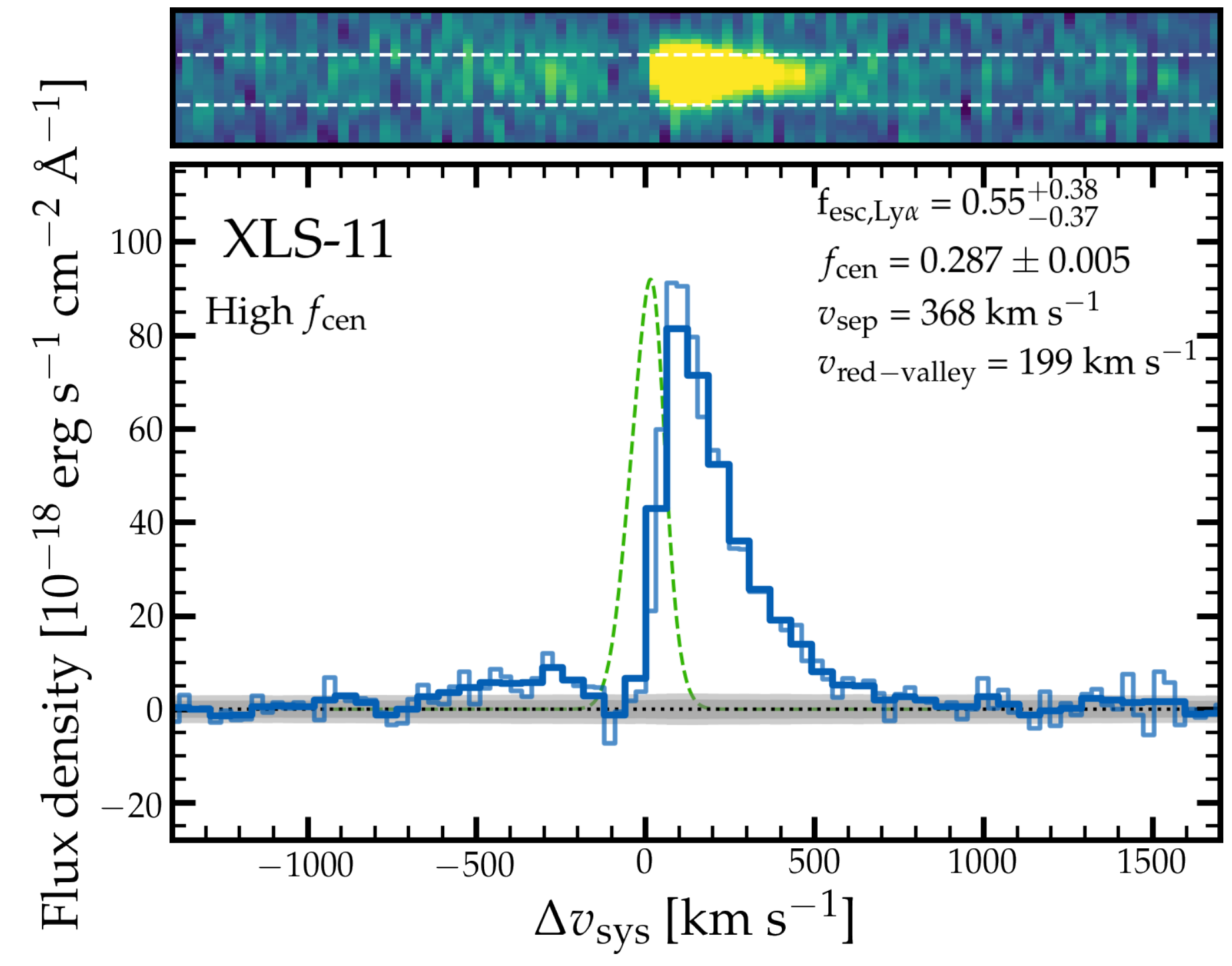} \\

    \includegraphics[width=5.3cm]{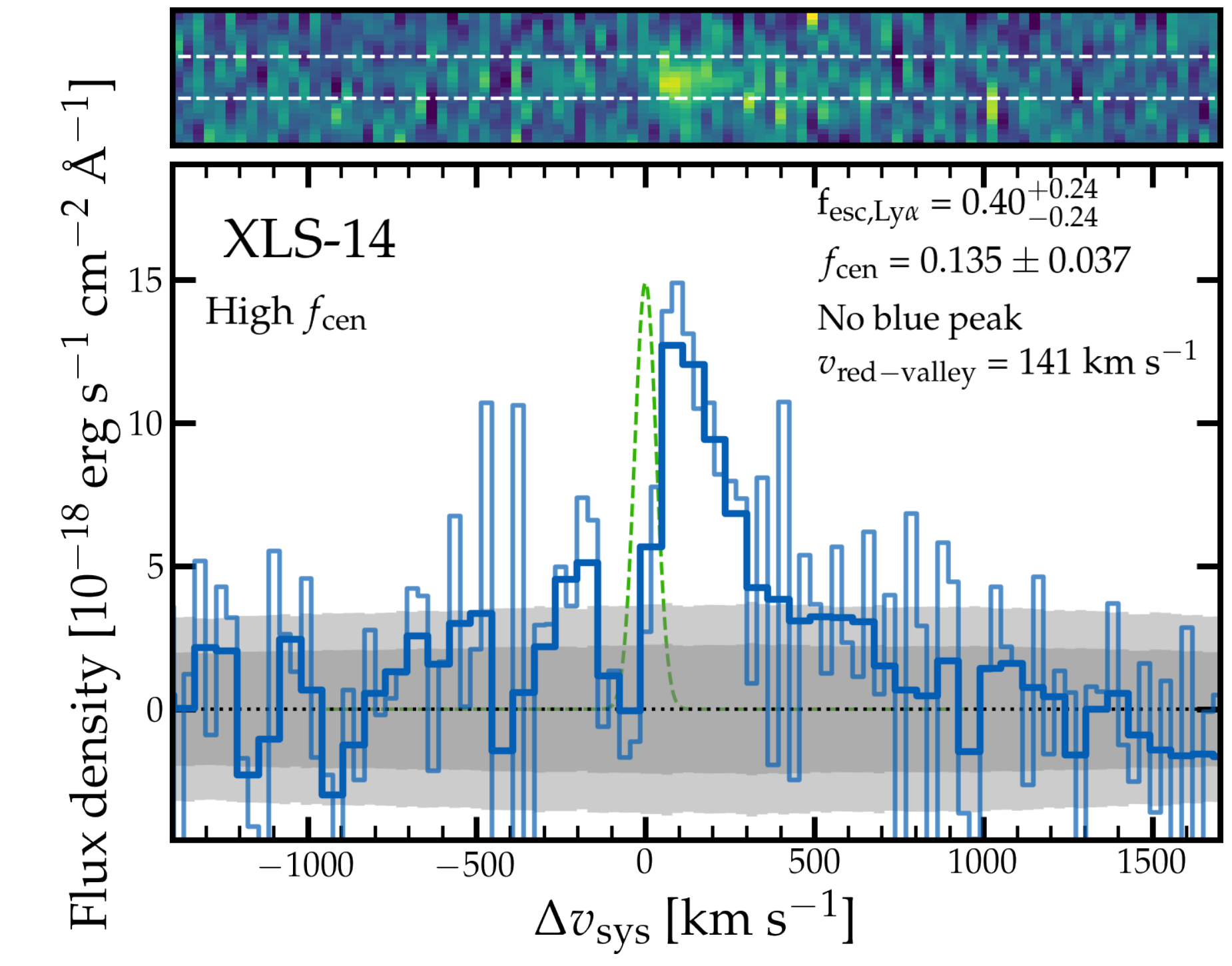} &
    \includegraphics[width=5.3cm]{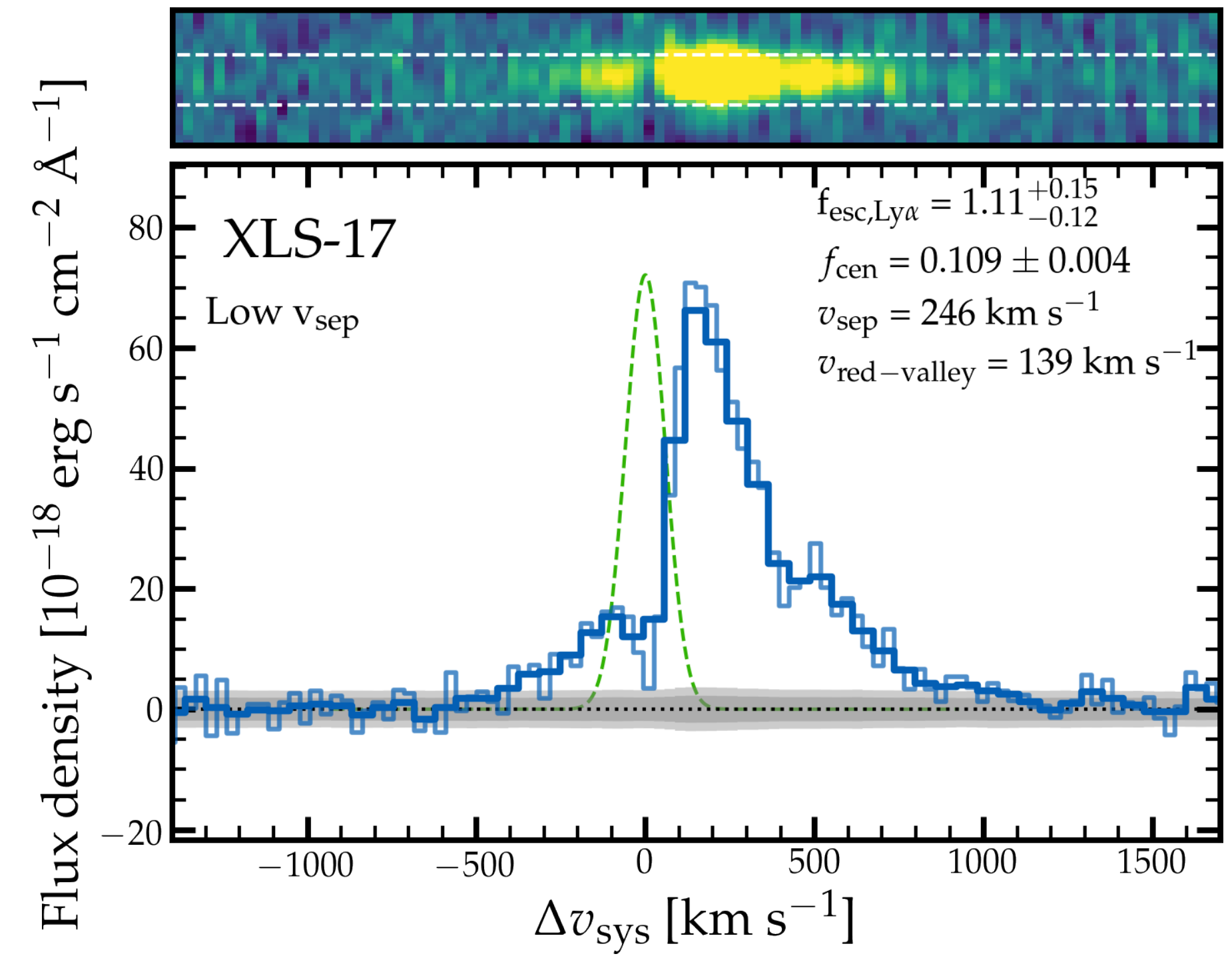} &
    \includegraphics[width=5.3cm]{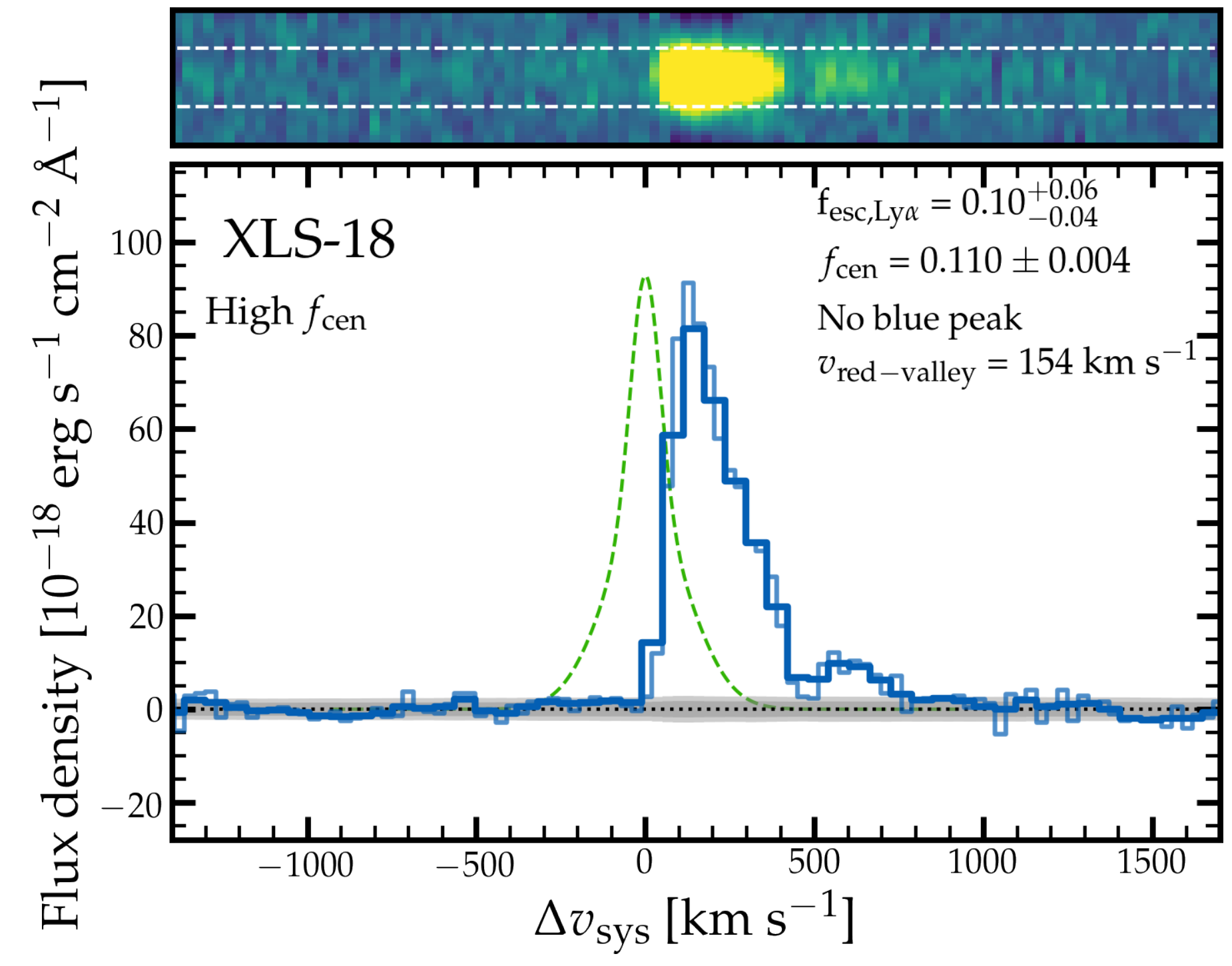} \\
    
    \includegraphics[width=5.3cm]{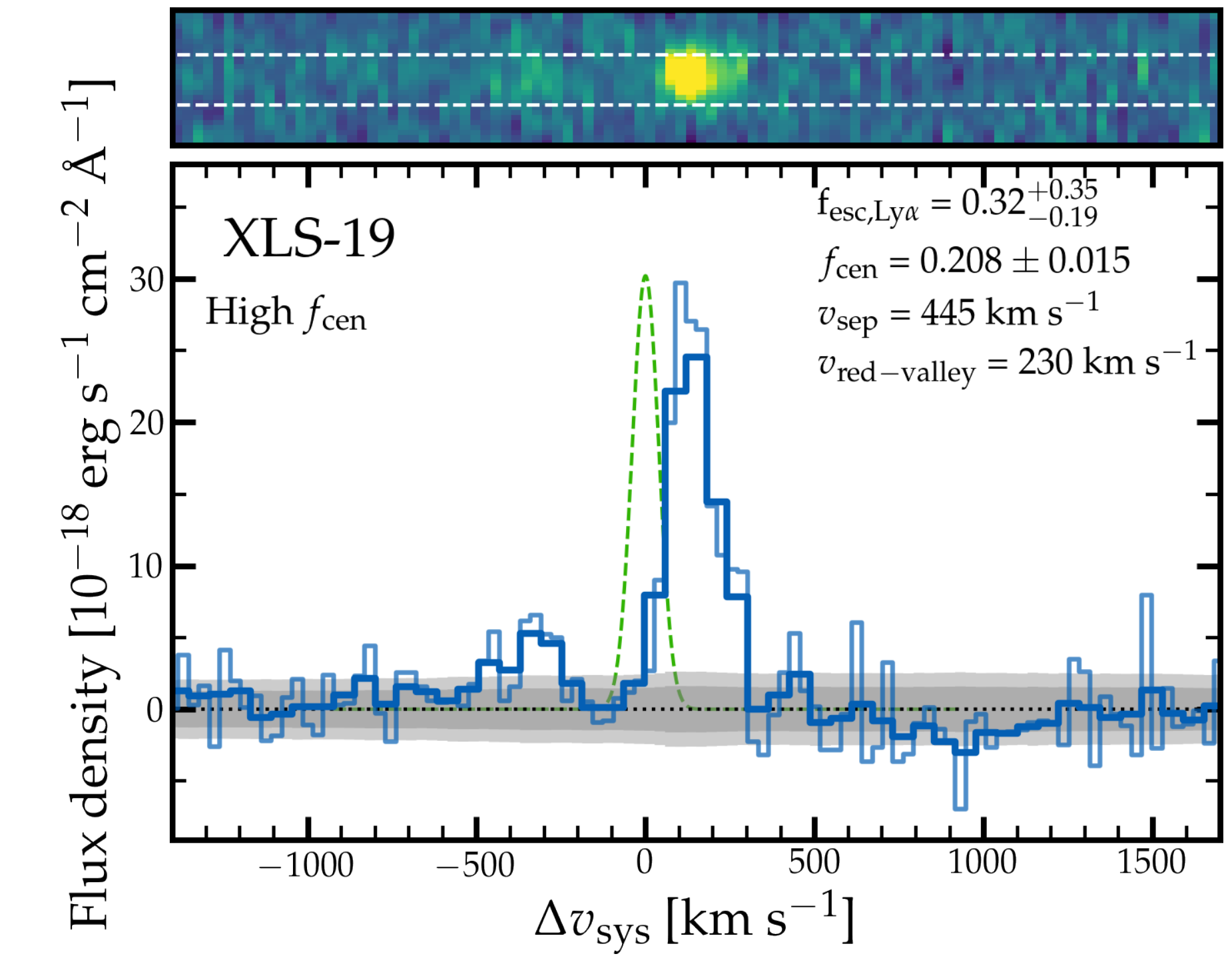} &
    \includegraphics[width=5.3cm]{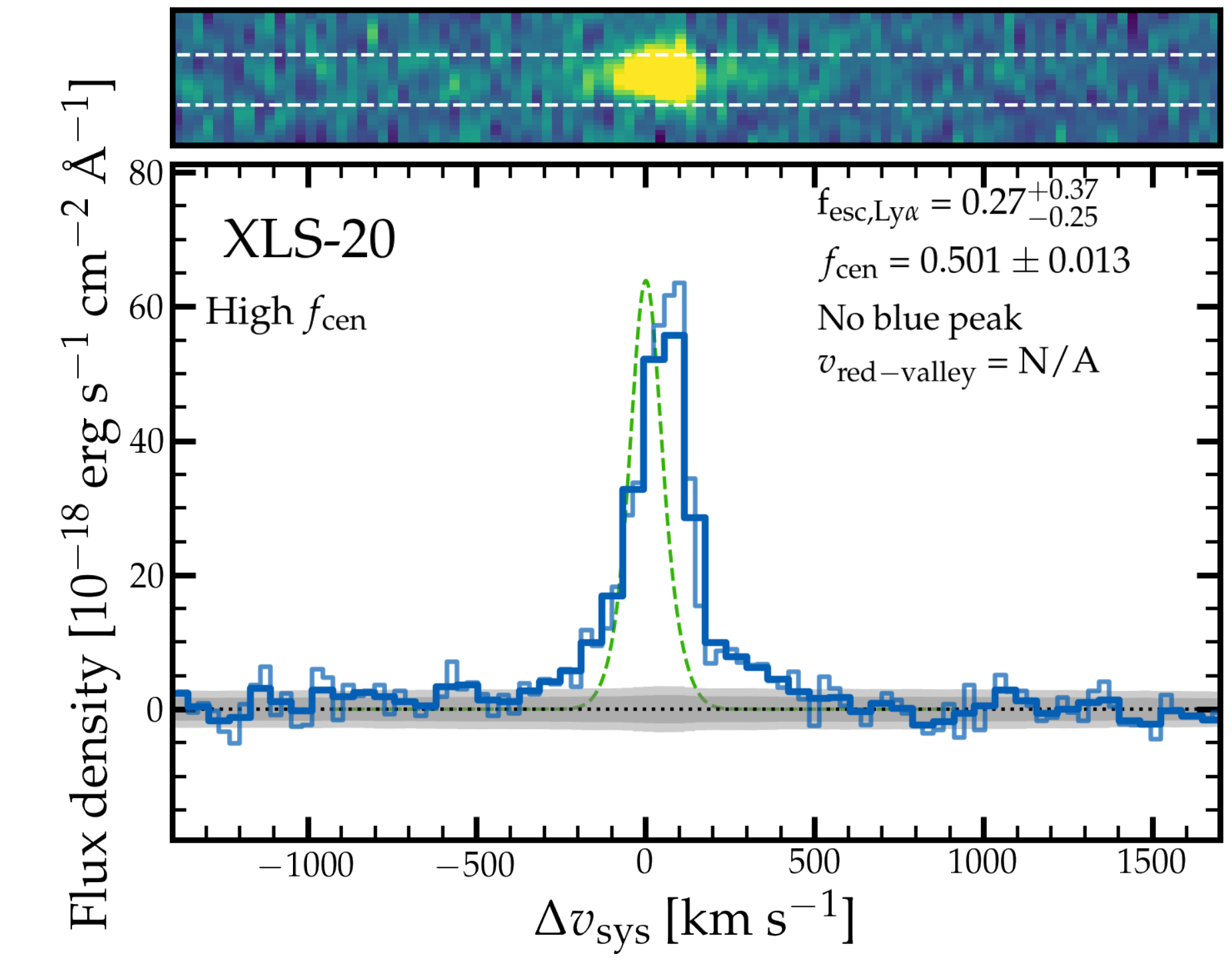} &
    \includegraphics[width=5.3cm]{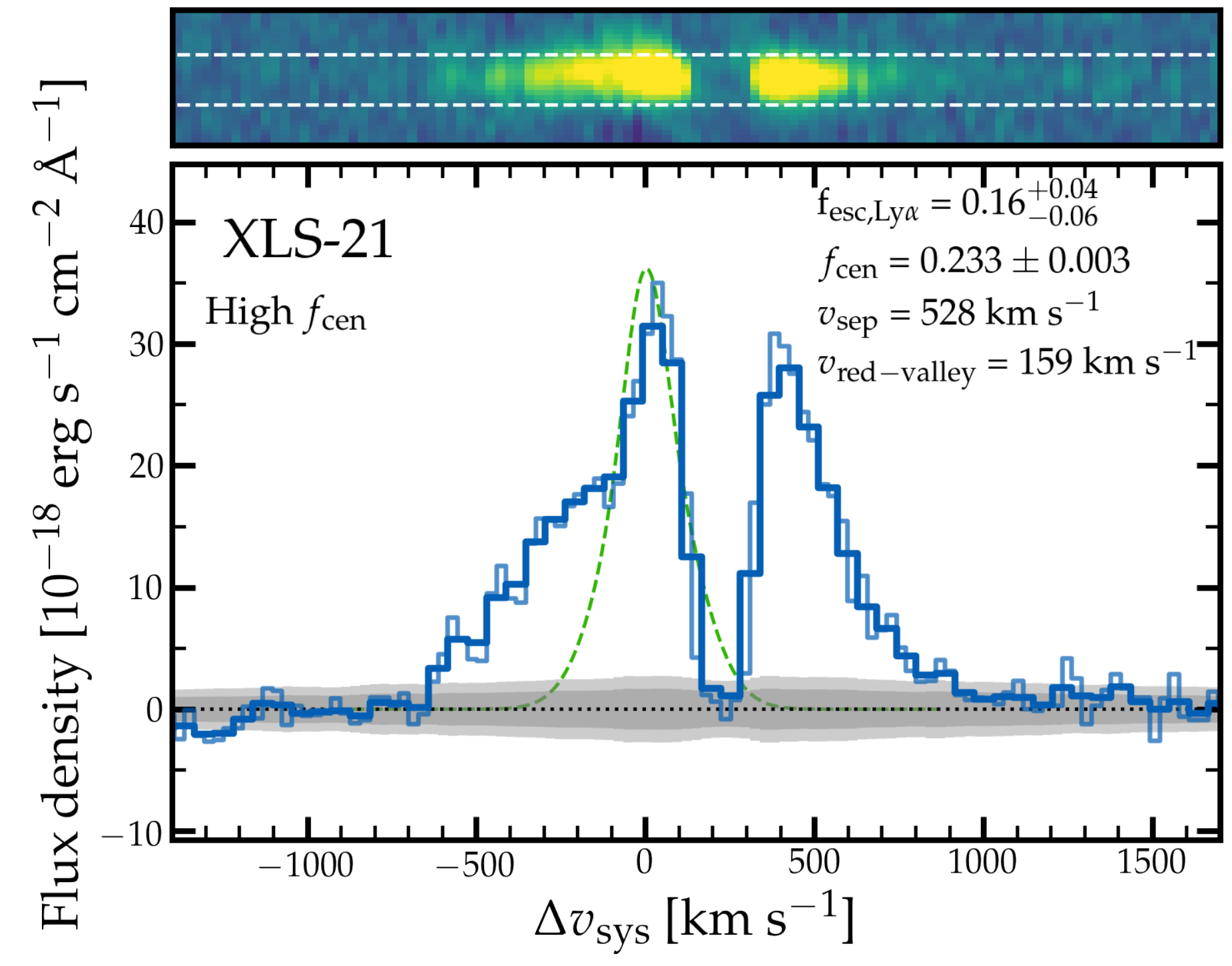} \\

    \includegraphics[width=5.3cm]{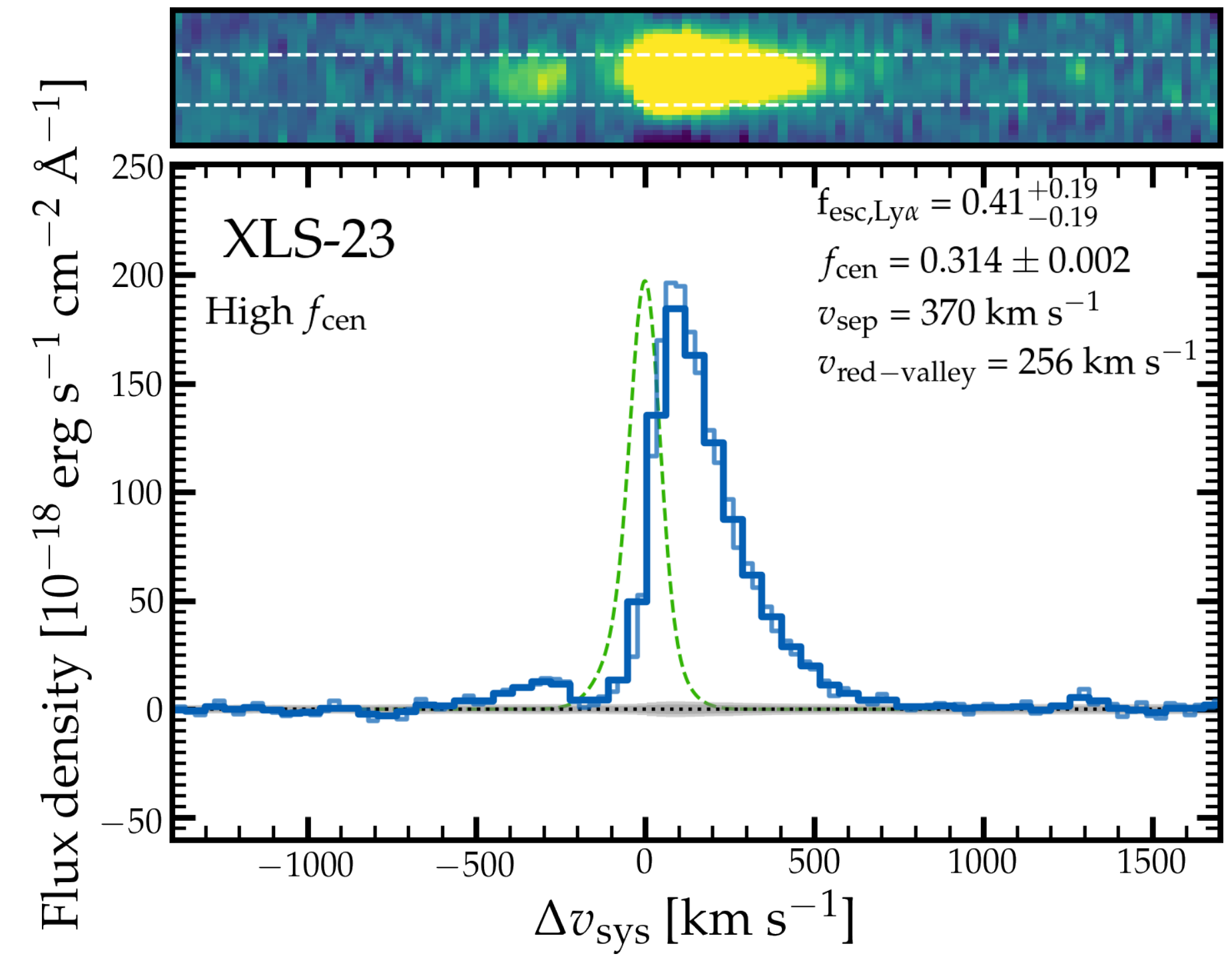} &
    \includegraphics[width=5.3cm]{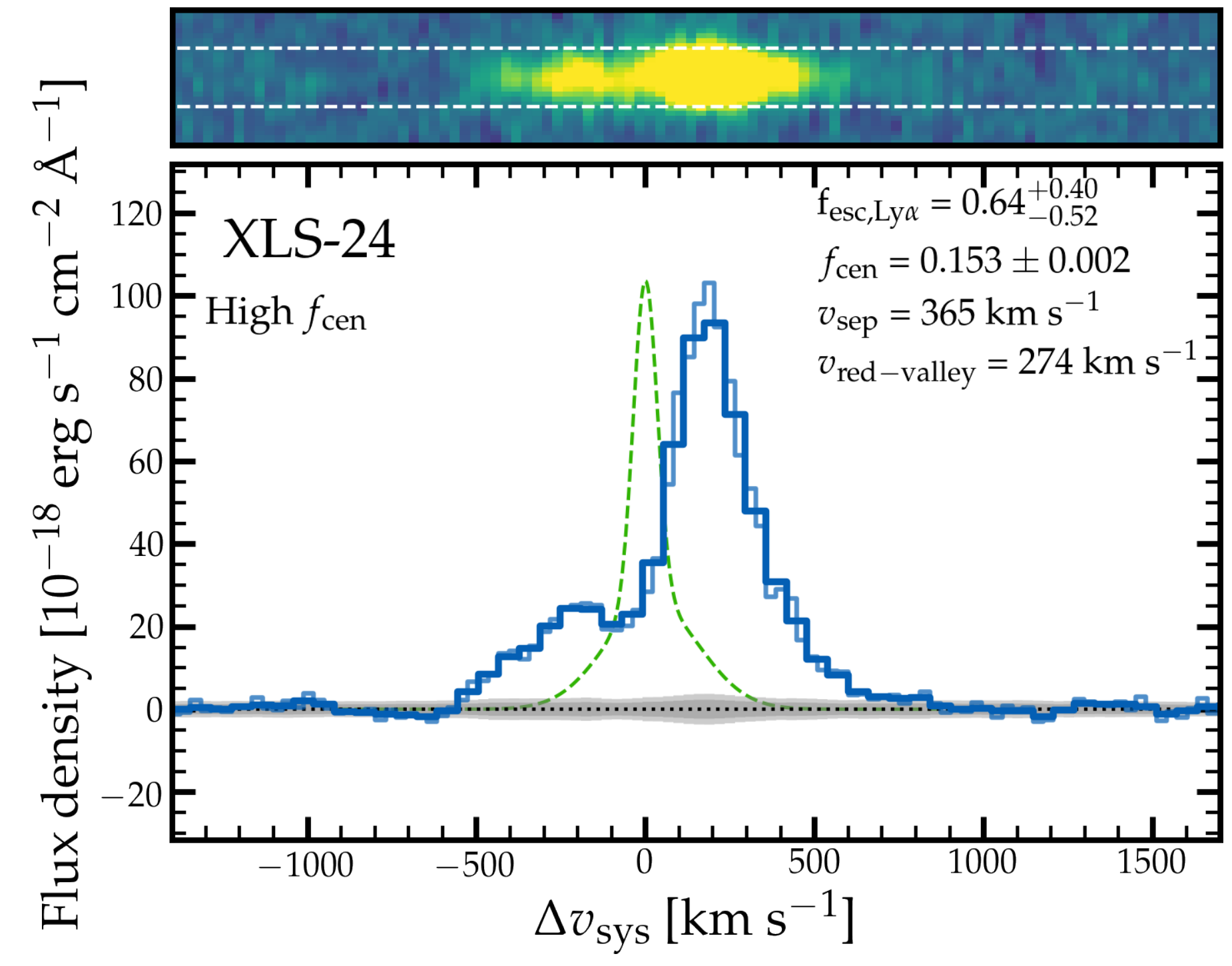} &
    \includegraphics[width=5.3cm]{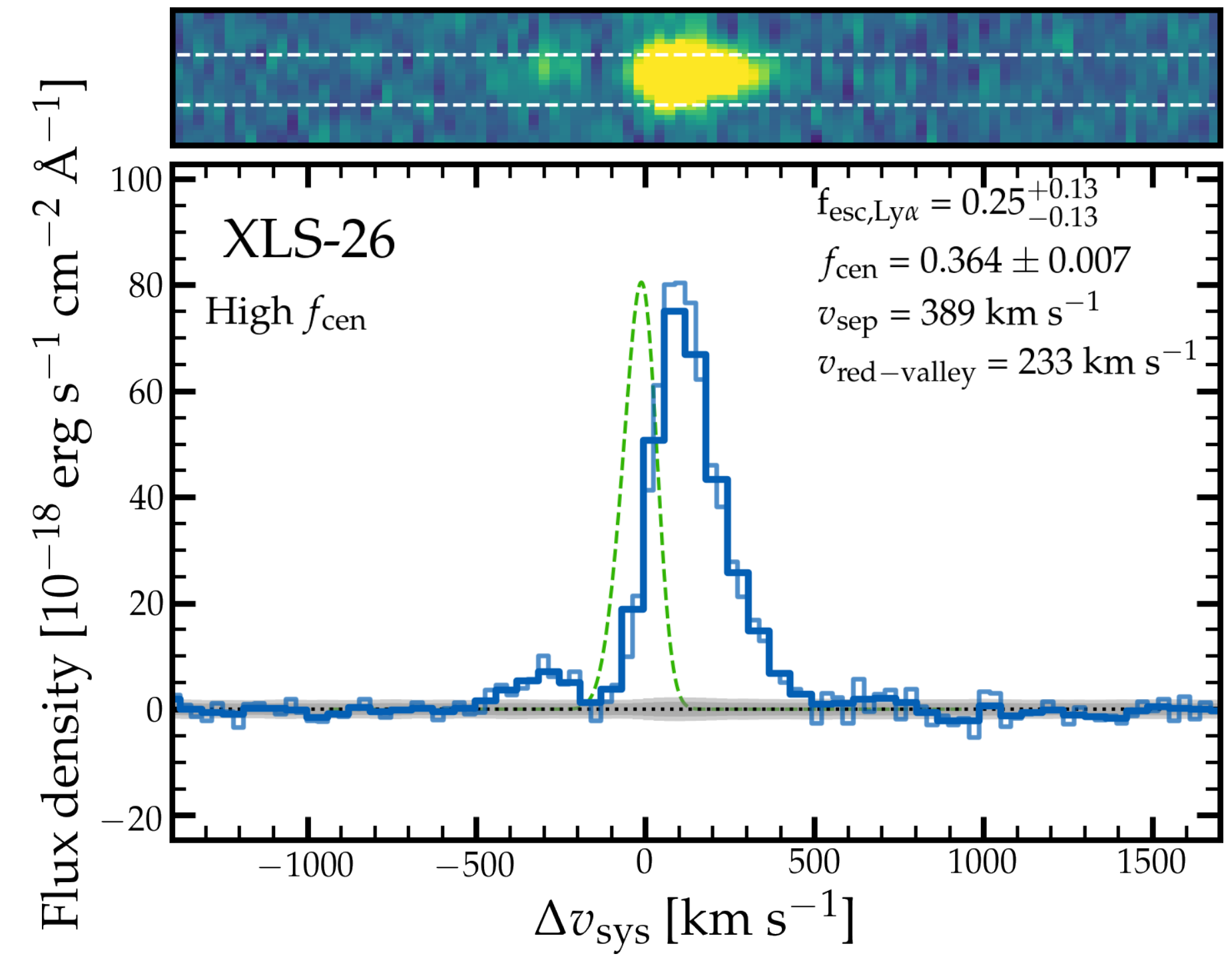} \\

    \includegraphics[width=5.3cm]{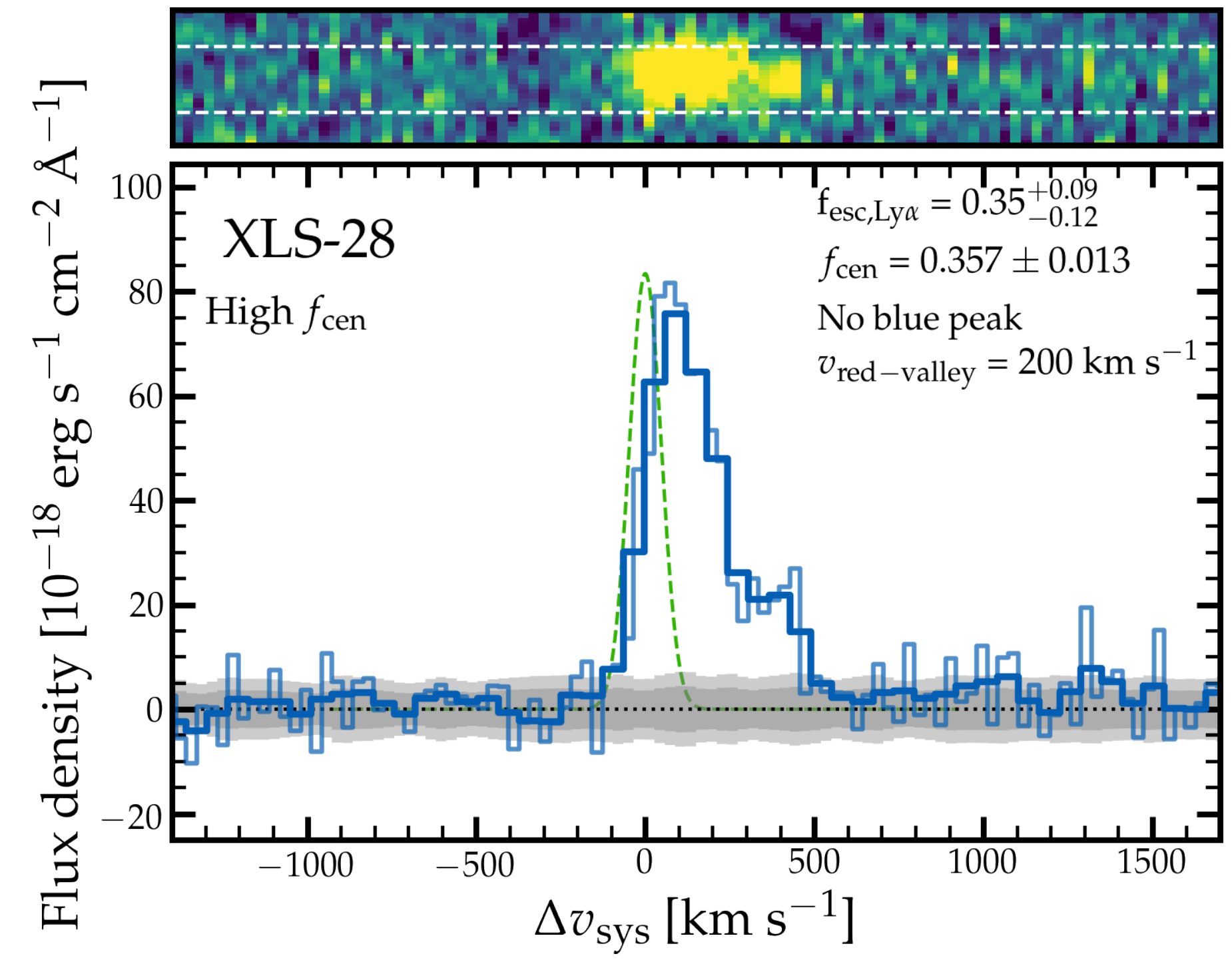} \\
    \end{tabular}
    \caption{Individual Lyman-$\alpha$ line-profiles of the LAEs classified in the High Escape subset. The velocity axis is centered on the systemic redshift. The thin and thick blue line shows the Ly$\alpha$ profiles with native and factor two binning, respectively. The green dashed line shows the line-profile of the rest-frame optical lines. The upper inset panels show the two-dimensional Ly$\alpha$ profiles, where the dashed white lines highlight the FWHM of the extraction window. }
    \label{fig:Lya_profiles_HIGH}
\end{figure*}

\begin{figure*}
    \begin{tabular}{ccc}
    \includegraphics[width=5.3cm]{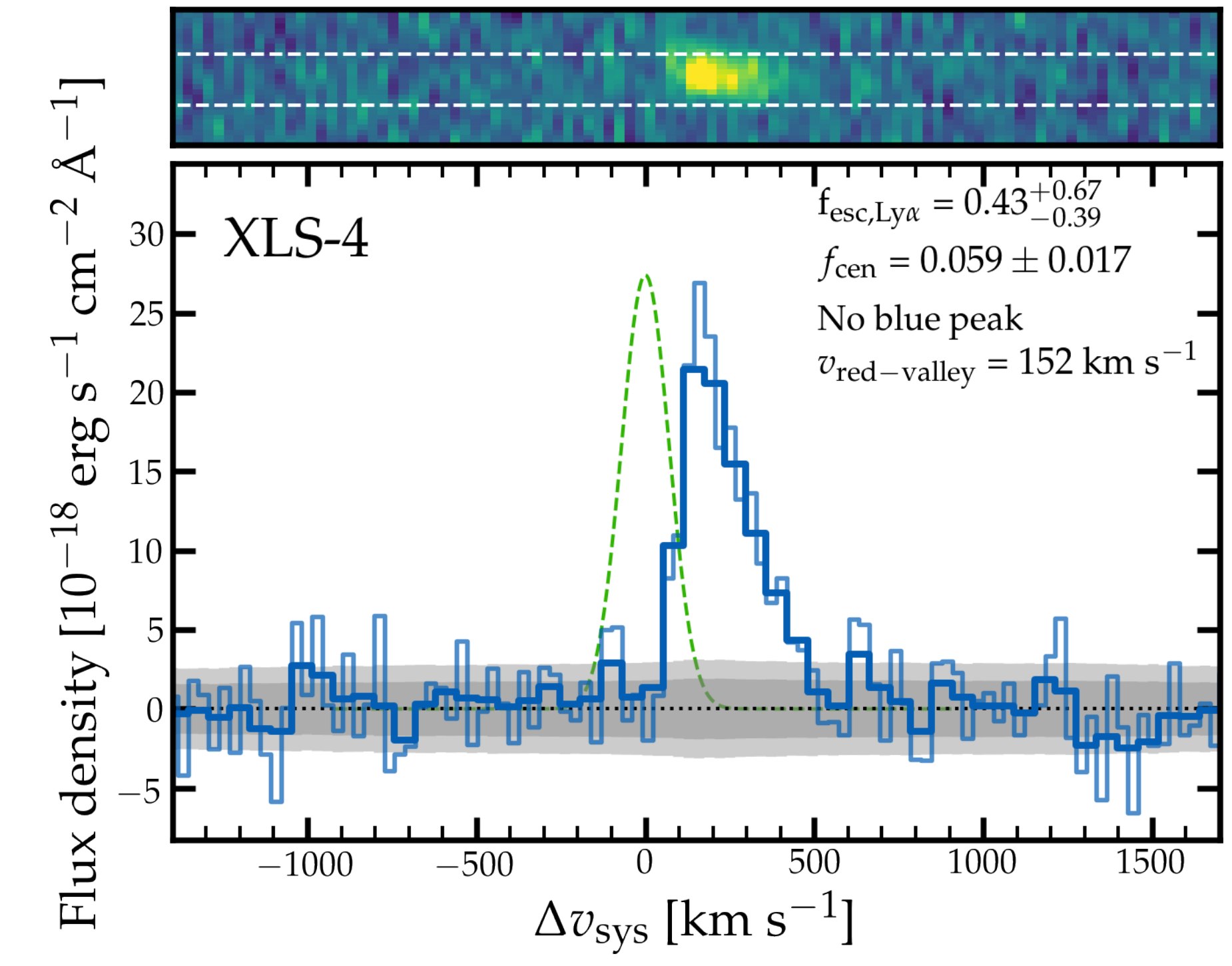} &
    \includegraphics[width=5.3cm]{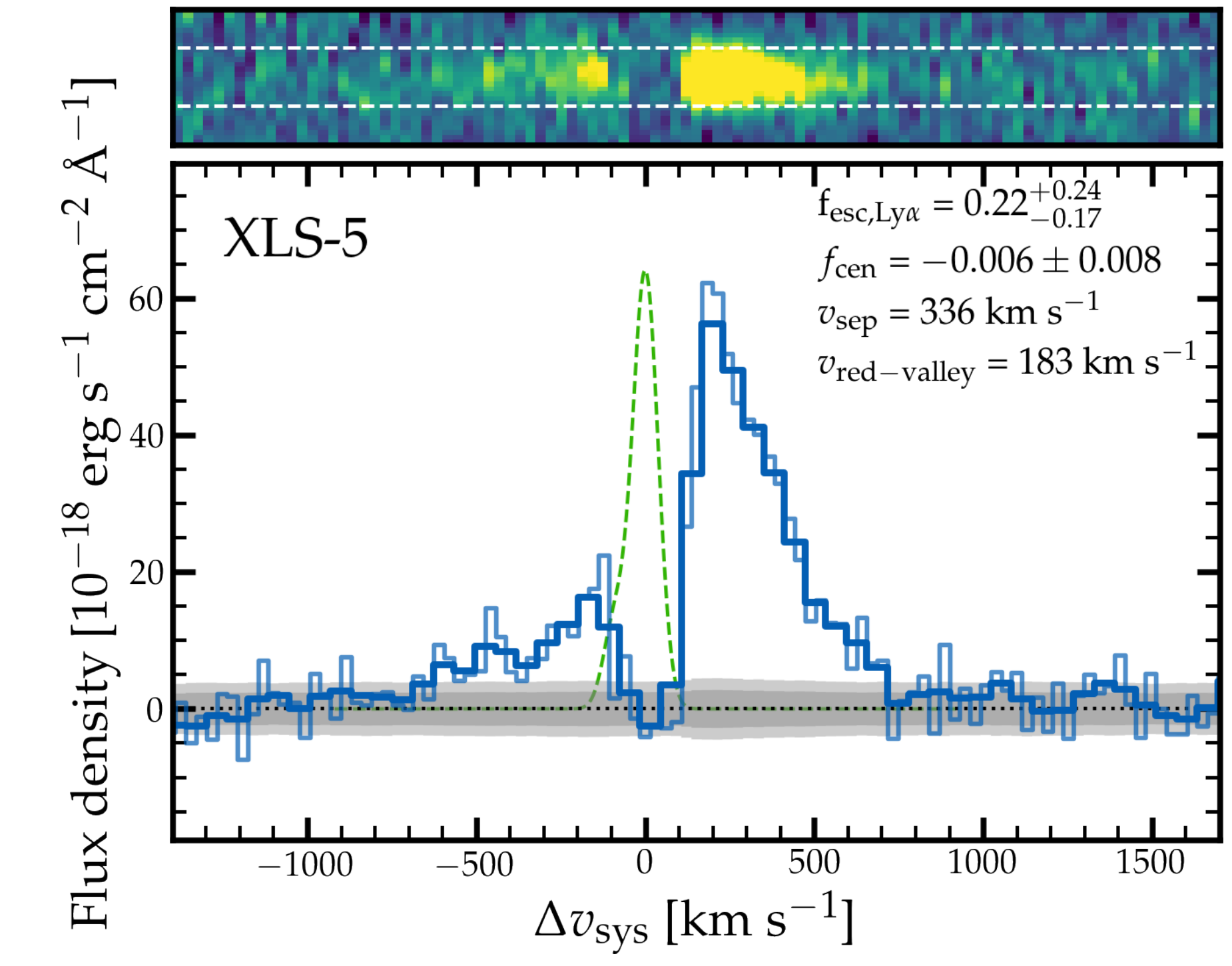} &
    \includegraphics[width=5.3cm]{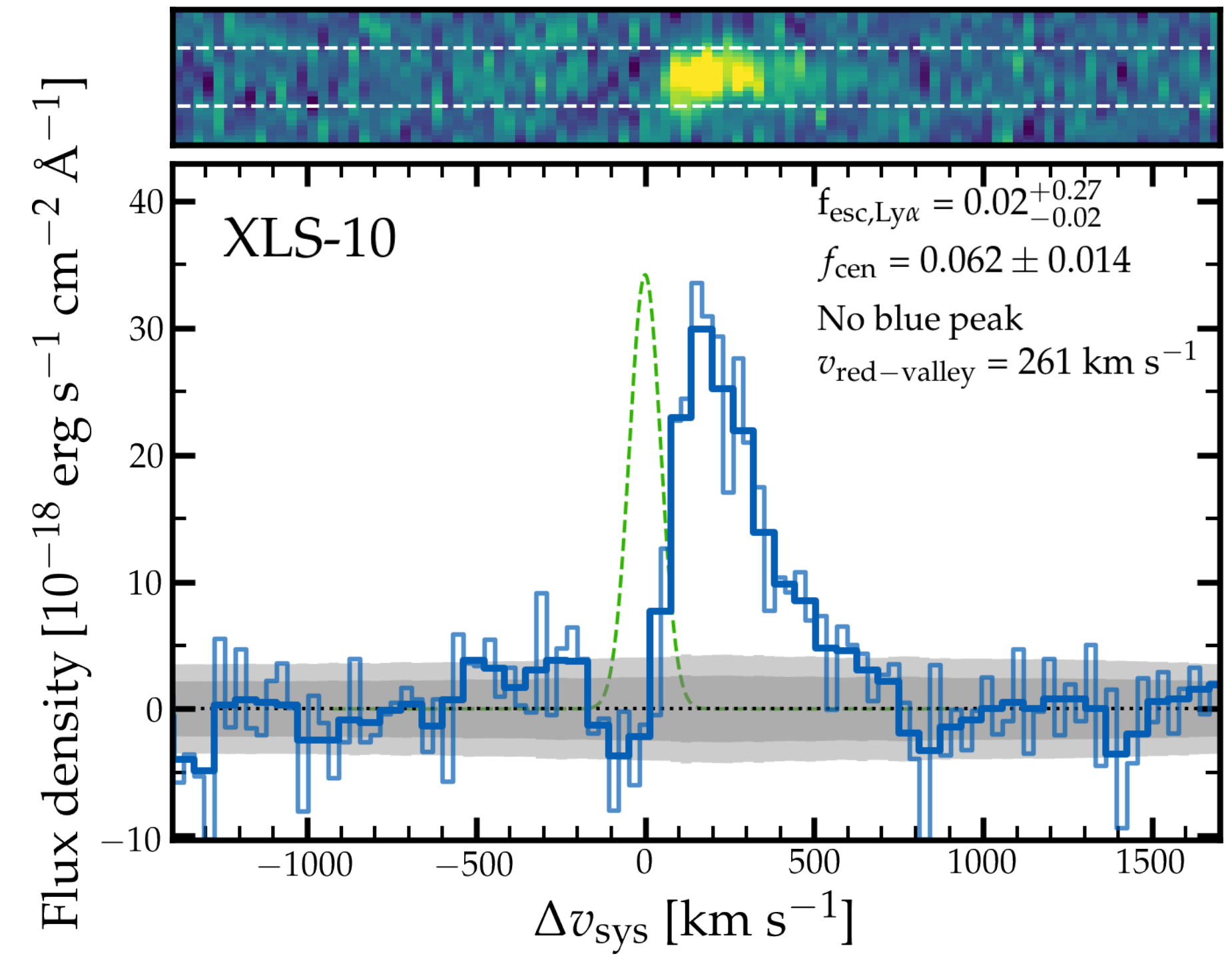} \\

    \includegraphics[width=5.3cm]{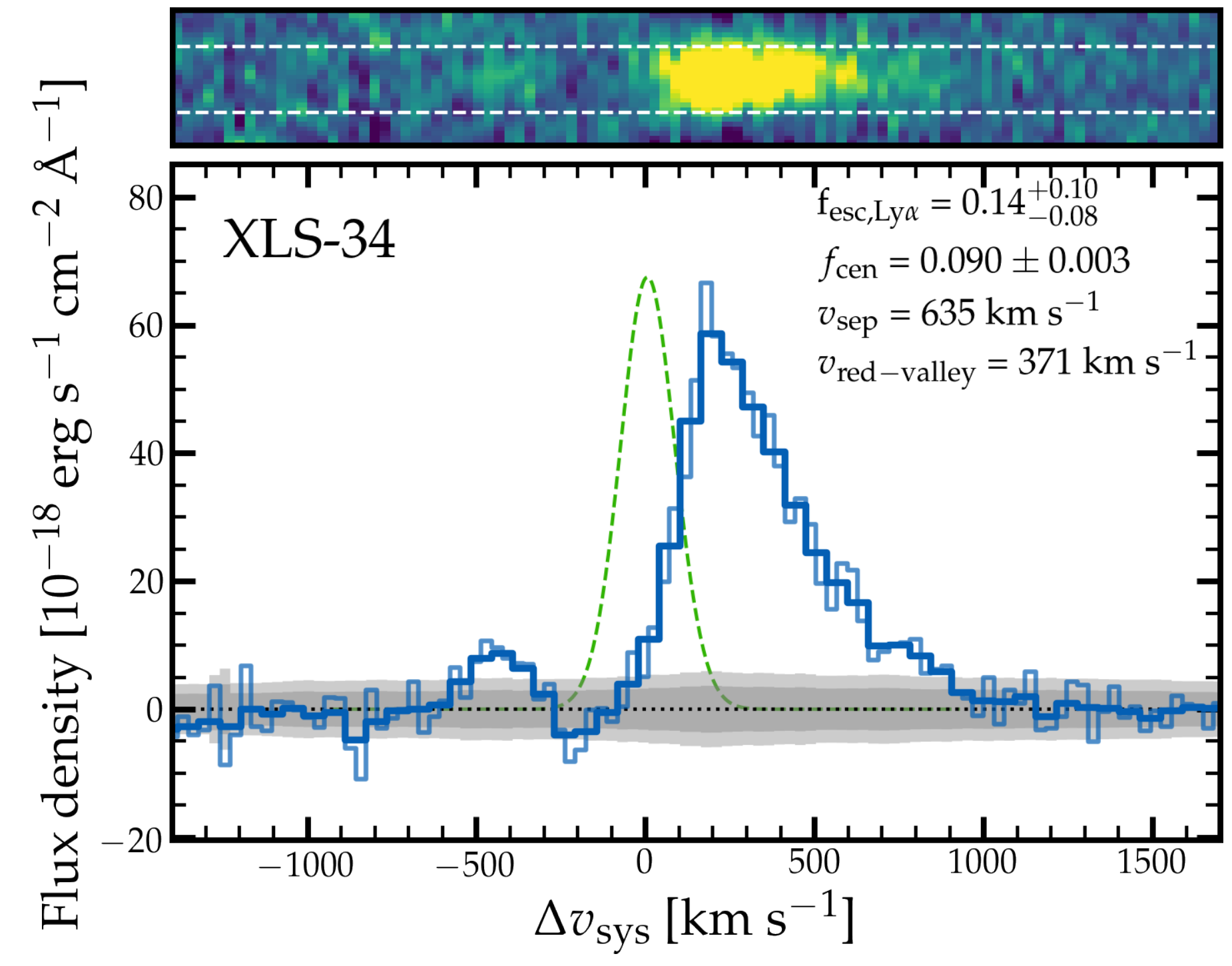} 
    \end{tabular}
    \caption{As Fig. $\ref{fig:Lya_profiles_HIGH}$, but now showing individual Lyman-$\alpha$ line-profiles of the LAEs classified in the Intermediate Escape subset. }
    \label{fig:Lya_profiles_Mid}
\end{figure*}

\begin{figure*}
    \begin{tabular}{ccc}
    \includegraphics[width=5.3cm]{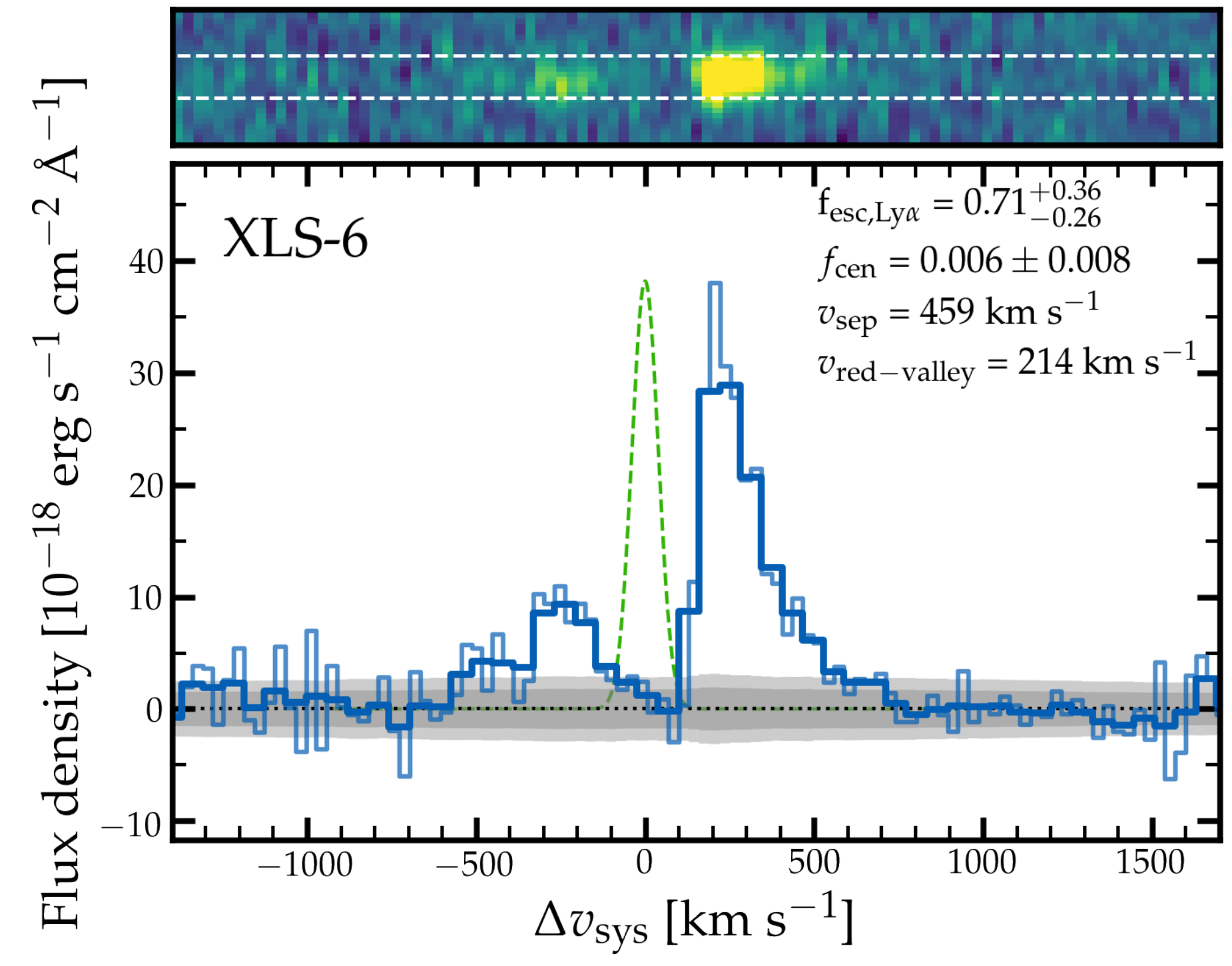} &
    \includegraphics[width=5.3cm]{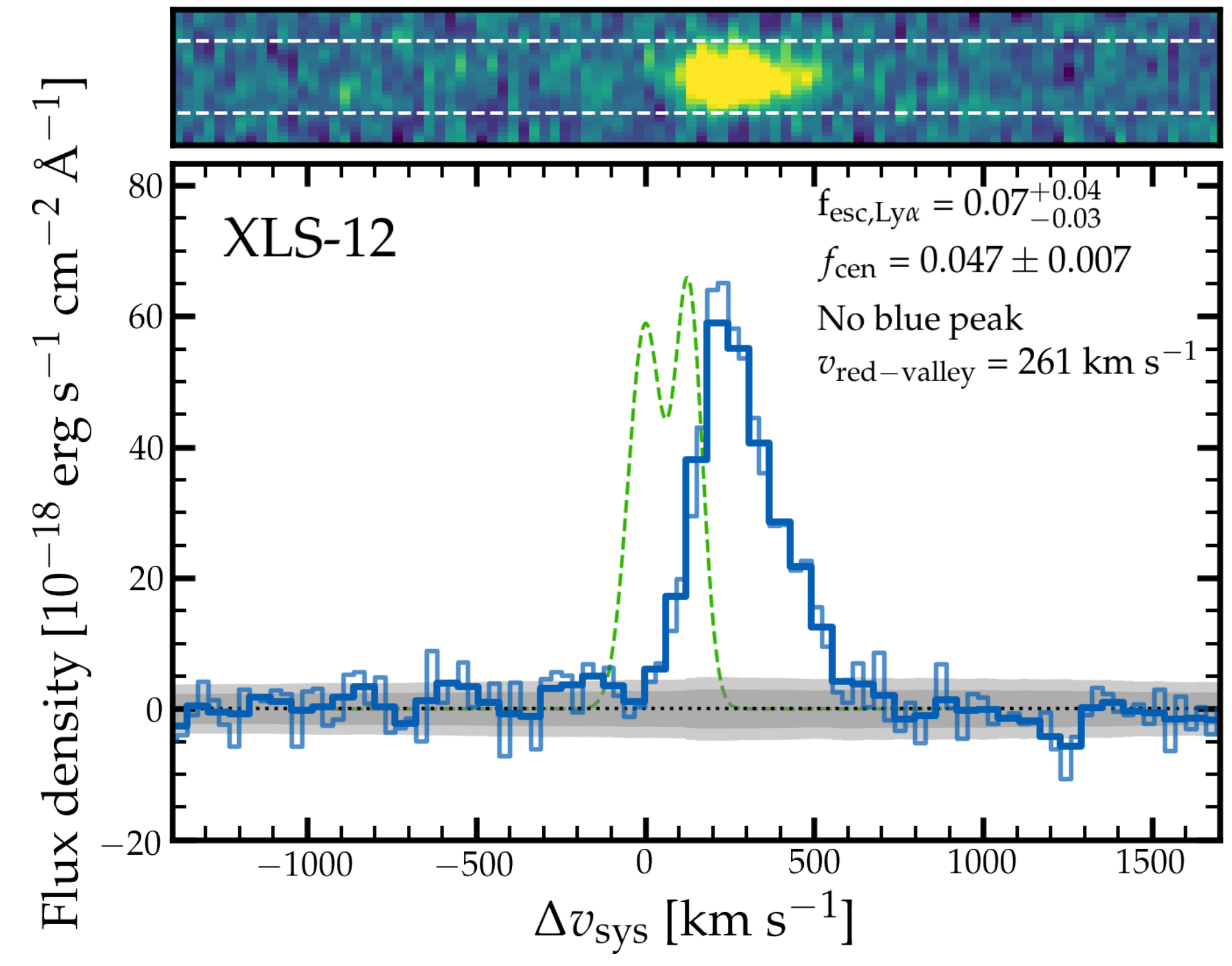} &
    \includegraphics[width=5.3cm]{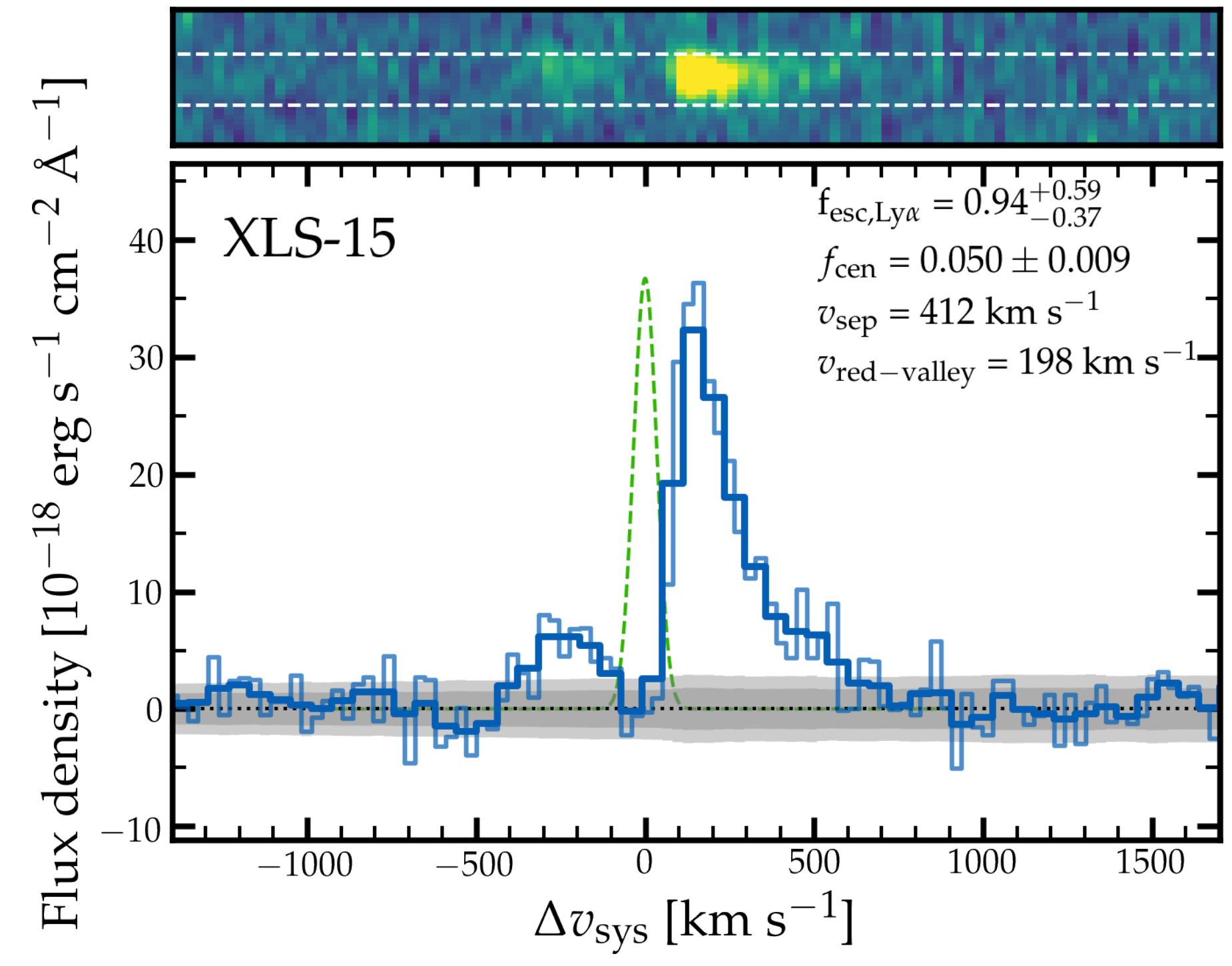} \\

    \includegraphics[width=5.3cm]{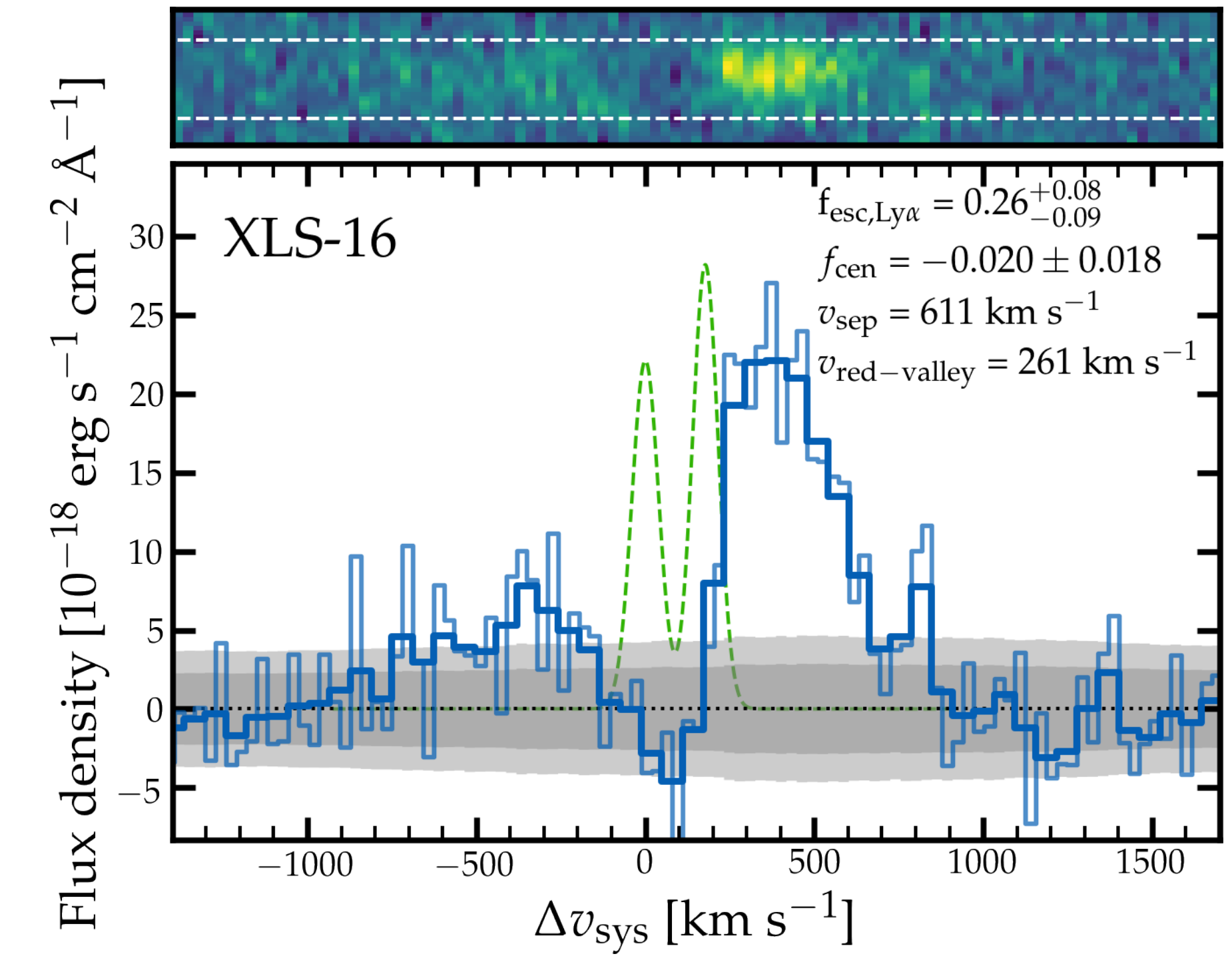} &
    \includegraphics[width=5.3cm]{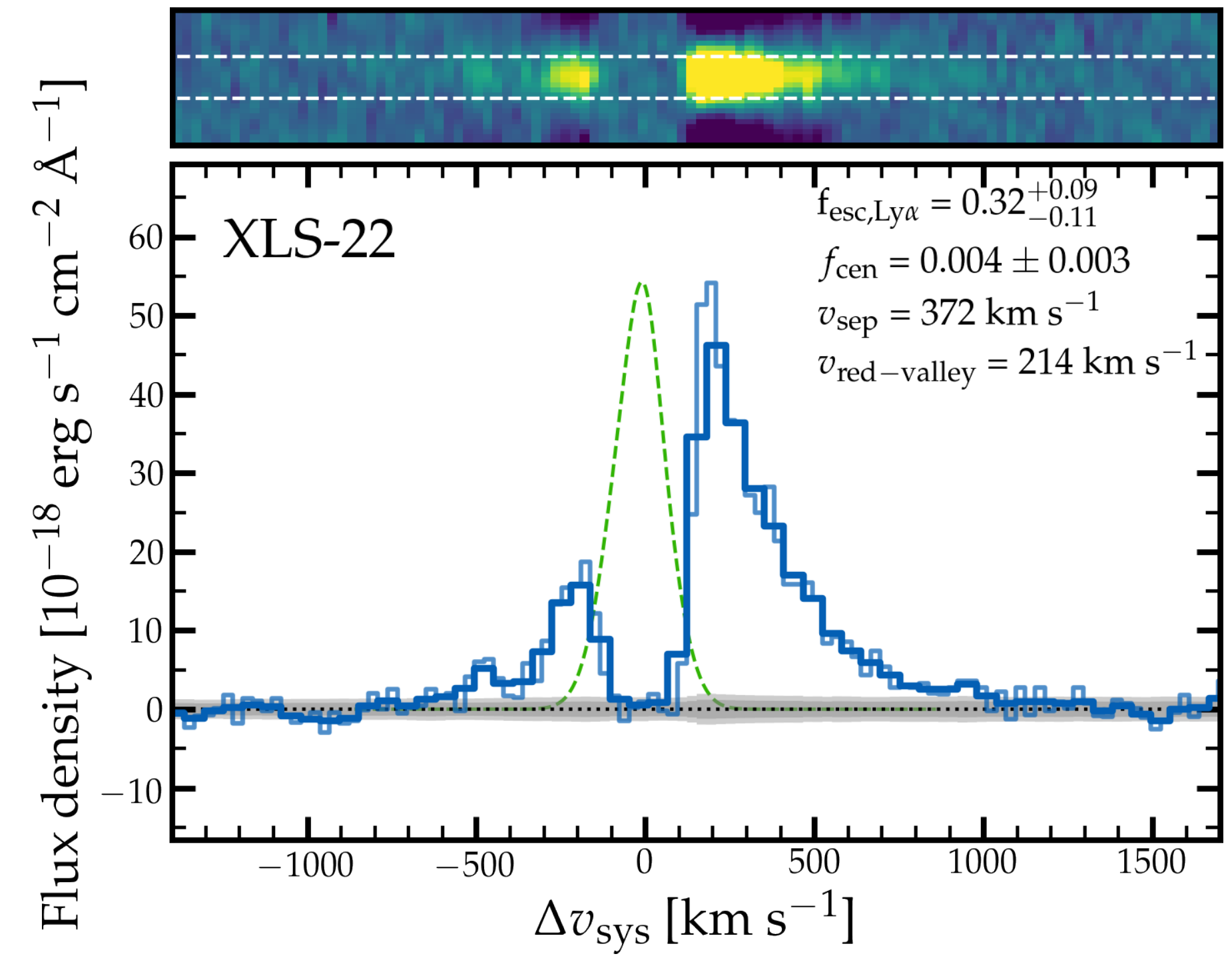} &
    \includegraphics[width=5.3cm]{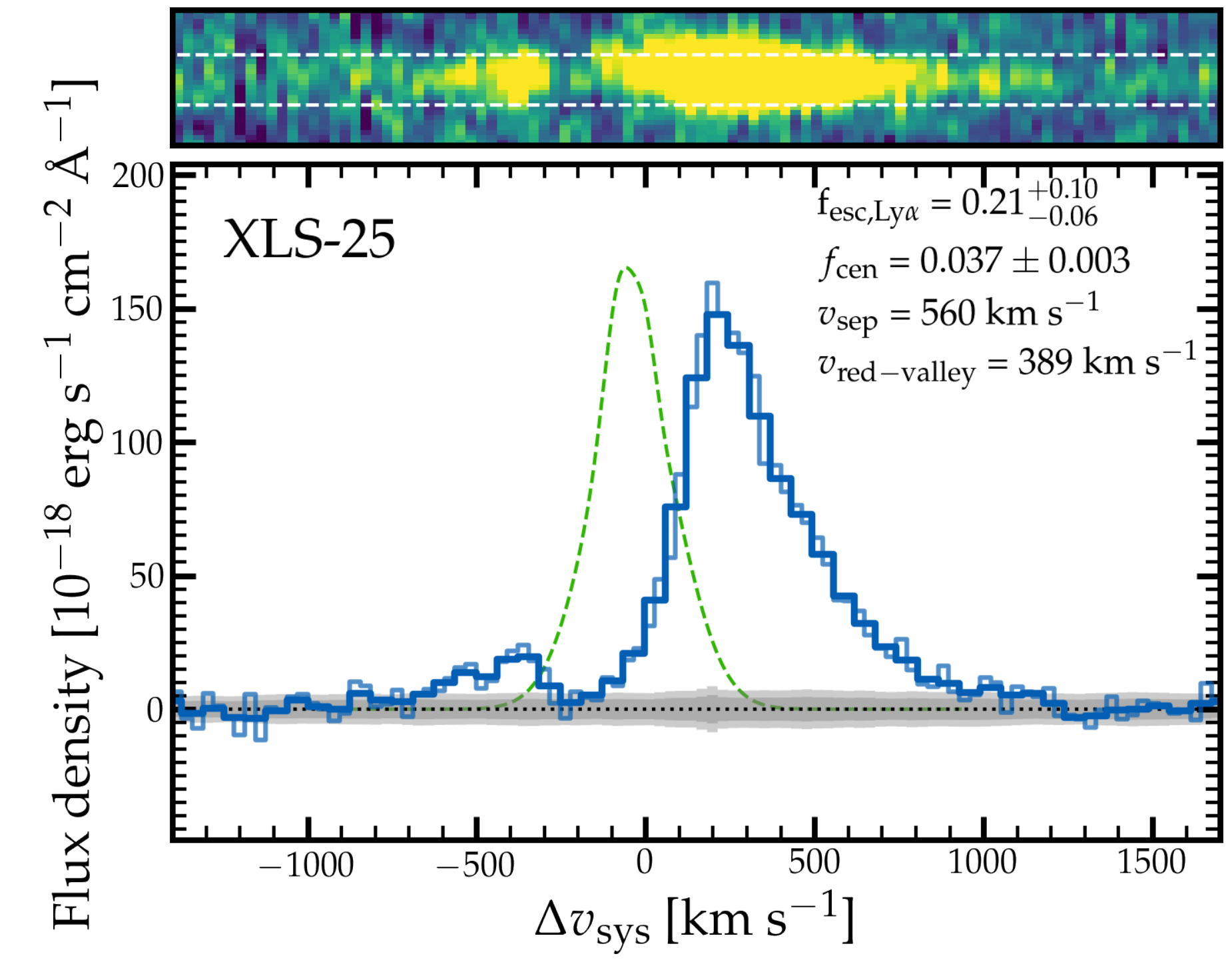} \\
    
    \includegraphics[width=5.3cm]{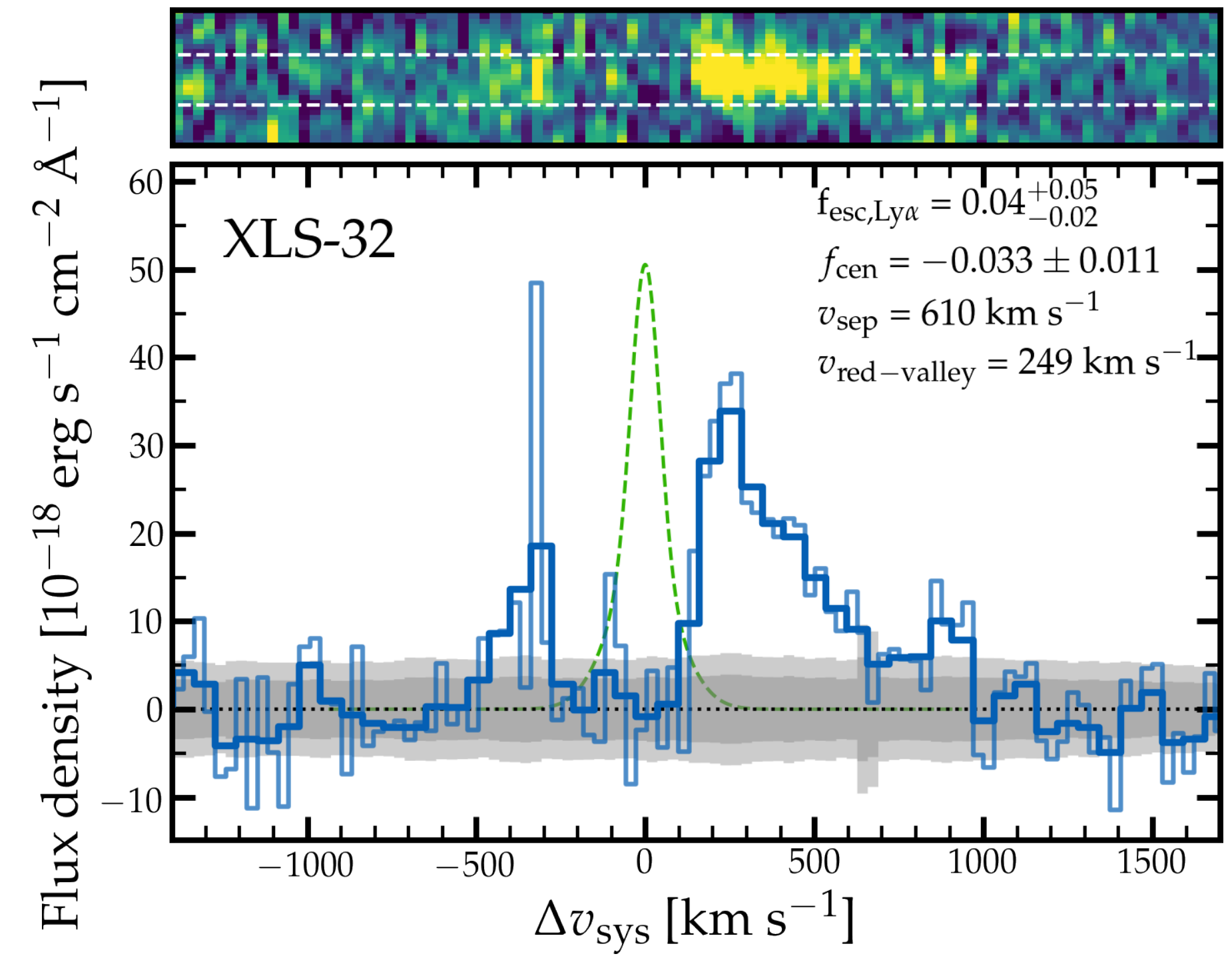} &
    \includegraphics[width=5.3cm]{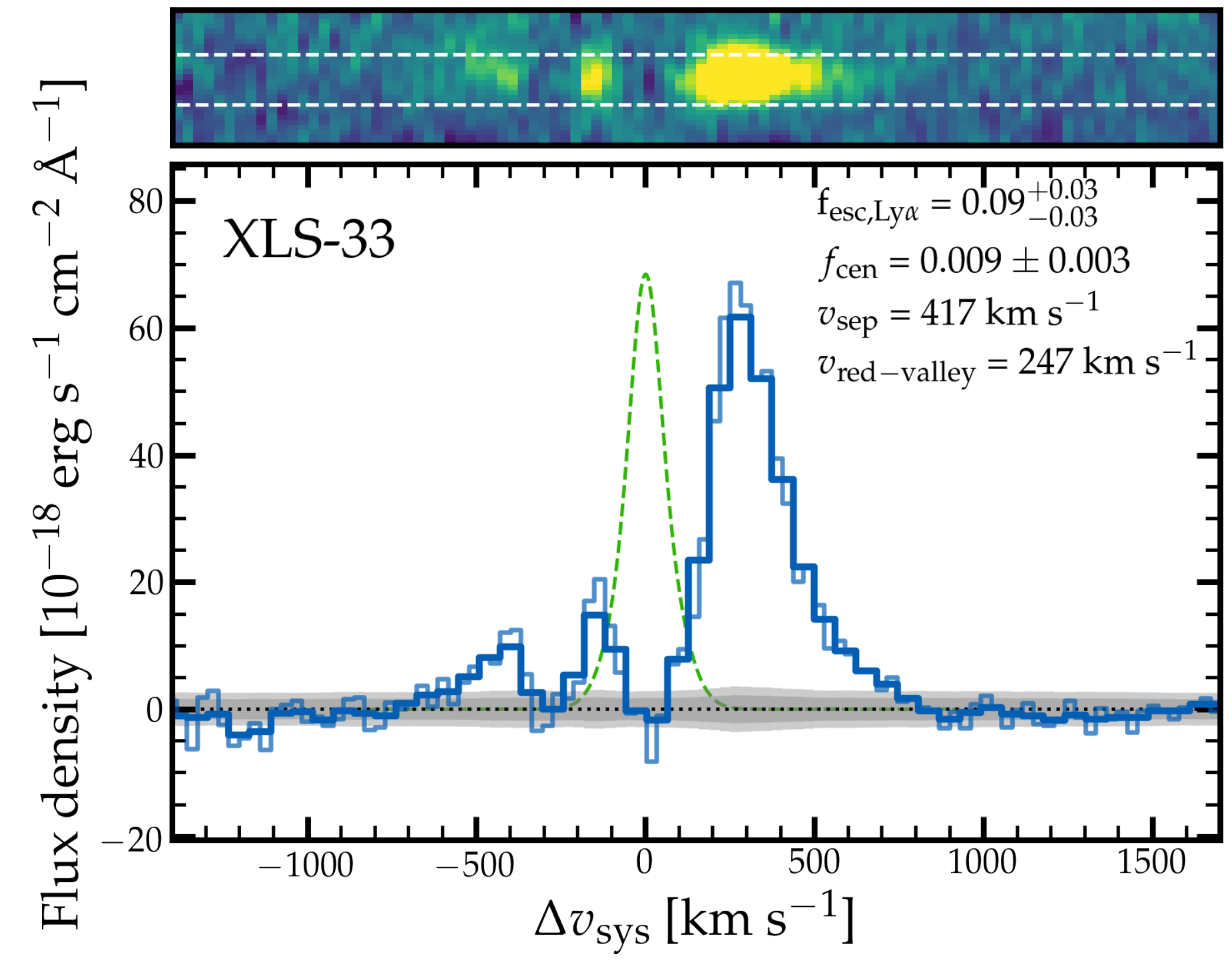} &
    \includegraphics[width=5.3cm]{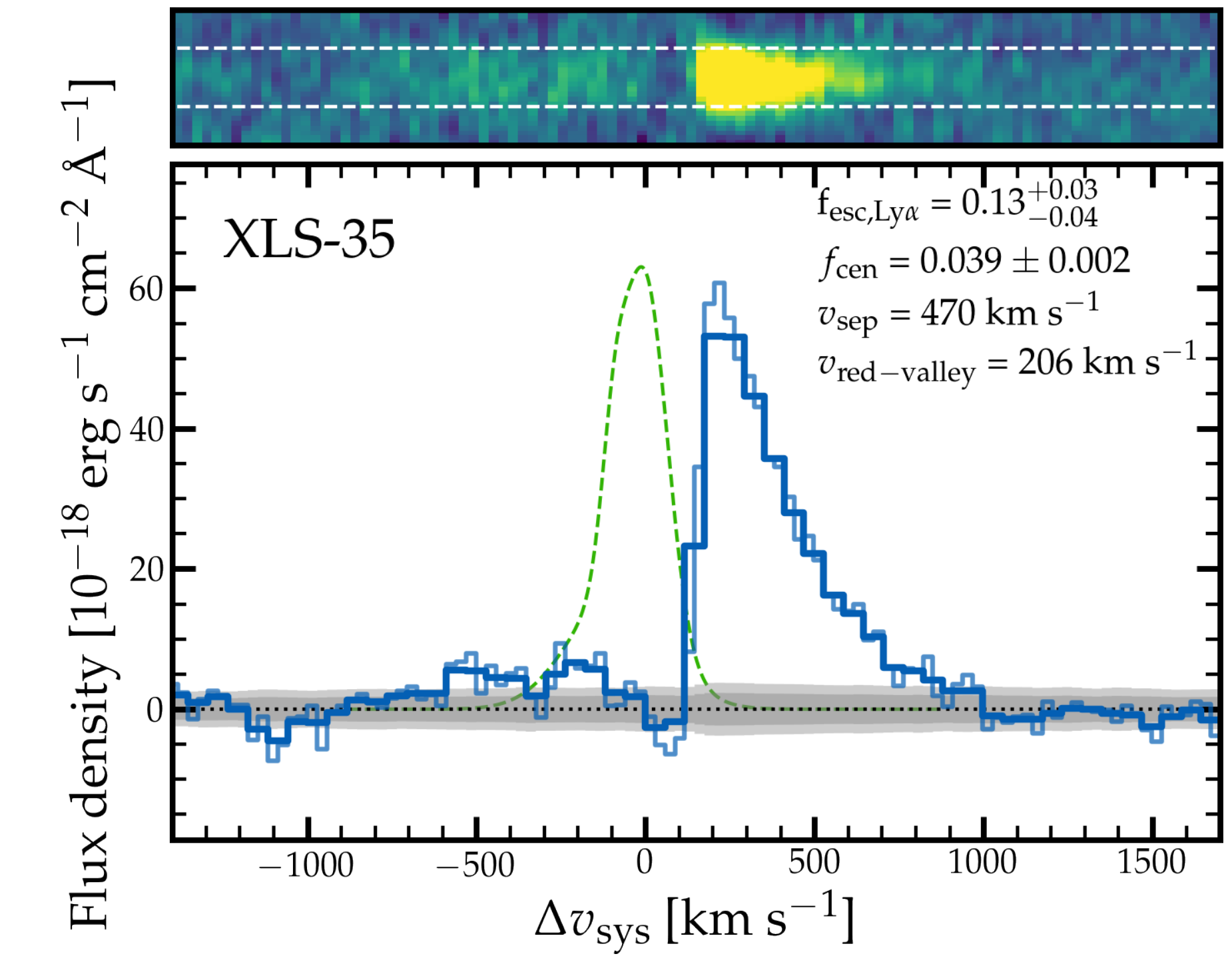} \\
    \end{tabular}
    \caption{As Fig. $\ref{fig:Lya_profiles_HIGH}$, but now showing individual Lyman-$\alpha$ line-profiles of the LAEs classified in the Low Escape subset. }
    \label{fig:Lya_profiles_Low}
\end{figure*}

Figure $\ref{fig:Lya_profiles_HIGH}$ shows the individual Ly$\alpha$ profiles of the XLS-$z2$ LAEs classified in the High Escape subset, while Figures $\ref{fig:Lya_profiles_Mid}$ and $\ref{fig:Lya_profiles_Low}$ show the profiles of the LAEs in the Intermediate and Low Escape subsets, respectively. In each panel we list the Ly$\alpha$ escape fraction of the LAE, the fraction of their Ly$\alpha$ emission that escapes within $\pm100$ km s$^{-1}$ of the systemic redshift ($f_{\rm cen}$), the peak separation (when applicable) and the velocity difference between the peak of the red line and the nearest minimum blueward of it (i.e. the valley; $v_{\rm red-valley}$). For XLS-20 we do not list $v_{\rm red-valley}$ because its Ly$\alpha$ line does not show a distinct asymmetric red peak. For the LAEs in the High Escape subset, we also list the reason why they have been classed in that subset.

We note that in some cases multiple velocity components are detected in the rest-frame optical spectra, such as a broad component (e.g. XLS-18, XLS-24) or a second narrow component (e.g. XLS-12, XLS-16, but also XLS-25 and XLS-35 where they are blended). In these cases, as described in \cite{Matthee21}, the systemic redshift is placed at the component that is spatially closest associated with the Ly$\alpha$ emission. XLS-33 shows triple-peaked Ly$\alpha$ emission, but this is likely due to an absorbing system at $\approx-400$ km s$^{-1}$ (see also XLS-18 for another such example), instead of a peak at the systemic velocity.

\renewcommand\thefigure{B.\arabic{figure}}

\section{Comparison between edge ($850-912\AA$) and total ($0-912\AA$) \fesc} \label{appendix:edgetotal}

Here we demonstrate the difference between the edge \fescs and total \fescs with SEDs matched to the High Escape stack's highly ionizing nature and with LyC transmission curves for $0-912\rm{\AA}$. The LyC transmission curves from \citet{McCandliss17} based on theoretical photionization cross-sections are a function of three parameters: the column density ($N_{\rm{HI}}$), the neutral Hydrogen fraction ($X_{\rm{HI}}$), and neutral Helium fraction ($X_{\rm{HeI}}$). In the bottom panels we see that even for column densities of $\approx10^{18}$ cm$^{-2}$ where the edge \fescs is $\approx0$, the 0-912$\AA$ \fescs can be $\approx20\%$. We also note that free-bound emission of H which peaks shortward of the LyC limit may also affect the determination of the LyC escape fractions, such that reported values may be currently overestimated since they do not account for this \citep[e.g.,][]{Inoue10}. Refined estimates for the edge-to-total correction will require future studies.

\begin{figure*}
\centering
\includegraphics[width=\linewidth]{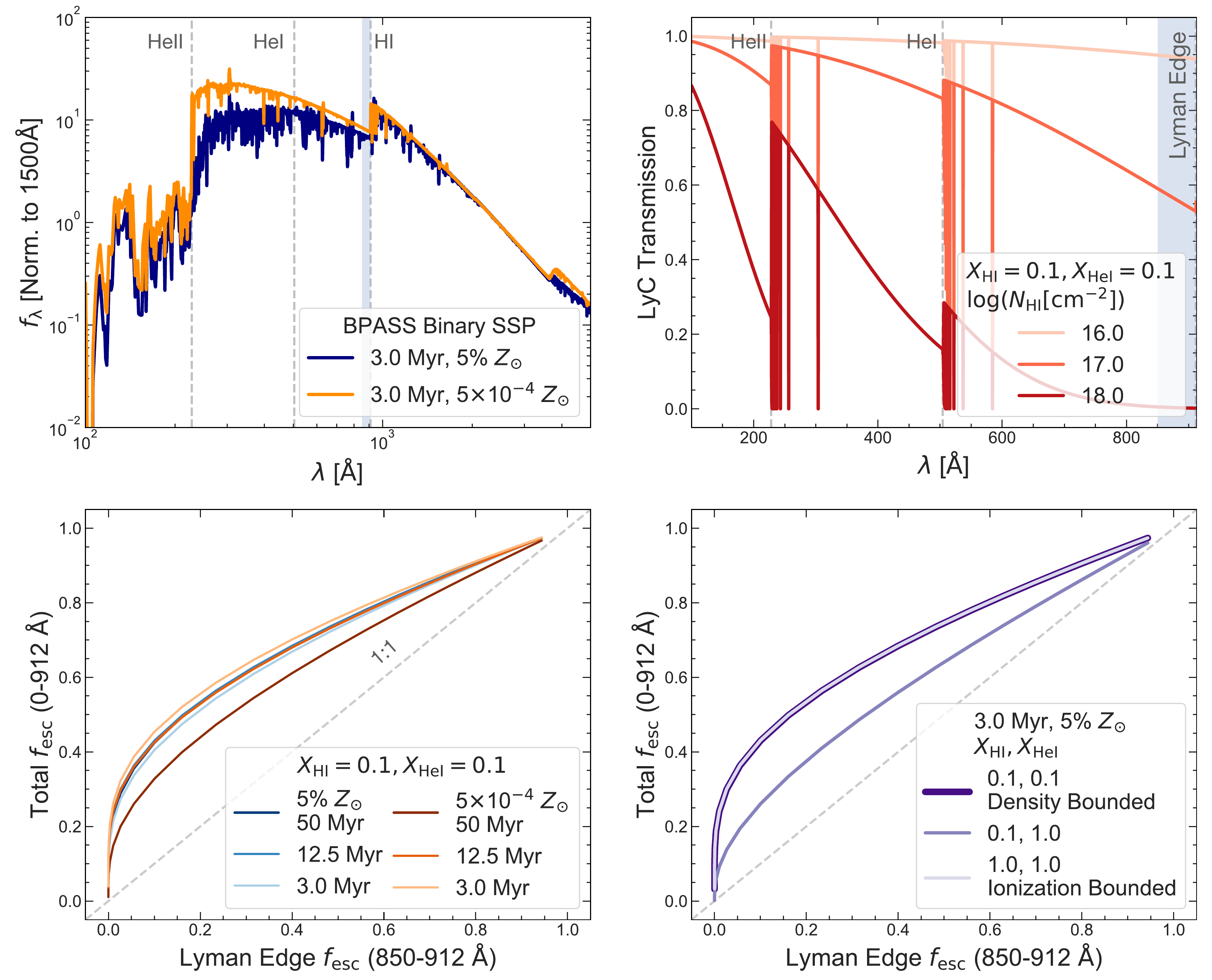}
\caption{The difference between Lyman edge \fescs ($850-912\AA$) and total \fescs ($0-912\AA$). \textbf{Top left:} BPASS SEDs selected to illustrate the hard ionizing nature of the High Escape stack.  \textbf{Top right:} LyC transmission curves from \citet{McCandliss17} for $\log{(N_{\rm{HI}})}$ and neutral fractions relevant to the semi-permeable portions of density bounded nebulae (see \S\ref{sec:feschard}). \textbf{Bottom left:} The LyC \fescs computed over the entire ionizing spectrum compared to the \fescs measured at the Lyman edge as a function of SED age and metalliicity. These curves are a result of convolving the SEDs with LyC transmission curves and comparing against the intrinsic SED. The total \fescs lies significantly above the $1:1$ line (dashed gray). Remarkably, even for meagre edge \fesc ($<5\%$), significant total \fescs ($20-40\%$) is possible. Varying metallicity and age has little effect on the edge to total \fescs conversion. \textbf{Bottom right:} Same as bottom-left panel, but exploring the effect of neutral fractions. Somewhat unintuitively, what matters to these curves is not the absolute Hydrogen and Helium neutral fractions, but the fractions \textit{relative} to each other (e.g., the ionization bounded and density bounded cases are indistinguishable).}
\label{fig:hardphotons}
\end{figure*}

\section{Author affiliations} \label{appendix:affiliations}
$^{3}$Department of Astronomy, University of Geneva, Chemin Pegasi 51, 1290 Versoix, Switzerland\\
$^{4}$Cosmic Dawn Center (DAWN), Niels Bohr Institute, University of Copenhagen, Jagtvej 128, K\o benhavn N, DK-2200, Denmark\\
$^{5}$Department of Physics, Lancaster University, Lancaster, LA1 4YB, UK\\
$^{6}$Kapteyn Astronomical Institute, University of Groningen, Landleven 12, 9747 AD Groningen, The Netherlands\\
$^{7}$Stockholm University, Department of Astronomy and Oskar Klein Centre for Cosmoparticle Physics,\\ AlbaNova University Centre, SE-10691, Stockholm, Sweden\\
$^{8}$Center for Gravitation, Cosmology, and Astrophysics, Department of Physics, University of Wisconsin-Milwaukee,\\3135 N. Maryland Avenue, Milwaukee, WI 53211, USA\\
$^{9}$Instituto de Investigaci\'on Multidisciplinar en Ciencia y Tecnolog\'ia, Universidad de La Serena, Ra\'ul Bitr\'an 1305, La Serena, Chile\\
$^{10}$Departamento de Astronom\'ia, Universidad de La Serena, Av. Juan Cisternas 1200 Norte,  La Serena, Chile\\
$^{11}$Department of Physics \& Astronomy, Johns Hopkins University, Bloomberg Center, 3400 N. Charles St., Baltimore, MD 21218, USA\\
$^{12}$Hubble fellow\\
$^{13}$Max Planck Institut fur Astrophysik, Karl-Schwarzschild-Strasse 1, D-85748 Garching bei M\"unchen, Germany\\
$^{14}$Department of Physics, Ulsan National Institute of Science and Technology (UNIST), Ulsan 44919, Republic of Korea\\
$^{15}$ Centro de Astrof\'{\i}sica e Gravita\c c\~ao  - CENTRA, Departamento de F\'{\i}sica, \\ Instituto Superior T\'ecnico - IST, Universidade de Lisboa - UL, Av. Rovisco Pais 1, 1049-001 Lisboa, Portugal\\
$^{16}$Instituto de Astrof\'isica de Canarias, E-38200 La Laguna, Tenerife, Spain\\
$^{17}$Departamento de Astrof\'isica, Universidad de La Laguna, E-38205 La Laguna, Tenerife, Spain\\
$^{18}$Leiden Observatory, Leiden University, PO\ Box 9513, NL-2300 RA Leiden, The Netherlands\\

\bsp	
\label{lastpage}
\end{document}